\documentclass{ws-ijmpa}

\newcommand{\be}{\begin{equation}}
\newcommand{\ee}{\end{equation}}
\newcommand{\ba}{\begin{eqnarray}}
\newcommand{\ea}{\end{eqnarray}}
\newcommand{\siml}{\lower4pt \hbox{$\buildrel < \over \sim$}}
\newcommand{\simg}{\lower4pt \hbox{$\buildrel > \over \sim$}}
\def\Mesz{M\'esz\'aros}

\begin{document}
%\tableofcontents

\markboth{B. Zhang \& P. \Mesz}
{Gamma-ray bursts: progress, problems \& prospects}

\catchline{}{}{}

\title{GAMMA-RAY BURSTS:\\ PROGRESS, PROBLEMS \& PROSPECTS
%INSTRUCTIONS FOR TYPESETTING MANUSCRIPTS\\
%USING \TeX\ OR \LaTeX\footnote{For the title, try not to use more
%than 3 lines. 
%Typeset the title in 10 pt roman, uppercase and 
%boldface.} 
}

\author{\footnotesize BING ZHANG\footnote{Also Canadian Institute of
Theoretical Astrophysics} and PETER M\'ESZ\'AROS\footnote{Also
Department of Physics, Pennsylvania State University; currently on
leave at The Institute for Advanced Study, Princeton NJ 08540}
%Typeset names in 8 pt roman, uppercase. Use the footnote to indicate the
%present or permanent address of the author.}
}

\address{Department of Astronomy \& Astrophysics\\
The Pennsylvania State University\\ 525 Davey Lab, University Park, PA
16802, USA
%\footnote{State completely without abbreviations, the
%affiliation and mailing address, including country. Typeset in 8 pt
%italic.}
}

%\author{SECOND AUTHOR}

%\address{Group, Laboratory, Address\\
%City, State ZIP/Zone, Country
%}

\maketitle

\pub{Received (Day Month Year)}{Revised (Day Month Year)}

\begin{abstract}

The cosmological gamma-ray burst (GRB) phenomenon is reviewed.  The
broad observational facts and empirical phenomenological relations of
the GRB prompt emission and afterglow are outlined.  A well-tested,
successful fireball shock model is introduced in a pedagogical
manner. Several important uncertainties in the current understanding
of the phenomenon are reviewed, and prospects of how future
experiments and extensive observational and theoretical efforts may
address these problems are discussed.

\end{abstract}

\section{Introduction} 

The study of astrophysical objects is overwhelmingly done by
astronomers using the temporal and spectral information contained in
the electromagnetic signals that these objects emit. Gamma-ray bursts
(GRBs) are, by definition, electromagnetic signals in the gamma-ray
band (in the spectral domain) with short durations (in the temporal
domain). They are, however, unusual in having most of their
electromagnetic output in gamma-rays, typically at sub-MeV energies,
and having most of it concentrated into a brief episode, typically
lasting tens of seconds.

The road leading to an understanding the nature of these objects has
been bumpy, mainly due to the limited information contained in these
abrupt gamma-ray episodes. For comparison, almost all the other
astrophysical objects are observed in a broader spectral band during a
much longer observation time. Historically, the lack (until recently)
of observational breakthroughs on GRBs had left ample freedom for
modelers to play around, so that the cumulative list of models
championed to interpret GRBs has been more numerous than for any other
astrophysical phenomenon.  As in other fields in science, experimental
(or observational) progress eventually constrains and eliminates most
predictive models.
At a certain stage of the development of a field, it is useful
to ask oneself questions such as what is it that we know, what is it
that we still do not know, and how could we find the answers for the
unknowns. This is the main emphasis of this review. There exist
already a number of comprehensive reviews on GRB
observations\cite{fm95,vkw00,kul00}, on theories of GRB\cite{p99,m02}
and on some general or selected
topics\cite{m01,cl01,wzh02,hsd02,dermer02,zmw02,ghis03,d03,w03,mkrz03}.
Our purpose here is to provide a review which differs from others in
several aspects. For example, we summarize the current observational
progress in an organized, itemized form, making a distinction between
solid facts and empirical laws. The introduction of the standard
theoretical model is pedagogical, the content of which overlaps other
reviews. However, we also dedicate a lot of space to discuss the
uncertainties in the current models and highlight some active debates
that are going on in the community.  We also attempt to reflect in an
unbiased manner the work from various groups, rather than focusing on
our own work, although some ``selection effects'' must still exist due
to our incomplete survey of the literature.

The structure of the review is as follows.  We start with a brief
summary of the GRB research history as well as the unique role of GRB
study in general astrophysics. We then compile the observational facts
and empirical laws (\S\ref{sec:prog1}), and introduce the standard
theoretical framework, i.e. the fireball shock model
(\S\ref{sec:prog2}). In \S\ref{sec:prob} we outline some outstanding
issues in the GRB field, which comprises most of the hot topics (and
sometimes hot debates) in the current GRB scene. These issues include
whether the fireball is dominated by the conventional (hadronic)
kinetic energy content or by the less-studied magnetic energy; whether
the GRB prompt emission occurs before the fireball deceleration
(internal) or at the deceleration radius (external); whether the GRB
jets are uniform or with a quasi-universal angular structure; whether
the GRB progenitors give rise to a one-step or to a (long delay)
two-step collapse, etc. After outlining these issues, we attempt to
foresee in \S\ref{sec:pros} developments in the upcoming new era for
GRB study, led by future missions such as {\em Swift} and {\em GLAST},
complemented by other space- and ground-based detectors for both
electromagnetic and non-electromagnetic signals.  We review some of
the theoretical predictions related to these observational
capabilities and discuss how future observational campaigns could
constrain and possibly settle some of the outstanding issues listed in
\S\ref{sec:prob}.

\subsection{Major Landmarks in the Study of GRB}

As a framework for the following discussions, we list below some of
the most important events in the development of the understanding of
GRB.  There may be many other important events which are not included
in this list, and we apologize for any omission. Most of the papers
selected below are marked by their high citation counts in the ADS
archives, an indication that the works are significant and widely
accepted. For the theoretical papers, we only include those that
survived the later observational data and correctly contributed to the
current knowledge of the GRB phenomenon, or those which, although not
fully tested yet, represent ``popular" views and draw broad attention
from the GRB community.  The papers extend up to October 2003, when
this review is completed.  They include both observational and
theoretical developments, which have mutually influenced, and have
benefited from, each other.

A list of key advances on the observational side is:
\footnote{Notice that all the afterglow-related breakthroughs are
for the so-called ``long'' GRBs (see \S\ref{sec:prog1} for
definition).  No afterglow from a ``short'' GRB has so far been firmly
identified.}.

\begin{itemlist}
\item GRBs were discovered serendipitously in the late 1960s, and the
data were reported several years later\cite{kso73,m74}; 
\item The {\em Compton Gamma-ray Observatory} ({\em CGRO}) was
launched in 1991, with the {\em Burst and Transient Experiment} ({\em
BATSE}) on board; {\em BATSE} provided evidence for an isotropic
spatial distribution of GRBs, giving significant support to a
cosmological origin interpretation\cite{m92};
\item Two categories of GRBs, ``long'' ($t_\gamma\simg 2$ s) and ``short'',
($t_\gamma \siml 2$ s) were identified\cite{k93};
\item Systematic analyses of GRB spectral data indicate that a
so-called ``GRB function'' (or ``Band function'') fits reasonably well
most of the GRB spectra (for both classes of bursts)\cite{b93};
\item Another detector on board of {\em CGRO}, the Energetic Gamma Ray
Experiment Telescope ({\em EGRET}), detected some bursts in the hard
gamma-ray band; one burst, GRB 940217, was detected to have long-lived
GeV emission extending for 1.5 hrs\cite{h94};
\item A signature consistent with cosmological time dilation was
detected\cite{norris94}; 
\item In 1997, the Italian-Dutch satellite {\em Beppo-SAX} pinpointed 
the first GRB low energy afterglow (for GRB 970228) in the X-ray
band\cite{c97}, facilitating optical\cite{van97} and radio\cite{f97}
detections;
\item The first measurments of the GRB redshifts were obtained (for GRB
970508\cite{m97} and GRB 971214\cite{k98}, giving a solid proof that
GRBs are at cosmological distances;
\item A likely GRB-supernova association (GRB 980425 vs. SN 1998bw)
was discovered\cite{g98,k98b};
\item A bright and prompt optical flash and a radio flare were
discovered to accompany the energetic burst GRB 990123\cite{a99,k99a};
\item An achromatic steepening break in the afterglow lightcurves was
found in several bursts, hinting that at least some GRB fireballs are
likely to be collimated\cite{k99,h99};
\item X-ray spectral features with moderate significance were
discovered in several GRB X-ray afterglows\cite{p00,r02};
\item Assuming the lightcurve steepening is due to the jet break, a
collection of the burst jet data revealed that GRBs have a
quasi-standard energy reservoir\cite{f01};
\item A class of so-called ``X-ray flashes'' (XRFs) were
identified\cite{heise,kippen}, which appear to be closely related to
GRB;
\item Prompt optical afterglows from two more GRBs, GRB 021004 and GRB
021211, were detected by ground-based robotic
telescopes\cite{f03a,f03b,li03};
\item The prompt gamma-ray emission was reported to be strongly polarized in 
GRB 021206\cite{cb03} (cf Refs.\refcite{rf03,bc03});
\item The first clear GRB-SN association was identified: an unambiguous
supernova light-curve bump and spectrum in the GRB 030329's optical
afterglow lightcurve was discovered\cite{sta03,h03};
\item A distinct high energy spectral component which appears to evolve
separately from the usual sub-MeV component was identified in GRB
941017\cite{g03};
\item Broadband observations of GRB 030329 suggest the there might
exist a multi-component jet structure in some bursts, the total energy
output being standard\cite{b03,sheth03}.
\end{itemlist}

On the theoretical side, some of the major steps in the development
towards the contemporary ``standard'' model are listed below.  This
list does not include many neutron-star-based Galactic GRB models (see
Ref.~\refcite{h91} for a review), many of which were proposed before
the proofs of a cosmological origin of GRBs became solid.

\begin{itemlist}
\item A generic argument about the ``compactness problem'' was
raised\cite{r75} in 1975;
\item A self-similar solution of the relativistic blast waves was
found in 1976\cite{bm76}. Although this study was not aimed at GRBs,
it laid a mathematical foundation for relativistic GRB blast wave
models, and has been extensively applied since the discovery of GRB
afterglows;
\item Some generic features of a gamma-ray fireball were discussed in
1978\cite{cr78};
\item A cosmological model of GRBs was formally proposed in 
1986\cite{p86,g86,p90}, involving neutron star - neutron star (NS-NS)
mergers as the power source. A pure photon-pair fireball was found to
expand relativistically, leaving a quasi-thermal photospheric
emission;
\item It was found that by including a small amount of baryon
contamination, the fireball thermal energy is essentially converted
into kinetic energy of the expanding gas\cite{sp90};
\item It was pointed out that the kinetic energy is re-converted to
heat and radiation via shocks\cite{rm92,mr93}. This suggestion laid
the foundation for the current fireball-shock models. The specific
scenario proposed in Refs.~\refcite{rm92,mr93} involves the shock
forming when the fireball is decelerated by the ambient interstellar
medium (ISM), which is known as the ``external shock scenario''.
Synchrotron radiation was proposed as the mechanism of GRB emission,
noting that this would soften and fade as a power law in time.
\item Around the same time, several major GRB progenitor models were
further developed or proposed: the NS-NS merger model was further
studied in some detail\cite{e89,npp92}; a ``failed'' supernova model
was proposed\cite{w93}, which is now known as the ``collapsar'' model;
and a millisecond magnetar (neutron stars with surface magnetic field
of order $10^{15}$ G) model was proposed\cite{u92};
\item Another fireball shock scenario involving collisions among the
individual shells in the fireball outflow, known as the ``internal
shock scenario'' was proposed\cite{rm94,px94}, as a way to resolve the
very short time variability problem.
\item The hydrodynamical evolution of a fireball was extensively
modeled both numerically and
analytically\cite{mlr93,mr93b,psn93,katz94,sp95,kps99,mr00,mrrz02};
\item The detailed reverse and forward shock synchrotron and inverse
Compton spectral components of external shocks were
explored\cite{mrp94,mr94}, indicating the presence of a reverse shock
optical synchrotron component (now identified as responsible for
prompt optical flashes) and a GeV component due to inverse Compton.
\item Based on the synchrotron radiation assumption, GRB radio and
optical afterglows were suggested\cite{pr93,k94}. A detailed
self-consistent afterglow calculation\cite{mr97} stressing also the
X-ray afterglow was performed shortly before the observational
discovery of afterglows.  The predictions include the power-law decay
of the optical lightcurves as well as prompt optical flashes arising
from the reverse external shock during the early stage of the
afterglows. All these were verified by the later
observations\cite{van97,a99};
\item Prompted by the discovery of the first GRB
afterglows\cite{c97,van97,f97} the fireball model was studied and
tested extensively, and found to be successful in interpreting the
observations\cite{wrm97,v97,w97a,w97b,t97};
\item It was suggested\cite{p98} that GRBs are likely originated from
``hypernovae'', and are associated with star forming regions;
\item The simplest standard afterglow model was set up\cite{mrw98,spn98}, 
and was further developed by incorporating additional
effects\cite{mrw98,rm98,dl98,dl99,cl99};
\item Stimulated by the increasing evidence that GRBs are associated
with supernovae, the GRB progenitor models involving collapses of
massive stars were further developed. The ``collapsar model'' was
extensively studied numerically\cite{mw99,a00,zwm03} and
analytically\cite{mr01,wm03}. A variant of the massive-star-collapse
model, invoking a delayed GRB with respect to the SN explosion, was
proposed\cite{vs98}, known as the ``supranova'' model;
\item The scenario involving GRB prompt gamma-rays from internal shocks 
and an afterglow from the external shock as a unified sequence
requiring a joint fit was analyzed\cite{sp97a,sp97b}. In parallel to
this, extensive efforts were made to model the GRB prompt emission
both within the internal shock
models\cite{kps97,dm98,psm99,spm00,gsw01} and the external shock
models\cite{pm98,cd99,dm99,dbc99};
\item The consequences of GRBs originating from collimated jets was 
proposed and studied\cite{r97,r99}, leading to an extensive campaign
of theoretical jet modeling\cite{sph99,pm99,msb00,hgdl00};
\item Prompted by the detection of the optical flashes associated with
GRB 990123\cite{a99}, GRB 021004\cite{f03a} and GRB
021211\cite{f03b,li03}, the reverse shock emission was studied in
greater
detail\cite{sp99a,sp99b,mr99,ks00,k00,kz03a,zkm03,wdhl03,kz03b,kmz03}. The
recipes to use early afterglow data to systematically infer fireball
parameters were recently proposed\cite{zkm03,kz03b,kmz03};
\item The possibility of producing Fe X-ray lines in supranova and
in collapsar stellar funnel models was
investigated\cite{lcg99,rm00,weth00,mr01,vietri01,lrr02,kmr03}.
\item Broadband modeling of GRB afterglows was carried out,
leading to constraints on some unknown model
parameters\cite{wg99,pk01,pk02,harrison01,yost03};
\item Prompted by the finding of a standard energy reservoir in long
GRBs\cite{f01}, it was proposed that long GRBs may have a
quasi-universal structured jet configuration\cite{rlr02,zm02b,zdlm03}.
\end{itemlist}

\subsection{The multi-disciplinary nature of the GRB field}

The GRB field is almost unique in astrophysics in its
multi-disciplinary nature. Involving stellar-scale events located at
cosmological distances, GRBs straddle the traditional distance scales,
and are a high energy phenomenon which emits a broad-band
electro-magnetic spectrum extending over at least fifteen decades, as
well some possible non-electromagnetic signals (such as cosmic rays,
neutrinos and gravitational waves). This makes the GRB field an
intersection of many branches in astrophysics.

GRBs were discovered at about the same time when quasars (or broadly
speaking active galactic nuclei, AGNs) and pulsars, two other major
developments in the 1960's, were discovered. The nature of those two
classes of objects was agreed upon soon after their discoveries. AGNs
are extra-galactic sources believed to be powered by gigantic black
holes, while pulsars are compact neutron stars located in our
Galaxy. Since the lack of GRB observational breakthroughs hampered the
progress in this latter field, scientists from both the AGN and the
pulsar community brought to bear their collective wisdom from their
own fields to tackle the GRB problem. The GRB neutron star models were
developed to fairly sophisticated levels\cite{h91}, especially
motivated by the reported detection of absorption and emission
features in some GRB spectra\cite{m81,m88}. It turned out that the
``classical" GRBs are of cosmological origin, so that the wisdom
borrowed from the AGN community (especially about blazars, the most
energetic type of AGNs) finally bore the most rewarding
fruit. Nonetheless, neutron star models eventually turned out to be
partially correct, i.e., one sub-class of such models that invokes
ultra-strong magnetic fields\cite{dt92} turned out to be useful in
interpreting the class of soft, repeating GRBs\cite{} (which
appear to observationally distinct from the ``classical" cosmological
GRBs). These sources, known as ``soft gamma-ray repeaters'' (SGRs),
together with another type of objects called ``anomalous X-ray
pulsars'' (AXPs), are now believed to be ``magnetars'', neutron stars
with super-strong ($\sim 10^{15}$ G at surface) magnetic
fields\cite{td95,td96}. Even within the current paradigm, pulsar
models can still provide helpful insights for the understanding of
cosmological GRBs. For example, the recent claim of strong
polarization of gamma-ray prompt emission in GRB 021206\cite{cb03}
(cf. Refs. \refcite{rf03,bc03}) suggests a strongly magnetized central
engine. A similar conclusion of a strong but less
extreme magnetization was reached through a combined reverse-forward
shock emission analysis for GRB 990123\cite{zkm03} and GRB
021211\cite{zkm03,kp03}. This hints that 
magnetic fields and/or Poynting flux may play an essential role in
GRBs, and that some insights from pulsar wind nebula theories may turn
out to be important to unveil the GRB mystery. We discuss this in more
detail in \S\ref{sec:mag}.

Besides its close intimacy with the AGN and the pulsar fields, the GRB
field is also broadly related to several other branches of
astrophysics.  First, in the stellar context, since long GRBs are
related to the deaths of massive stars, as supported by many
observational facts, GRB study is closely related to the fields of
stellar structure and evolution\cite{heger03},
supernovae\cite{wzh02b,i98}, and supernova remnants (e.g. the external
shock invoked in GRB models is the relativistic version of the
supernova remnant shock\cite{a01,ed02,lp03}). Progenitor studies have
stimulated stellar population studies\cite{fwh99}. Central engine
studies, on the other hand, have greatly promoted studies about
mechanisms for extracting energy from accretion disks or spinning
black holes\cite{bz77,mr97b,pwf99,lwb00,npk01,van01}.  Second, within
the galactic context, GRB afterglow lightcurves\cite{cl99,wl00,dl02}
and spectral features\cite{p00,r02,s03} probe the properties
(e.g. density profile and chemical abundance) of the ambient
interstellar medium (ISM) or the prestellar wind. Studying the
properties of the GRB host galaxies as well as the GRB locations
within the host galaxies brings valuable information about the global
star forming history of the universe and the nature of GRB
progenitors\cite{bloom98,f99,bloom02}. Third, GRBs are also important
objects within the cosmological context. Not only can they play a role
similar to AGNs in probing the observed low-redshift universe, but
they may also form and can be detectable at much higher
redshifts\cite{lr00,cl00,bl02,gou03}. Thus, detecting high-$z$ GRBs
would allow us to see into deeper and earlier epochs of the universe,
and to probe how the universe ends its dark age and gets
re-ionized\cite{m-e03}. Fourth, in high energy astrophysics, the
origin of ultra high energy cosmic rays (UHECRs) has remained a
mystery for decades. Among many other models, a GRB
origin\cite{w95,v95} is a leading model for the so-called ``bottom-up
models''. In this scenario, cosmic ray protons are believed to be
accelerated in GRB shocks. The same processes can also generate high
energy neutrinos\cite{px94,wb97,bm00,mw01}, so that GRBs are currently
regarded by many as the top potential high energy neutrino sources for
large area neutrino telescopes being built. Finally, several GRB
progenitor scenarios are believed to generate gravitational wave
signals\cite{kp93,km03}, and GRBs are also one of the major targets
for large gravitational wave detectors.  On balance, it would appear
that hardly any other field has such a broad range of
interactions with other branches of astrophysics.

\section{Observational Progress: facts \& empirical
relations\label{sec:prog1}} 

GRB observational facts have been extensively reviewed, and here we
summarize them in an organized, itemized manner. We will try to
separate relatively solid observational facts from the empirical 
statistical laws which are derived from these facts. We divide our 
discussion into prompt emission and afterglow emission. 

\subsection{Prompt emission\label{sec:grb}}

\subsubsection{Solid facts\label{sec:grb1}}

The main characteristics of GRB prompt emission have been collected in
Ref. \refcite{fm95} (and references cited in that paper). Below we 
list the solid facts in itemized form, including also some recent 
discoveries in the field\footnote{We discuss here only the prompt 
gamma-ray (and X-ray) emission. Prompt optical flashes have been detected in
some bursts accompanied with the prompt gamma-ray emission. However,
since they are generally interpreted as due to emission from the
external reverse shock, we discuss the optical flashes in the
early afterglow sub-section below.}.

%\begin{romanlist}
\begin{itemlist}
\item Temporal properties:
\begin{itemlist}
\item Durations ($T$, technically defined according to different
criteria, e.g. $T_{90}$ or $T_{50}$ being the time interval within
which $90\%$ or $50\%$ of the burst fluence is detected): they span 5 
orders of magnitude, i.e. from $\simg 10^{-2}$ s to
$10^3$ s, typical values: $\sim 20$ s for long bursts and $\sim 0.2$ s
for short bursts; 
\item Lightcurves: very irregular. Some bursts consist of very 
erratic, spiky components, while others are smooth with one or a
few components. Some bursts contain distinct, well-separated emission
episodes. Figure \ref{fig:grb-lc} is a typical GRB lightcurve. A
compilation of different types of burst profiles can be found in
Ref.\refcite{fm95}; 
\begin{figure}
\centerline{\psfig{file=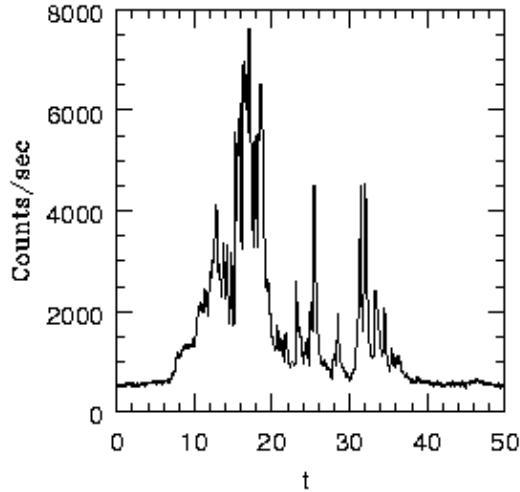,width=8cm}}
\vspace*{8pt}
\caption{Typical lightcurve of a BATSE GRB, showing photon count as a
function of time (in unit of second), from Ref.1. }
\label{fig:grb-lc}
\end{figure}
\item Widths of individual pulses ($\delta t$) vary in a wide
range. The shortest spikes have millisecond or even sub-millisecond
widths, and $\delta t/T$ could reach as low as $10^{-3}-10^{-4}$;
\item The vast majority of individual pulses are asymmetric, with
leading edges steeper than the trailing edges, although only a small
fraction can be visually descerned. By integral means GRBs are
asymmetric on all timescales\cite{nemiroff94}, and on
average\cite{norris02}. Smooth single peak bursts typically have a 
fast-rise--exponential-decay (FRED)-type lightcurve. This comprises
about $7\%$ of all bursts\cite{bhat94}.
\end{itemlist}
\item Spectral properties:
\begin{itemlist}
\item The continuum spectrum is non-thermal. Thermal (Planck-like) 
spectra are ruled out for the great majority of bursts. For most, 
the spectrum is well described by a smoothly-joining broken power law, 
known as a ``Band-function''\cite{b93} (see
Fig. \ref{fig:band}). Three independent 
spectral parameters are involved, i.e., a low energy photon spectral
index ($\alpha$), a high energy photon spectral index ($\beta$), and
the transition energy ($E_0$) or peak of the energy spectrum for
$\beta <-2$ ($E_p$). This spectral shape is valid both for the integrated
emission over the whole burst duration, and for the emission during a
certain temporal segment of the burst;  
\begin{figure}
\centerline{\psfig{file=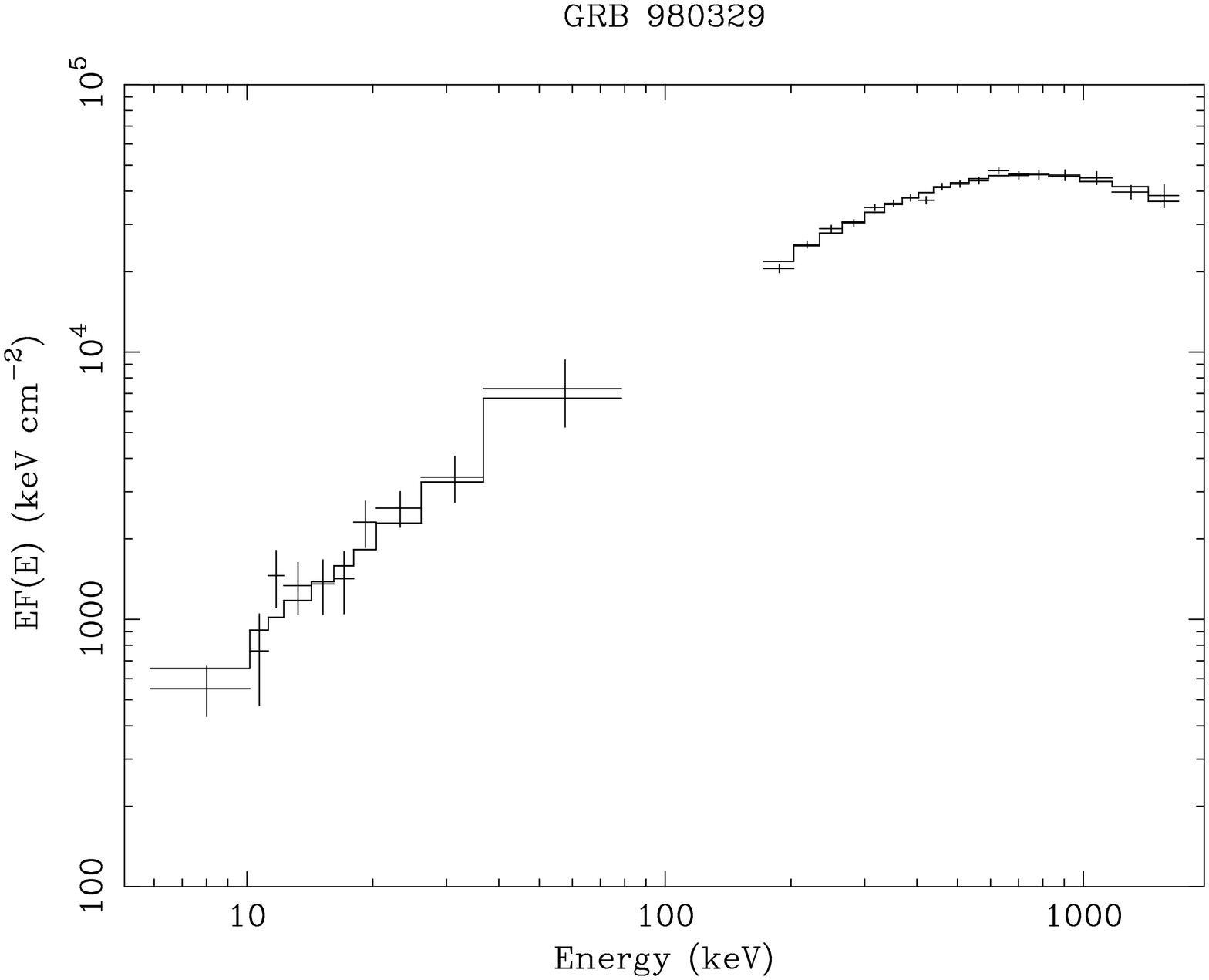,angle=-90,width=5cm}
            \psfig{file=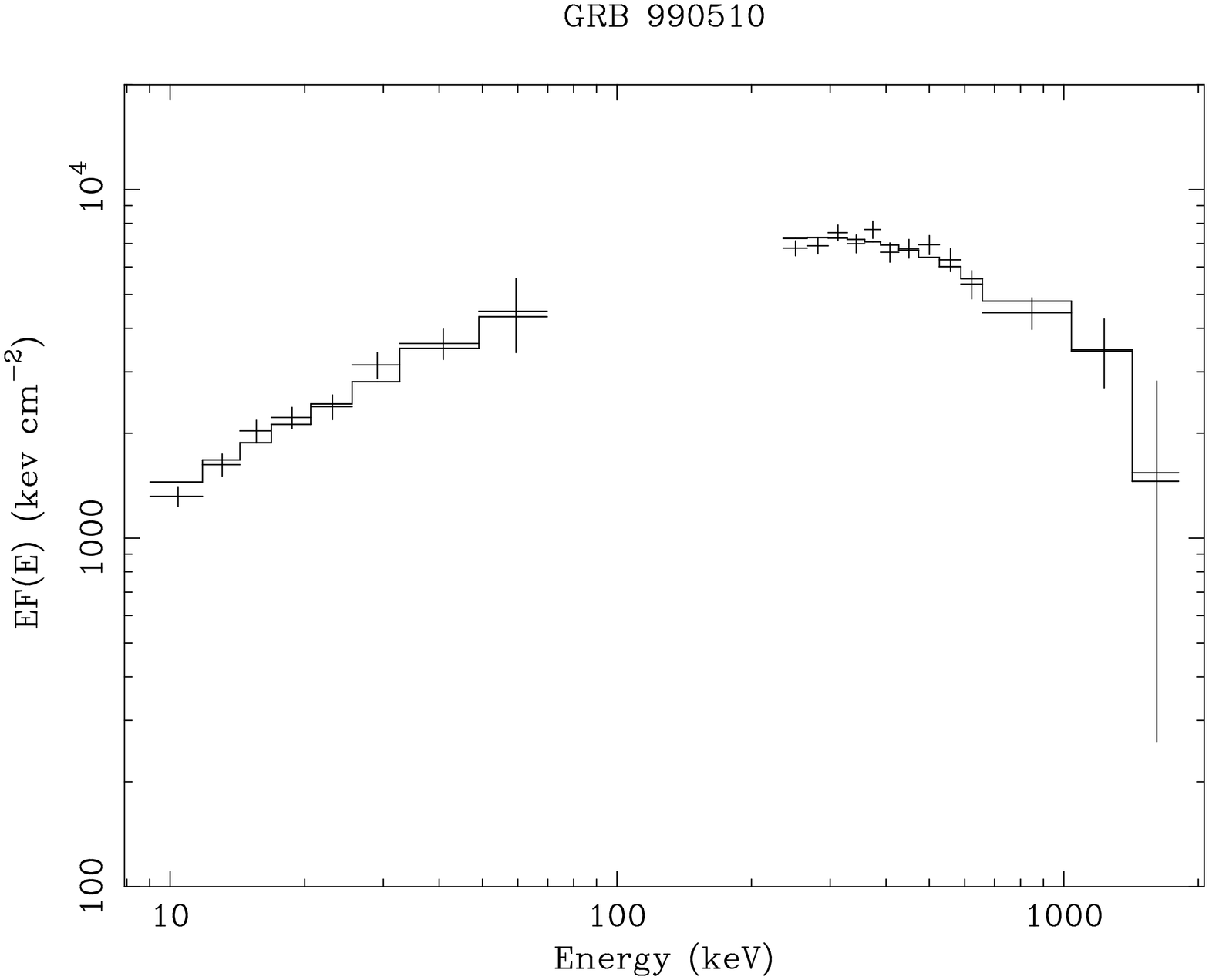,angle=-90,width=5cm}}
\vspace*{8pt}
\caption{Examples for the GRB prompt emission spectrum (GRB 980329 and
GRB 990510, from Ref.215).}
\label{fig:band}
\end{figure}
\item For a sample of 156 bright BATSE bursts with 5500 total 
spectra\cite{preece00}, it is found that $\alpha \sim -1\pm 1$, 
$\beta \sim -2^{+1}_{-2}$, and the $E_p$ distribution is {\em lognormal}, 
centered around $\sim 250$ keV with a full-width at half-maximum less 
than a decade in energy. The distributions of $\alpha$ and $\beta$ are 
also generally suitable for describing fainter and softer bursts. 
The ``narrow'' $E_p$ distribution among different bursts, however, is 
likely to be influenced by selection effects. Various investigations 
indicate that the lack of high $E_p$ bursts is likely intrinsic\cite{hs98}. 
In the low energy regime, however, the narrowness of the distribution 
function is mainly due to the ``bright'' flux-truncation in the sample. 
Lately, a group of X-ray transient events, which resemble normal GRBs 
in many aspects but with $E_p$ around or below 40 keV has been 
identified\cite{heise,kippen}.  These bursts, named ``X-ray flashes'' 
(XRFs), extend the $E_p$
distribution to the softer regime and are typically fainter. Globally, 
it seems that a ``narrow'' distribution of $E_p$ may be obtained only
when one is dealing with a sample of bursts within a narrow peak-flux 
range\cite{mallozzi95,sch03a}. On the other hand, since the bright 
BATSE sample\cite{preece00} contains many spectra for each burst (at
least eight), the results point to an intriguing conclusion, i.e.,
within a particular burst, the $E_p$ distribution is indeed very
narrow. Also bursts tend to soft in time\cite{norris86,ford95}, see
more discussions in \S\ref{sec:grb2};
\item High energy spectral components: Dozens of {\em BATSE} GRBs have 
also been detected at higher energies, e.g. by {\em EGRET}\cite{dingus01} 
and {\em Solar Maximum Mission} ({\em SMM})\cite{hs98}.  Most of these 
detections are consistent with a Band spectrum extended to high
energies without further breaks. Recently, however, a distinct high energy
component was reported in the time-dependent spectra of GRB
941017\cite{g03}. This component sticks out at the high energy end
of the conventional sub-MeV component, and extends to $\sim$ 200 MeV
(due to instrumental upper cut off) with a photon number index of -1
that can not be attributed to an extrapolation of the MeV
spectrum. More 
intriguingly, this component does not soften in time together with the 
low energy component, hinting an independent origin, i.e., perhaps 
a different emission site. At even higher energies (TeV), an excess of 
events coincident in time and space with GRB 970417 has 
been reported by the Milagrito group\cite{atkins00}. In the temporal
domain, GeV emission was detected in GRB 940217 lasting 1.5
hours\cite{h94}. However, this may be categorized as a high energy
afterglow\cite{zm01b} rather than prompt emission.
\item Spectral features: 
Absorption and emission features in GRB prompt emission spectra were
reported by the Soviet satellites {\em Venera 11} and 
{\em Venera 12}\cite{m81}, and by the Japanese mission {\em Ginga}\cite{m88}. 
The significance of these features was in the $2-3
\sigma$ range, with one claimed at $4\sigma$. However, these were not
confirmed by the {\em BATSE} team\cite{fm95}. 
The only spectral feature in prompt emission reported in recent years 
was a 3.8 keV absorption feature in GRB 990705\cite{amati00} detected 
by {\em BeppoSAX}. 
\end{itemlist}
\item Polarization properties: \\
Another important piece of information for electromagnetic radiation is 
its polarization, which is difficult to measure, especially in the 
gamma-ray band. Recently, a breakthrough in this direction was reported
with the {\em Ramaty High Energy Spectroscopic Imager (RHESSI)}. It was 
found that the prompt emission of the bright GRB 021206 was strongly 
polarized, and the claimed polarization degree is $80\% \pm
20\%$\cite{cb03} (cf. Refs. \refcite{rf03,bc03}). 
Unfortunately, due to its close angular distance from the Sun (which 
actually facilitated the detection by {\em RHESSI}) prevented the
detection of the optical afterglow. A typical radio afterglow was
however detected for this source\cite{frailgcn}.
\item Global properties:
\begin{itemlist}
\item Angular distribution: Isotropic for all bursts, or for either
long or  short bursts, respectively\cite{m92,paciesas99}. There might
be a small group of the so-called ``long-lag'' bursts\cite{norris02}
whose distribution is not isotropic but follows the super-galactic
plane. Also, a sub-group intermediate between long and short bursts
may show departures from isotropy\cite{attilameszaros00}; 
\item Intensity distribution: Generally there are two ways to quantify 
the intensity distribution. One way is through the
peak flux distribution (i.e. the so-called $\log N - \log PF$
plot)\cite{fm95}. This results in a -3/2 power law 
slope at high fluxes, as expected in a homogeneous Euclidean model,
and a shallower distribution at lower flux regimes that deviates from this
simple homogeneous model. A second way is to find the average value of 
$V/V_{\rm max}$, where $V$ and $V_{\rm max}$ are the volume of space
enclosed by the distance of the source and the maximum volume of space
enclosed by the distance at which the source could be still detected,
respectively. It is found that $<V/V_{\rm max}>$ is less than 1/2 (the 
value expected for the homogeneous Euclidean model), in the {\em BATSE}
data\cite{fm95}. Such an intensity distribution, along with the
isotropic angular distribution, are completely consistent with
GRBs being cosmological events\cite{schmidt01};
\item Event rate: BATSE detected GRBs at a rate of about 1 per day. 
By correcting for sky coverage and other factors, the actual event
rate is $\sim 600$ per year. Averaging over the whole Hubble volume
(for $H_0 \sim 70~{\rm km~s^{-1}~Mpc^{-1}}$), this corresponds to an
average birth rate of $\sim 7.5~{\rm Gpc^{-1}~yr^{-1}}$ or $\sim
0.4~{\rm Myr^{-1}~galaxy^{-1}}$, where an average galaxy number
density $n_g \sim 0.02~{\rm Mpc}^{-3}$ is adopted. The local GRB birth
rate is presumably smaller due to the drastic decrease in the star
forming rate at low redshifts, and a widely quoted local ($z=0$) rate
is\cite{schmidt01} $\sim 0.5~{\rm Gpc^{-1}~yr^{-1}}$, or $\sim 
0.025~{\rm Myr^{-1}~galaxy^{-1}}$. We now know that there are related
cosmic events such as X-ray flashes (XRFs, see next) that may increase 
the total event rate by a factor of 2 or 3. This would also increase
the GRB-XRF birth rate by a factor 2 or 3. We also know that GRBs are
very likely collimated (see \S\ref{sec:ag}). The beaming factor is
uncertain dependent on the possible jet structure. Assuming uniform
jets, the true-to-observed beaming correction is\cite{f01,vanp03} $\sim
500$ for GRBs. For structured jets, this number is generally
smaller. For a quasi-universal Gaussian-type structured jet which 
fits reasonably well the current GRB-XRF data\cite{zdlm03}, the
true-to-observed beaming correction factor is 14 for the combined
GRB-XRF sample. The real GRB-XRF birth rate (which corresponds to the birth
rate of their progenitor) should be increased by a same factor. {\em
Swift} will help to determine both the observed GRB-XRF event rate and
the geometric correction factor more precisely. 
\end{itemlist}
\item Taxonomy:
\begin{itemlist}
\item The duration distribution is clearly bimodal
(Fig. \ref{fig:bimodal}), leading to a 
classification of GRBs into two types, i.e. ``long''
bursts with $T_{90} > 2$ s, and ``short'' bursts with $T_{90} < 2$
s\cite{k93}. Such a classification is confirmed in the ``hardness''
domain, with long bursts typically being soft and short bursts
typically being hard\cite{k93}. The long and short categories roughly
consist of $75\%$ and $25\%$ of the total GRB population. It is
commonly speculated that the two types of GRBs may have 
different origins. All our current knowledge of GRB counterparts
(afterglows and host galaxies, etc) are for long bursts. The nature of 
short bursts is still a mystery. There have been extensive searches
for a third type of GRBs with intermediate
durations\cite{mukherjee98,horvath98,horvath02}, but the case is not
conclusive;
\begin{figure}
\centerline{\psfig{file=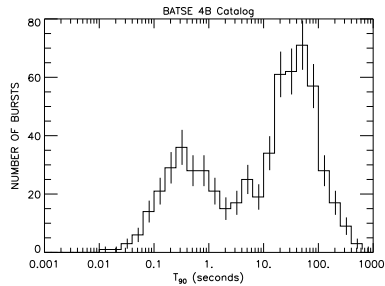,width=5cm}}
\vspace*{8pt}
\caption{The bimodal distribution of durations of the BATSE
GRBs, from Ref.189.}
\label{fig:bimodal}
\end{figure}
\item Based on temporal characteristics (lightcurves), it is difficult 
to categorize GRBs into well-defined sub-types. Nonetheless, some 
phenomenological classes have been suggested\cite{fm95}. These
classifications are however not fundamental. The differences among
different types are likely caused by different behaviors of the
central engine, different emission sites, or different viewing angles,
not necessarily reflecting different types of progenitor;
\item Based on the spectral hardness, recently the so-called ``X-ray
rich GRBs'' and ``X-ray flashes'' (XRFs)\cite{heise,kippen} have been
widely discussed as forming an apparently new type of transient event
with respect to the conventional GRBs (Fig. \ref{fig:XRF}). Whether
they are the product of intrinsically different mechanisms or are simply a 
natural extension of GRBs towards softer and fainter regimes\cite{barraud03}
is still unclear, but evidence is mounting that XRFs and GRBs are
related events;
\begin{figure}
\centerline{\psfig{file=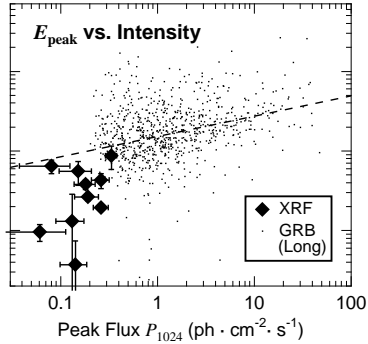,width=5cm}}
\vspace*{8pt}
\caption{The $E_p$-fluence distribution of the conventional long GRBs
as well as the recently identified X-ray flashes, from
Ref.38.}
\label{fig:XRF}
\end{figure}
\item The ``long-lag'' bursts such as GRB 980425/SN 1998bw may belong
to a sub-type of long bursts at closer distances (associated with the
super-galactic plane)\cite{norris02}.
\end{itemlist}
\end{itemlist}
%\end{romanlist}

\subsubsection{Empirical laws\label{sec:grb2}}

\begin{figure}
\centerline{\psfig{file=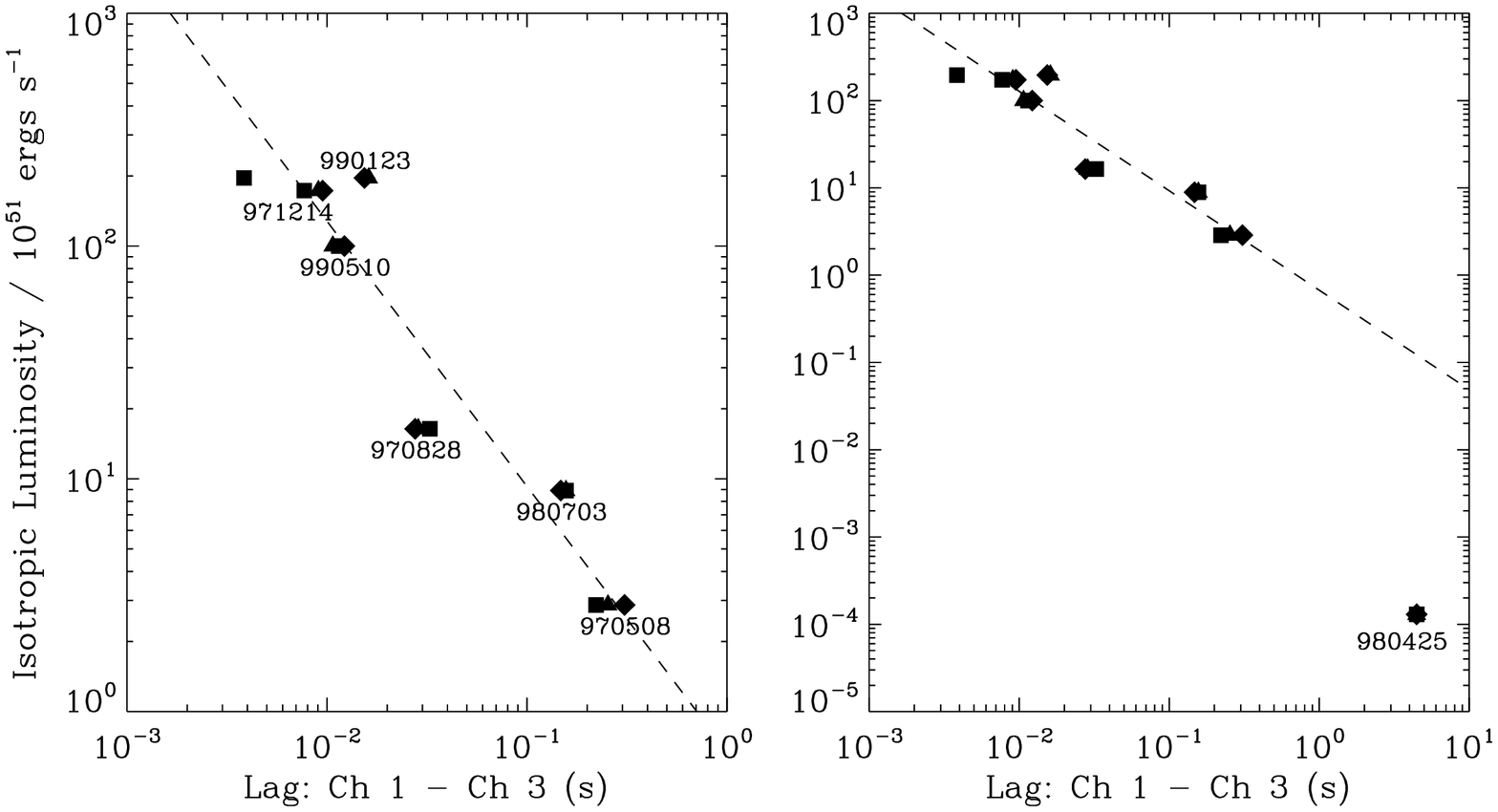,angle=90,width=10cm}}
\centerline{\psfig{file=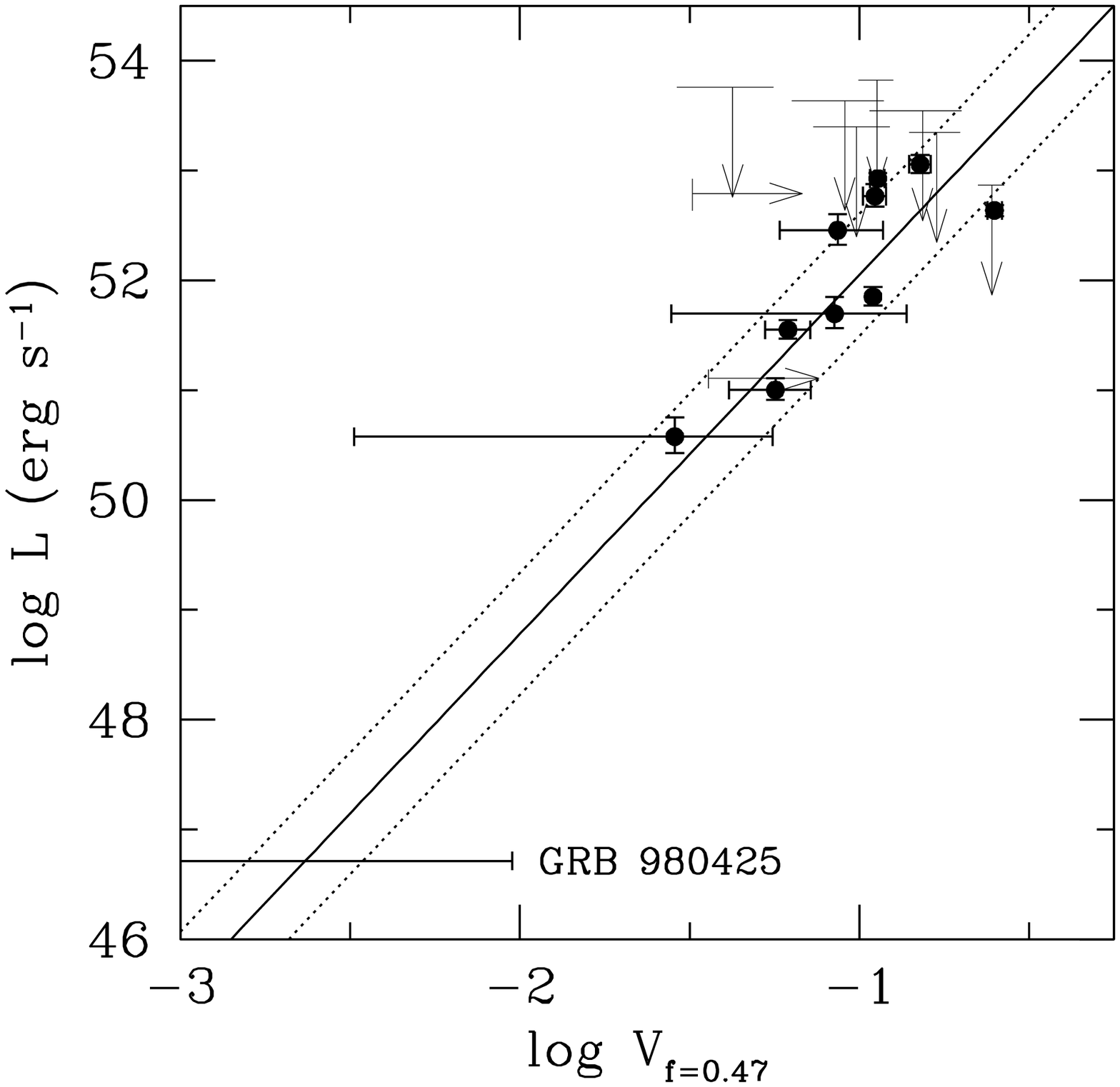,width=5cm}
            \psfig{file=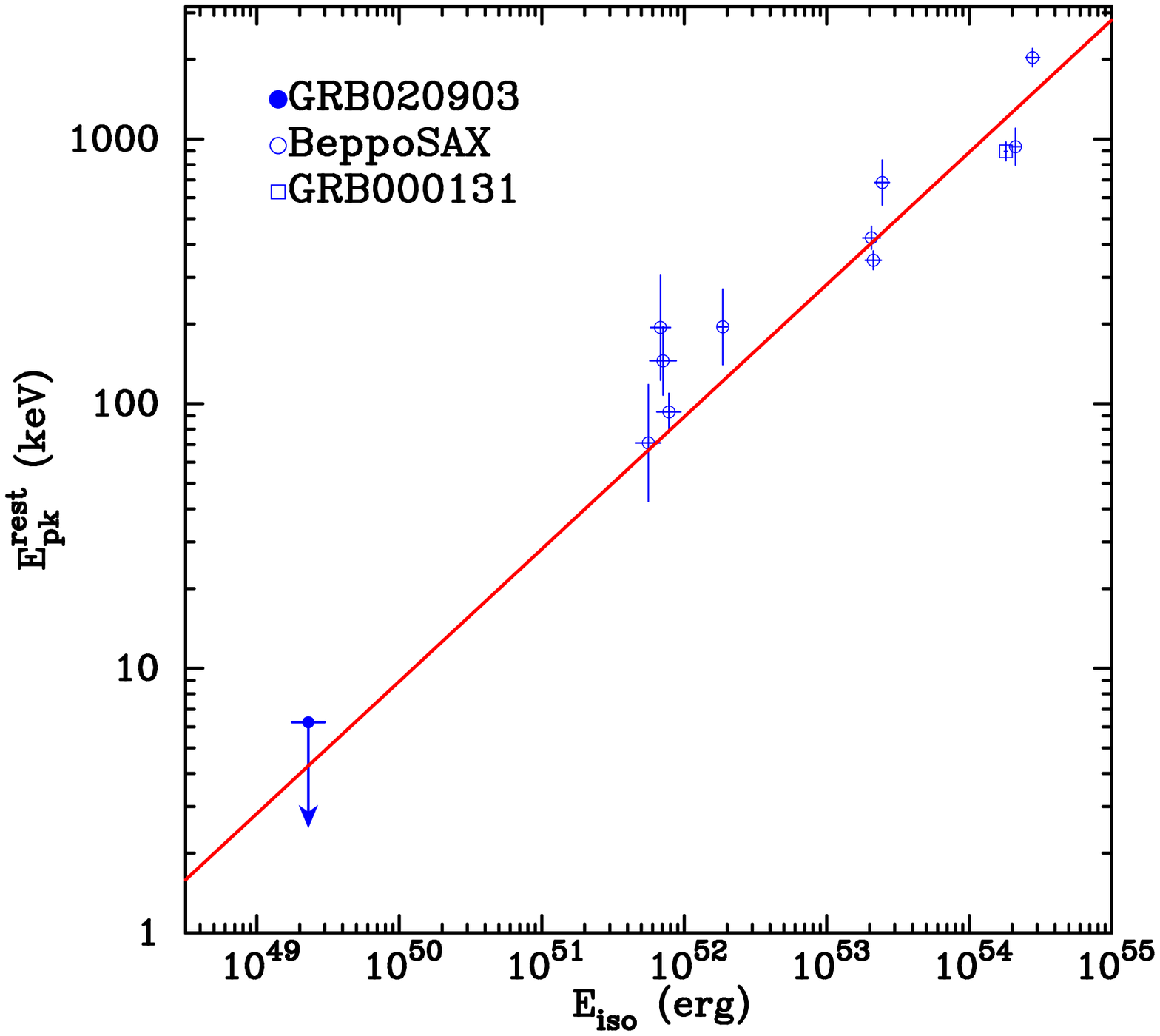,width=5cm}}
\vspace*{8pt}
\caption{Three ``Cepheid''-like correlations found in GRB data, which
could be adopted as rough distance indicators and redshift estimators.
(1) Top: the spectral lag - luminosity correlation (from
Ref.210); (2) Bottom left: the variability - luminosity
correlation$^{202}$ (from Ref.203); and (3)
Bottom right: the $E_p$ - luminosity correlation$^{215}$ which
extends to XRFs (from Ref.217).}
\label{fig:cepheid}
\end{figure}

Besides the above relatively solid observational facts, some ``secondary''
empirical relations have been derived through various statistical
analyses. These relations are potentially useful, posing important
constraints on the theoretical models. Some are even helpful for
deriving some important but unknown parameters. Below is a
non-exhaustive list. The items 1-4 are for the temporal information
alone, while the items 5-8 include the combined temporal and spectral
information. 
\begin{itemlist}
\item Power density spectra (PDSs) are powerful tools to quantify GRB
temporal characteristics. It is found that the averaged PDS for many
long bursts follows a power law of index -5/3 over almost two decades
in Fourier frequency, with a break around $\sim 1$ Hz\cite{bss98,bss00} 
at the higher end. This indicates that variabilities shorter than 
$\sim 1$ s are smeared out, for reasons as yet unclear; 
\item By tracking individual pulses, the pulse width distribution for
long bursts is found to be lognormal\cite{lf96,np02}, with the peak variability
timescale $\delta t_{\rm pk} \sim 1$ s. In some GRBs, there are so-called
``quiescent times'' where no gamma-rays are emitted are identified, 
and the durations of such quiescent episodes are positively correlated 
with the duration of the emission episode immediately following the 
quiescence\cite{rr01}. By subtracting the quiescent times, the 
intervals between pulses also have a lognormal distribution\cite{lf96,np02};
\item The temporal behavior can be also quantified through a
``variability'' parameter, defined by some appropriate statistical
definitions\cite{frr00,reichart01}. An intriguing empirical relation 
is that the more variable bursts (with larger $V$ parameters) tend to have
higher intrinsic isotropic luminosities $L$ (derived from afterglow
spectroscopic measurements), i.e. $L\sim V^{3.3}$. Although such a
correlation is derived from a small parent sample, it exhibits a
``Cepheid''-like correlation and makes ``variability'' a possible
distance indicator of GRBs (Fig. \ref{fig:cepheid}, bottom left); 
\item The cumulative lightcurves can reveal the global energy release rate 
of the central engine. For most bursts, the cumulative lightcurve has
a constant slope as a function of time\cite{mcbreen02a}. There are two 
smaller groups (comprising $\sim 4.8\%$ and $\sim 2.8\%$ of the
whole population studied) whose cumulative lightcurves increase
with time more rapidly or more slowly than the linear increase,
respectively\cite{mcbreen02b}; 
\item In many bursts (although not exclusively) there is a clear trend
of spectral evolution. There are two types of evolution. One is
``hard-to-soft''\cite{norris86,ford95}, in which hard emission leads
the soft emission, and another is ``tracking''\cite{golen83}, for which 
the spectral hardness tracks the intensity. In either case, the $E_p$
(derived by 
fixing both $\alpha$ and $\beta$) decay is found to be exponential with
photon fluence, and within a burst, the decaying constant is invariant 
from pulse to pulse\cite{lk96};
\item For asymmetric pulses, the pulse peak times migrate towards
later times at lower energies, and the pulse width becomes
wider\cite{norris96,cheng95}. For 6 GRBs with spectroscopic redshift
measurements, it is found that the spectral lag is correlated with the
isotropic luminosity as $L_{53} \approx 1.3\times(\tau / 0.01~{\rm
s})^{-1.14}$, making spectral lag another possible ``Cepheid-like''
luminosity indicator\cite{norris00} (Fig. \ref{fig:cepheid}, top);
\item The low energy spectral index $\alpha$ is found to correlate
with the $E_p$ of time-resolved GRB spectra, although the trend is still
controversial\cite{crider97,lp02}; 
\item The $E_p$'s are found to be positively correlated with the
isotropic luminosity\cite{lpm00,lr02}. 
With the spectroscopic redshifts of 12 BeppoSAX bursts, a firm correlation,
i.e., $E_p \propto E_{\rm rad}^{0.52\pm 0.06}$ (where $E_{\rm rad}$ is 
the total isotropic energy radiated in the X-ray and gamma-ray range,
i.e. 1 keV - 10 MeV) is found\cite{amati02}, which may be taken as yet 
another ``Cepheid-like'' luminosity indicator\cite{atteia03}
(Fig. \ref{fig:cepheid}, bottom right). The relation is found to
extend from the hard GRB to the soft XRF
regime\cite{sakamoto03,lamb03}. However, due to the small sample
effect, the issue is not conclusive, yet. Furthermore, GRB 980425 had
an $E_p \sim 70$ keV and $E_{iso} \sim 10^{48}$ ergs, which apparently
does not fit into the simple correlation; 
\item Another redshift indicator is found in the gamma-ray spectral
data by taking into account of excesses from the exact power-law
dependence\cite{bagoly03};
\item Luminosity function: Although the present afterglow
spectroscopic redshift measurements provide too small a sample to
perform a direct measurement, several attempts have been made to
derive the luminosity function of GRBs. Although without full
consensus, it is likely that the luminosity is a power-law, 
with a possible steepening at high
luminosities\cite{schmidt01,schaefer01,lloyd02b,norris02,stern02}; 
\item Using the $V-L$ correlation and the spectroscopic redshift
data\cite{amati02}, it is found that there might exist a cosmological
evolution effect for GRB properties, such as luminosity (or even
luminosity function), total energy, $E_p$, duration, etc., although
more spectroscopic redshift measurements are needed and selection
effects need to be further clarified\cite{lloyd02b,wg03}. 
\end{itemlist}

\subsection{Afterglow emission\label{sec:ag}}

\subsubsection{Observational facts\label{sec:ag1}}

The afterglow observations have been extensively reviewed in
Refs. \refcite{vkw00,kul00,hsd02} and references therein. Here we 
outline some of the main results in itemized forms. These include
\begin{itemlist}
\item Global properties:
\begin{itemlist}
\item Afterglows are (quite) broad-band, having been detected in the X-ray,
the optical/infrared and the radio bands. In each band, the lightcurve
generally displays a power-law decay behavior. X-ray afterglows are 
exclusively decaying when they are detected. Optical afterglows are
generally decaying, with an initial early rising lightcurve having been
caught in only a few bursts (e.g. GRB 990123\cite{a99}). In contrast, an
initial rising lightcurve in the radio band has been detected in
many GRBs (followed by a canonical decay), and the typical timescale for 
the radio afterglow to reach the peak is $\sim 10$ days\cite{frail_03}.
Figure \ref{fig:ag-lc} gives several examples of broadband afterglow
lightcurves; 
\begin{figure}
\centerline{\psfig{file=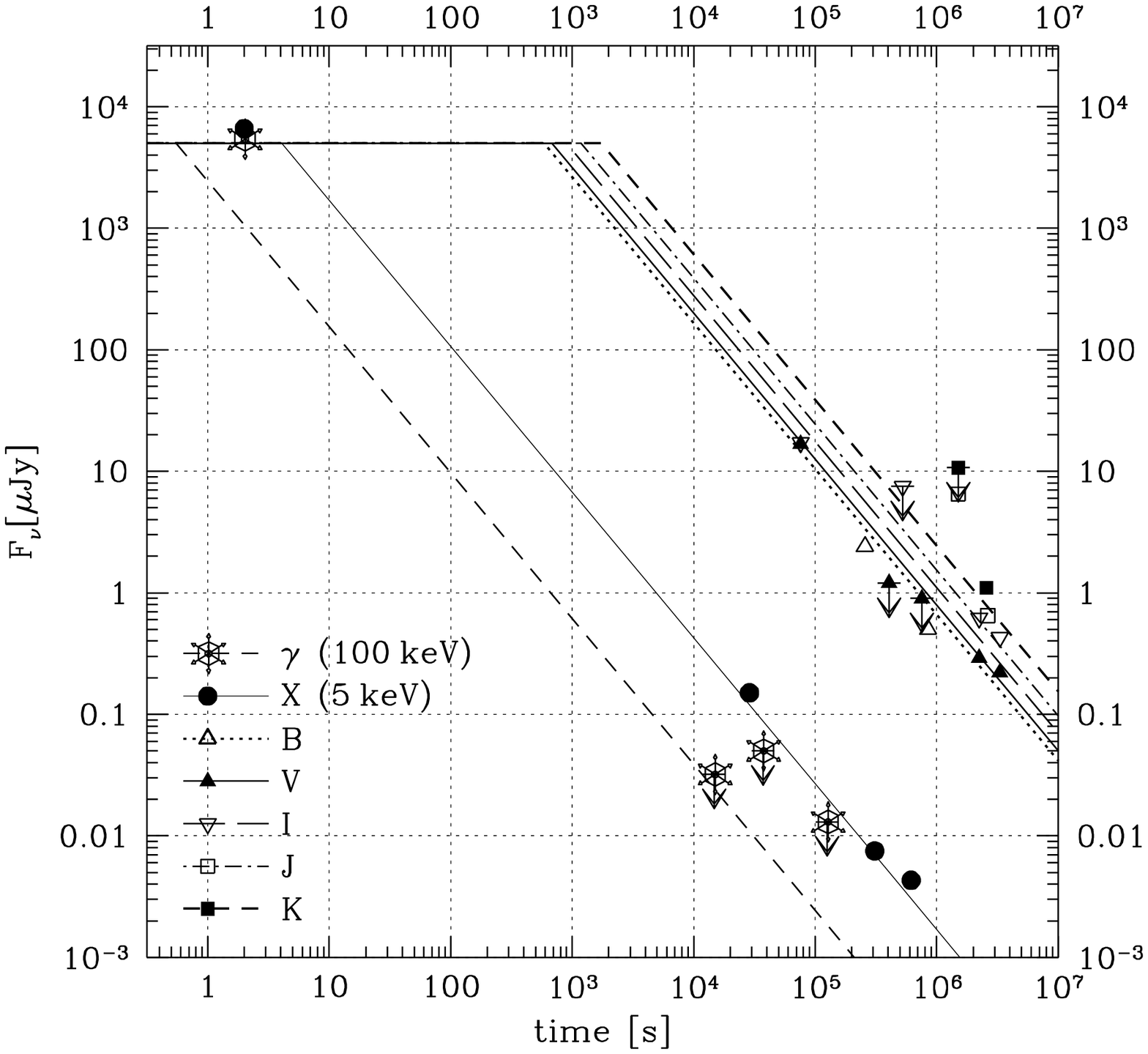,width=5cm}
            \psfig{file=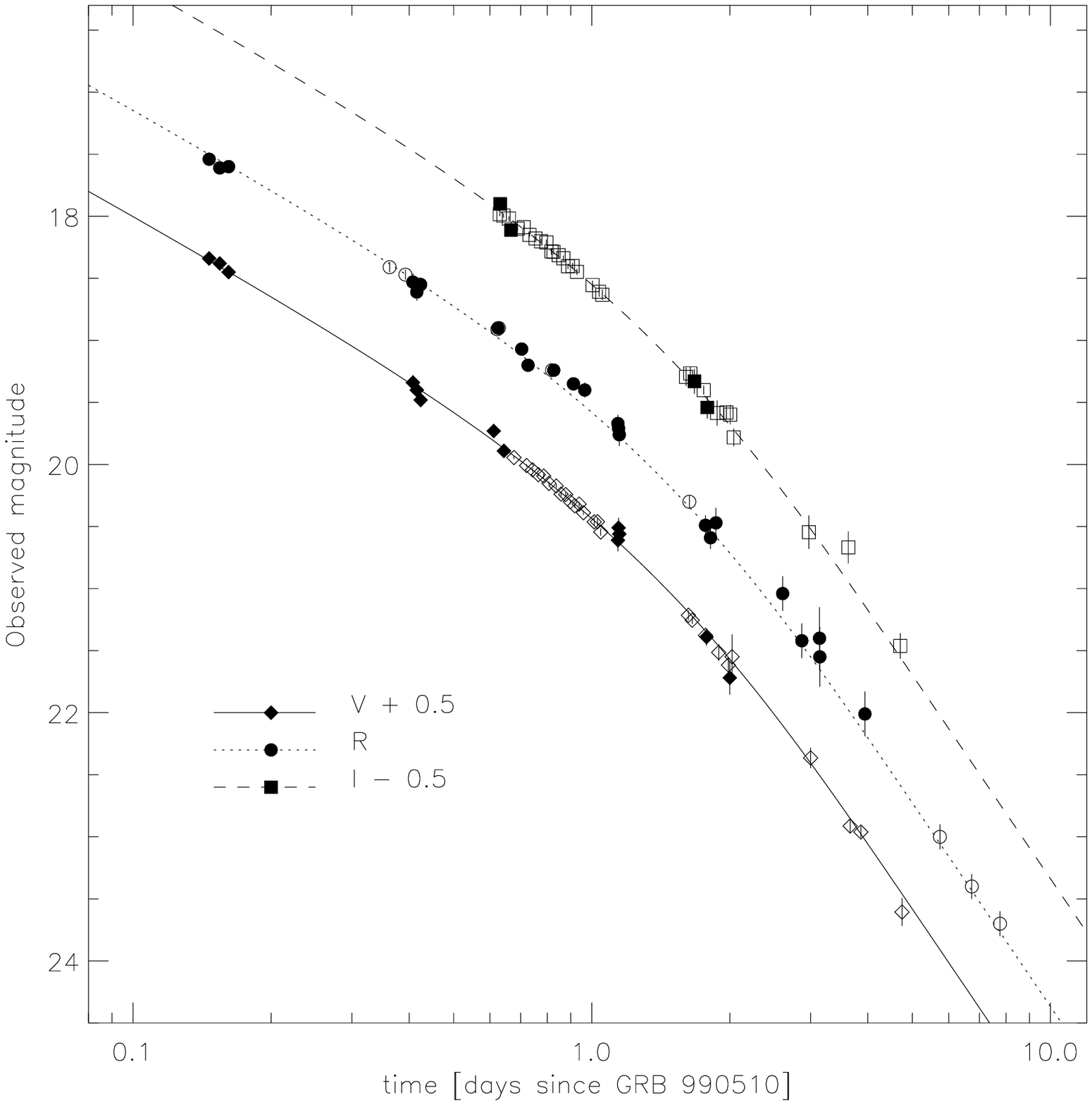,width=5cm}}
\centerline{\psfig{file=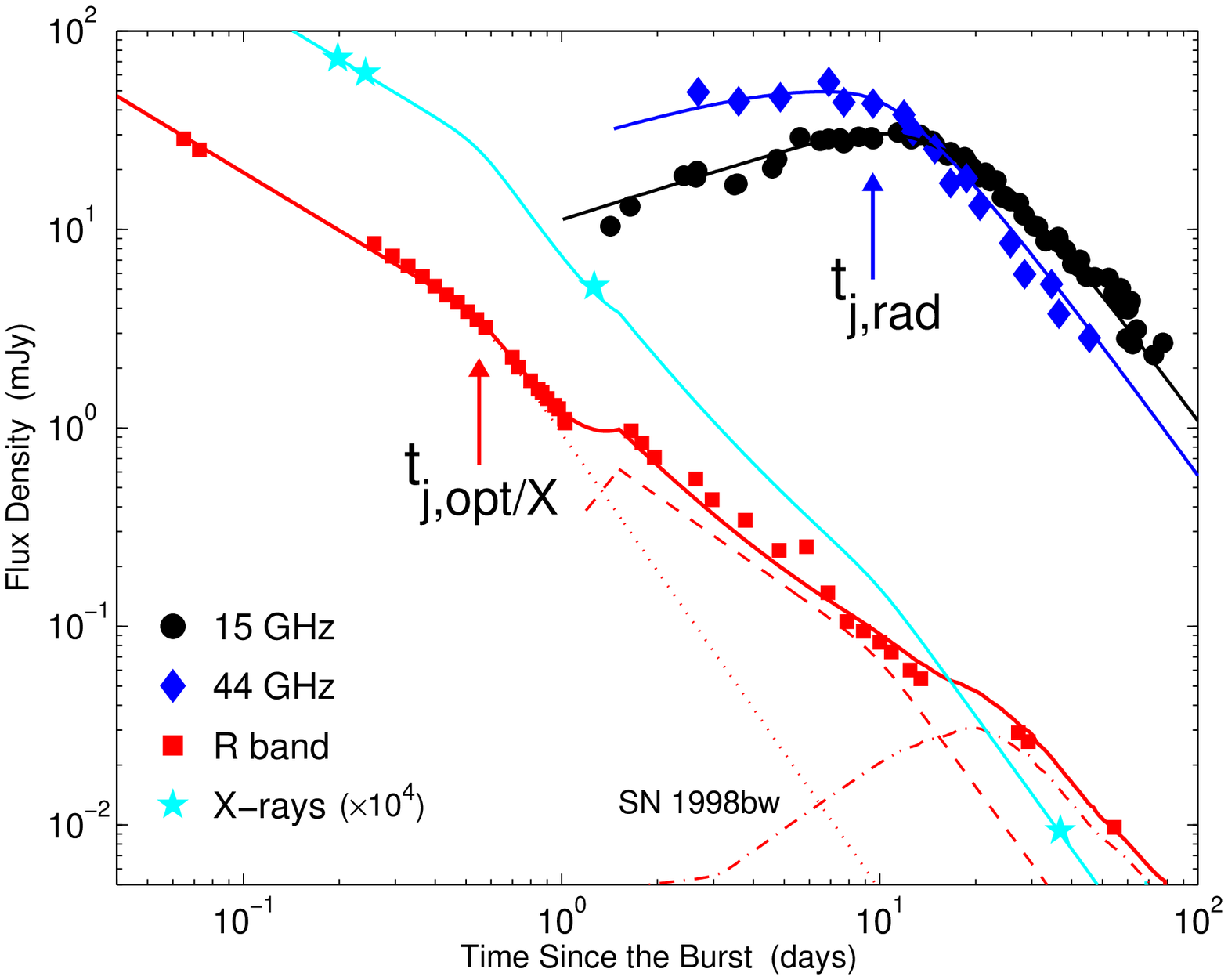,width=5cm}}
\vspace*{8pt}
\caption{Several well-monitored broad-band afterglow lightcurves.
Upper left: GRB 970228 (Ref.79 and references therein);
Upper right: GRB 990510 (from Ref.33), notice the
achromatic jet break; Bottom: GRB 030329 (from Ref.48),
Notice irregularities in the lightcurves and the association of a
supernova bump.}
\label{fig:ag-lc}
\end{figure}
\item Very often there are various types of deviations from the simple 
power law decay. These include steepenings, bumps and wiggles
(e.g. GRB 021004; GRB 030329);
\item Not all bursts have afterglows detected in all of the three main
bands. X-ray afterglows are the most commonly detected. About $60\%$ of
BEppoSAX bursts with X-ray afterglow detections are detected in the
optical band. The other $\sim 40\%$ of GRB afterglows are optically
dark, i.e., there are no optical transients identified with sufficient
exposure. These are dubbed as ``dark bursts''. In the HETE era, the
fraction of dark bursts gets smaller, e.g. $10\%$\cite{ricker03}.
Radio afterglows are detected in about half of all GRB afterglows.
\item Essentially every GRB with an afterglow detection has an underlying
host galaxy. The GRB host galaxy properties (magnitude, redshift distribution,
morphologies, etc) are typical of normal, faint, star forming
galaxies\cite{d03}. The GRB afterglow's positional offsets with respect to 
the host galaxy are consistent with GRBs being associated with the star 
forming regions in the galaxies\cite{bloom98}; 
\item GRBs are at cosmological distances. Their redshifts are usually
measured from the emission features of the host galaxies or the
absorption features imposed on the afterglow continuum. As of October
2003, there are about 33 redshift measurements, and detected redshifts
range from 0.168 for GRB 030329 (or 0.0085 for GRB 980425) to 4.5 for
GRB 000131. See Ref.\refcite{greiner} for a compilation of the
afterglow data, including redshifts; 
\item At least some GRBs are associated with supernova explosions. A
famous example was GRB980425/SN1998bw association\cite{g98,k98b}. SN
1998bw was a peculiar, energetic Type Ib/c supernova. Using it as a
template, other possible associations have been claimed by
identifying a reddened bump in the optical afterglow lightcurves of 
GRB 980326\cite{bloom99}, GRB 970228\cite{reichart99,galama00}, 
GRB 000911\cite{lazzati01}, GRB 991208\cite{castro01}, 
GRB 990712\cite{sahu00}, GRB 011121\cite{bloom02a}, and 
GRB 020405\cite{price03}, which may be attributed to supernova remnant 
contribution. Very recently, an unambiguous supernova signature has been 
detected in the $z=0.168$ GRB 030329, firmly establishing the GRB-SN 
association in this object\cite{sta03,h03} (Fig. \ref{fig:GRB-SN}).
\end{itemlist}
\begin{figure}
\centerline{\psfig{file=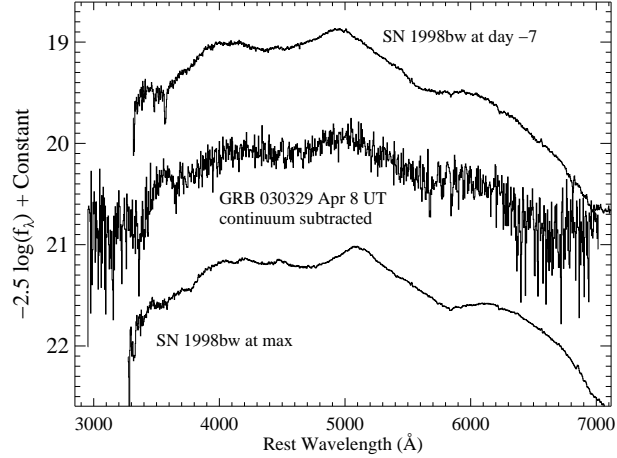,angle=90,width=8cm}}
\vspace*{8pt}
\caption{The optical afterglow spectrum of GRB 030329 clearly reveals
the spectral signature of Type-Ib/c supernovae (e.g. SN 1998bw) at
day-10 since the burst trigger. This establishes a firm association
between long GRBs and SNs (from Ref.45).}
\label{fig:GRB-SN}
\end{figure}
\item X-ray afterglows: 
\begin{itemlist}
\item The continuum spectrum is essentially a power law. By writing
$F_{\rm x}(t,\nu) \propto t^{\alpha} \nu^{\beta}$, one
has\cite{piro01} $\alpha \sim -0.9$ and $\beta \sim -1.4$, both with a
range of scatter. Late time lightcurve flattening was observed in a
few bursts such as GRB 000926\cite{harrison01};
\item As of October 2003, X-ray emission line features have been claimed
to be detected in the X-ray afterglows of 6 bursts with moderate
significance ($< 4.5 \sigma$, typically $3\sigma$), i.e., 
GRB 970501\cite{piro99}, GRB 970828\cite{yoshida01}, GRB 991216\cite{p00}, 
GRB 000214\cite{antonelli00}, GRB 011211\cite{r02}, and
GRB 020813\cite{butler03}; 
\item Analyses of BATSE data reveal soft tails in some GRBs which are
likely to be early X-ray afterglows\cite{connaughton}.
\end{itemlist}
\item Optical afterglows:
\begin{itemlist}
\item Expressing the optical flux as $F_{\rm opt}(t,\nu) \propto t^{\alpha}
\nu^{\beta}$, one has\cite{greiner} $\alpha \sim -1$ (at early times)
and $\beta \sim -0.7$, both with a range of scatter;
\item In some bursts, a clear lightcurve steepening is seen after
some time $t_{break}$ of order of days. The break is achromatic
over different bands, and the temporal index after
the break is typically $-2$ with some scatter. This break is
typically attributed to the presence of a jet, and is termed the ``jet
break''. At later times, the decay rate gradually slows down until
finally reaching a constant level due to the contribution of the host
galaxy; 
\item Other irregular temporal features are occasionally seen in
various bursts\cite{gcn}. These include\cite{greiner}, for example, a
substantial re-brightening in GRB 970508, an achromatic bump signature
in GRB 000301C, wiggles in GRB 021004 and step-like features in GRB
030329; 
\item Polarization: As of October 2003, 8 GRB optical afterglows have
been detected to be polarized\cite{bjornsson03,covino03,greiner03},
i.e., GRB 980425, GRB 990510, GRB 990712, GRB 010222, GRB 020405, GRB
020813, GRB 021004 and GRB 030329. The degree of polarization is small, 
typically several per cent. Large polarization angle changes (by
nearly 90 degrees) were found in GRB 021004 and GRB
030329\cite{rol03,greiner03};
\item Early optical flashes: Optical afterglow observations typically
have started hours after the burst trigger due to technical reasons. 
However, early optical afterglows of two GRBs, i.e., GRB 990123\cite{a99} and
GRB 021211\cite{li03,f03b}, were detected within 100 seconds
after the GRB trigger. In both cases, the early afterglows indicate a
steep decay with index $\sim 2$. For GRB 990123, a sharp rising
lightcurve was also detected, and the V-magnitude is $\sim 9$ at the
peak\cite{a99}. The early optical afterglow of a third burst, GRB
021004, was detected about 3 minutes after the trigger, which displays
a very shallow initial lightcurve decay\cite{f03a}. 
\end{itemlist}
\item Radio afterglows:
\begin{itemlist}
\item In the radio band, the spectral index is generally positive for
the observations which are typically in the 5 and 8.5 GHz bands. The
lightcurves 
usually do not follow a simple power law decline\cite{frail03}. Some
sources can be observed on timescales of years, and a late-time
flattening (with respect to the standard fireball model) is often
observed\cite{frail03b};
\item Prompt, short-lived radio flares have been detected in several
bursts\cite{kul99b,frail00}; 
\item At early times, radio afterglows show strong
fluctuations which are suppressed at later times\cite{frail00b}. This
can be interpreted as being due to interstellar scintillation
effects\cite{goodman97}. The detection of such an effect in GRB 970508 
clearly suggested that the source was expanding super-luminally, which 
gave a solid observational proof for the fireball shock model\cite{wkf98}.
\end{itemlist}
\item Taxonomy: According to their afterglow data, long GRBs may be further
classified into several sub-types. Below are two such suggested
classifications. They do not necessarily reflect intrinsically
different groups 
(e.g. with different progenitors), but might be caused by
environmental effects. So far there is no attempt to categorize GRB
sub-types using combined prompt emission and afterglow data, but
such efforts should be useful for identifying real sub-types.
\begin{itemlist}
\item Optically dark bursts: A fraction of GRBs with precise
localizations do not have bright enough optical afterglows to be
detectable. Possible reasons include dust extinction, high redshift,
or intrinsically faint nature. Evidence for dust extinction has been
collected for some bursts\cite{djorgovski01}. On the other
hand, observations for GRB 020124 and GRB 021211 indicate that at
least some optically dark bursts are simply bursts with intrinsically
faint afterglows\cite{berger02,li03}. Indirect evidence for high
redshift is also available for the recent dark burst GRB
031026\cite{031026gcn}.
\item Fast-fading GRBs: Several bursts (e.g. GRB 980519, GRB 980326)
show a steep afterglow decay ($\alpha \sim -2$) in their early phase. They do
not fit into the ``standard energy reservoir'' scenario\cite{f01}, and
may constitute a peculiar class of GRBs\cite{bfk03}. 
\end{itemlist}

\end{itemlist}

\subsubsection{Empirical laws\label{sec:ag2}}

As in the prompt emission studies, some secondary empirical laws
have been discovered in the afterglow data. Below is a non-exhaustive
list. 

\begin{itemlist}
\item Perhaps the most intriguing finding is that long GRBs seem to
have a standard energy reservoir. This conclusion is based on
the model that GRBs are collimated and that lightcurve steepenings are
due to a jet. From a simple toy model, one can derive the
so-called jet opening angle $\theta_j$ using the jet break time. For
those bursts whose redshifts, and hence the total gamma-ray
energies (assuming isotropic radiation) in the prompt phase,
$E_{\gamma,iso}$, are measured, it is found that $E_{\gamma,iso} 
\theta_j^2$ is essentially a constant, which means that the
geometrically corrected gamma-ray energy for different bursts is
essentially an invariant\cite{f01,bfk03} (Fig. \ref{fig:standardE}); 
\begin{figure}
\centerline{\psfig{file=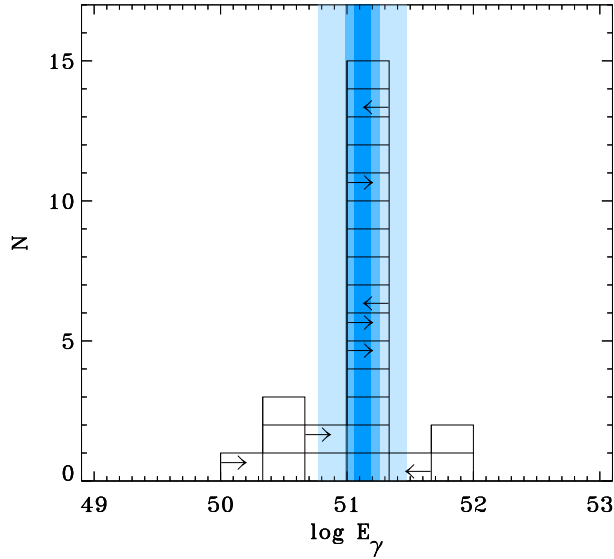,width=8cm}}
\vspace*{8pt}
\caption{The geometry corrected gamma-ray energy (i.e. $E_\gamma 
\sim E_{\gamma,iso} \theta_j^2/2$, where $E_{\gamma,iso}$ is the total 
energy emitted in gamma-rays assuming isotropic radiation, and 
$\theta_j$ is the jet opening angle inferred from afterglow 
lightcurves) is found to be a constant in many bursts, referring to a 
standard energy reservoir of long GRBs$^{36}$. Shown is the 
distribution of $E_\gamma$ with the latest data (from 
Ref.256).}
\label{fig:standardE} 
\end{figure} 
\item The total GRB fireball energy should be $E_{tot} \geq E_\gamma + 
E_K$, where $E_\gamma$ is the energy released as gamma-rays in the
prompt phase, and $E_K$ is the kinetic energy which is left over after
the prompt phase and is dissipated in the afterglow phase. The latter
component ($E_K$) may be directly measured using broadband afterglow 
fits\cite{pk01,pk02} or via analyses of the X-ray afterglow data for
different GRBs at the same afterglow epoch\cite{fw01,pkpp01,bkf03}. Both
approaches lead to the same conclusion that $E_K$ is also a standard
value for different bursts. Combining this and the previous argument
suggests that long GRBs have a standard energy reservoir; 
\item An anti-correlation has been derived between the GRB spectral lag 
and the jet opening angle\cite{sg02}. This relation, when coupled with
the Cepheid-like correlation for the spectral lag\cite{norris00}, is
another manifestation of the anti-correlation between the isotropic
luminosity and the jet opening angle, which is the direct consequence 
of the standard energy reservoir relation\cite{f01};
\item Other model parameters can also be inferred from broadband
afterglow fits. Consensus is emerging about the values of some of these
parameters. For example, the environment of most GRBs appears to be an
interstellar medium (ISM) with an approximately constant density, although 
the afterglows of a small group of GRBs are consistent with a stratified
wind-type medium (with a density profile $n \propto
r^{-2}$)\cite{pk01,pk02}. The ISM density appears not to vary greatly
among bursts\cite{pk02}. In the shock region, the portion of the energy
which goes into the electrons (denoted as $\epsilon_e$) is typically
$\sim 0.1$ with some scatter, while the energy portion that goes into
the magnetic fields (denoted as $\epsilon_B$) is typically $\sim 0.01$
or less\cite{pk01,pk02}. 
\end{itemlist}

\section{Theoretical Progress: Standard fireball shock model\label{sec:prog2}}

In this section, we briefly introduce the key ingredients of a
generic standard GRB fireball shock model in a pedagogical manner. 
This theoretical framework is the most widely used for interpreting the
current GRB and especially afterglow observations. Its ``standard''
nature is the product of its predictive power and success in passing
various observational tests. This section focuses on those aspects of 
the model which are robust and have the least uncertainties. Discussion 
of some of the less certain aspects of the fireball model are deferred to
\S\ref{sec:prob}. The contents of this section overlap to some significant
degree several other reviews such as Refs. \refcite{p99,m02,hsd02}.

\subsection{Relativistic bulk motion\label{sec:comp}}
A first ingredient of the standard model is that the emitting material 
responsible for GRBs and afterglows must be moving relativistically. This 
is a consensus of {\em all} cosmological GRB models (even if these
models are non-standard and differ considerably in technical details)
and even for the old galactic halo models. The arguments 
leading to this conclusion are straightforward\cite{r75,p86,g86,p90}, as 
discussed by, e.g. Refs. \refcite{p99,m02,cl01,w03}. For a typical GRB
gamma-ray fluence $F \sim 10^{-6}~{\rm erg~cm^{-2}}$ and distance $D
\sim 3$ Gpc, the total isotropic gamma-ray energy released is typically
$E=4\pi D^2 F \sim 10^{51}$ ergs. Naively (without relativistic
motion), the scale of the emission area is $c \delta t =3\times
10^8~{\rm cm}~(\delta t/ {\rm 10 ms})$. Assuming that a fraction $f_p$
of photons is above the two-photon pair production
($\gamma\gamma\rightarrow e^{+}e^{-}$) threshold
[$\epsilon_1 \epsilon_2 (1-\cos\theta_{12}) \geq 2 (m_e c^2)^2$, where
$m_e$ is the electron mass, $\epsilon_1$ and $\epsilon_2$ are the
energies of two photons, and $\theta_{12}$ is the angle between the
momenta of the two photons], and using an approximate pair production cross 
section of the order of the Thomson cross section $\sigma_T = 6.25\times
10^{-25}~{\rm cm}^2$, the pair-production optical depth is huge,
i.e. $\tau_{\gamma\gamma} =f_p \sigma_T F D^2 / (c \delta t)^2 m_e c^2 
\sim 10^{15} f_p (F/10^{-6}~{\rm erg~cm^{-2}}) (D/3~{\rm Gpc})^2$
$(\delta t/10~{\rm ms})^{-2}$. Thus, the gamma-rays should have been
attenuated in the source before traveling through the universe and 
reaching the earth. The only way to get rid of this apparent paradox 
is by invoking relativistic bulk motion, i.e., the GRB emitting region 
as a whole moves towards us observer with a high Lorentz factor. 

The relativistic motion eases this ``compactness problem'' in two
ways. Suppose that the emitting region (e.g. an ejected shell of relativistic
material) flies towards the observer with a bulk Lorentz factor $\Gamma$. 
First, the photon energy is blue-shifted by a factor of $\Gamma$, so that in 
the shell comoving frame, the bulk of gamma-rays as observed are actually
X-rays. This greatly decreases the number of photons above the pair
production threshold, i.e. $f_p$ drops by a factor of
$\Gamma^{2(\alpha-1)}$, where $\alpha \sim 2$ is the observed photon
number spectral index, i.e., $N(E_{ph}) dE_{ph} \propto
E_{ph}^{-\alpha} dE_{ph}$. Here the factor $(\alpha-1)$ arises from the
integration leading to the above-threshold photon numbers, and the factor 
2 takes into account the contributions of both the test photons and the 
target photons for pair production\cite{ls01}. A second effect
introduced by considering 
relativistic motion is that the real physical
scale of the emission region is $\Gamma^2 c \delta t$ for an observed
timescale of $\delta t$ (see \S\ref{sec:times} below). So altogether
the pair optical depth drops by a factor of\footnote{Notice that the
expressions of the optical depth in some of the previous
reviews\cite{p99} is incorrect\cite{ls01}, given the
above definition of $\alpha$.} 
$\Gamma^{2+2\alpha} \sim \Gamma^6$. For typical bursts, $\Gamma \geq 100$ is
required to have $\tau_{\gamma\gamma} <1$. This is only a rough
estimate; more sophisticated analyses result in various lower limits
of $\Gamma$ in different bursts\cite{kp91,feh93,wl95,bh97,ls01}. 
The typical $\Gamma$ required to satisfy the observations is of order 
$\sim 100$. Hence, GRBs involve the fastest bulk motions known so far 
in the universe. 
 
There are other evidence of relativitic motion of the fireball. One is 
from the radio afterglow data that initially shows large interstellar
scintillation but gets supressed later on. This presents a clear
evidence of superluminal expansion of the fireball caused by
relativistic motion\cite{wkf98}. Another is from the analyses of early 
afterglow reverse shock data\cite{sp99a,zkm03}. These analyses
directly point to a large initial Lorentz factor of the fireball (see
\S\ref{sec:reverse} for more discussions). 

\subsection{Two reference frames, three timescales\label{sec:times}}

\begin{figure}
\centerline{\psfig{file=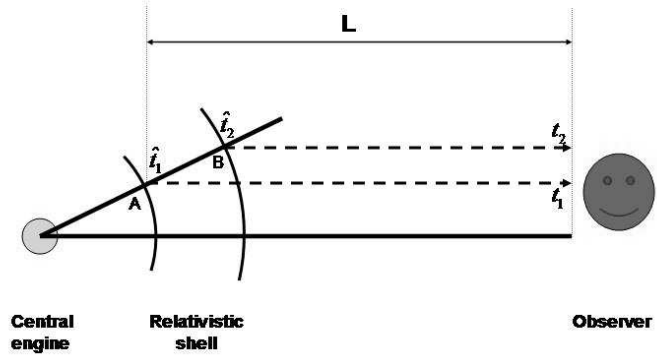,angle=0,width=8cm}}
\vspace*{35pt}
\caption{The geometric configuration among the GRB central engine,
relativistic emitting shells and the observer.}
\label{fig:geometry}
\end{figure}

A typical GRB problem involves three major physical elements
(Fig. \ref{fig:geometry}): 
a central engine, a relativistically moving shell (ejected by the central
engine) which produces the GRB emission, and an observer. There are essentially
only {\em two} inertial frames. One is the rest frame of both the
central engine and 
the observer (aside from a cosmological redshift factor, which is small 
compared to special relativistic effects), and the other is the rest 
(comoving) frame of the relativistic shell or ejecta. The physical quantities 
(e.g. scale length and time) as viewed in the two inertial frames are
different,  
and are related through special relativistic Lorentz transformations. For 
example, a length scale $\Delta'$ in the comoving frame is converted
to $\Delta = \Delta'/\Gamma$ along the shell's moving direction in the
rest frame of the central engine. 

Similarly there are only two sets of clocks attached in both inertia
frames, so that $dt'= d \hat t / \Gamma$, where $d\hat t$ and $dt'$ are 
the time intervals elapsed for {\em the same pair of events} in the central
engine/observer frame and the comoving frame, respectively. However,
in the GRB problem, there is a third relevant time scale involved. The
complication comes from the propagation effect. In general, we can
consider a shell emitter moving with a dimensionless speed $\beta$,
at an angle $\theta$ with respect to the line of sight of the
observer (Fig. \ref{fig:geometry}). In the rest frame of the central
engine/observer, 
the shell emits a first photon towards the observer at the time $\hat
t_1$ at the location A (the radius $r$), and emits a second photon
towards the observer at time $\hat t_2$ at the location B (the radius
$r+dr$), as recorded by clocks precisely adjusted in this inertia
frame. The time interval for emitting these two successive photons is 
$d \hat t = \hat t_2 - \hat t_1 \simeq dr/c$. Let's assume that the
distance between the observer and the location A is $L$. The first photon 
arrives at the observer at $t_1=\hat t_1 + L/c$, while the second photon 
arrives at the observer at $t_2 =
\hat t_2 + (L/c - \beta \mu d \hat t)$ where $\mu=\cos\theta$. These
two times are also measured by clocks precisely adjusted in the
same inertial frame. The time interval for the observer to receive the
two adjacent photon signals is simply 
\be
dt = (1-\beta \mu) d \hat t \simeq d \hat t/2\Gamma^2 = dr/(2\Gamma^2c),
\label{time}
\ee
where in deriving the second part of the equation, we have assumed
$\Gamma = (1-\beta^2)^{-1/2} \gg 1$ and $\theta \ll 1$ (so that $\mu
\sim 1$). Notice that eq. (\ref{time}) is a pure propagation effect,
which is also valid for the non-relativistic case. It is not noticeable in 
the Newtonian case, simply because $d \hat t \simeq dt$ when $\beta
\ll 1$. In the relativistic regime, this effect becomes very
important, and eq.(\ref{time}) is one of the fundamental arguments used to
solve the ``compactness problem'' as discussed above in \S\ref{sec:comp}. 

It is worth noticing that $d \hat t$ and $dt$ are two different times in 
{\it the  same inertial frame}, which describe two different pairs of
events. $d \hat t$ describes 
the time for the shell to {\em emit} two photons, while $dt$ describes 
the time for the observer to {\em receive} the same two photons. Since 
$d \hat t$ also describes the actual time of the shell behavior
(e.g. when the shell moves to the location A or location B), the best
way to refer to these two times is to call $d \hat t$ the ``time in 
the rest frame of the central engine'' or ``the time in the fixed 
(or lab) frame'', while calling $dt$ the ``observer's time''. In the 
literature, the latter is sometimes called ``the time in the observer's 
frame'', which is in principle right, but this is not the exact 
distinction between $dt$ and $d \hat t$. Sometimes in the literature
$d \hat t$ is called ``the time in the burster's frame''. This can be
confusing to some readers, since the ``burster" may be understood either
as the central engine powering the burst (in which case the definition 
is correct), or it may be understood as the flying shell which emits 
the burst (in which case the definition is incorrect). Also notice
that in some articles, the $d \hat t$ discussed here is denoted as $dt$, 
while the $dt$ discussed here is denoted as $dT$, $dt_{obs}$ or
$dt_\oplus$. Since the observer's time is the most relevant one to
describe the phenomenon, it is more convenient to define it as $dt$,
the most straightforward notation. In the rest of this review, we will 
adhere to such a notation system.

A variant of $dt$ which is also widely discussed in GRB problems
is the so-called ``angular time'' $dt_{ang}$. Instead of fixing
$\theta$ and varying $r$ (or $\hat t$), in some problems one needs to
fix $r$ (and hence $\hat t$) but vary $\theta$. This is relevant for
the problem in which a shell-like emitter (with a fixed radius $r$) 
is illuminated instantaneously at the same engine-frame $\hat t$, while 
the observer sees a long-duration emission caused by the time delay 
of radiation coming from higher latitudes. In a similar manner as for
the derivation of $dt$, one has $dt_{ang} = (r/c) \sin\theta d\theta$, 
or $t_{ang} = (r/c) (1-\cos\theta)$. For relativistic motion
($\Gamma\sim 1/\theta \gg 1$), one has $\theta \sim \sin\theta \sim 1/\Gamma$,
so that $t_{ang}\simeq (r/c \Gamma^2)$, which is of the same order as
$t$ (eq.[\ref{time}])\cite{fmn96,sp97a}. Notice again that $t_{ang}$ is
also a pure propagation effect and is also applicable for the
Newtonian case. 

The third time scale is the comoving time of the shell, measured by
another set of clocks. It is related to the engine-frame time $d \hat t$ 
through 
\be
d t' = d \hat t / \Gamma = dt/ \Gamma (1-\beta\mu) = {\cal D} dt \simeq
2\Gamma dt, 
\label{doppler}
\ee
where ${\cal D} = [\Gamma(1-\beta\mu)]^{-1}$ is the Doppler factor 
(since the observed radiation frequency is boosted by the
same factor with respect to the frequency in the comoving frame, i.e., 
$\nu={\cal D} \nu'$). The final approximation in eq.(\ref{doppler})
is again for $\Gamma \gg 1$ and $\mu \sim 1$.

\subsection{Fireball evolution, characteristic radii and
times\label{sec:radii}} 

\begin{figure}
\centerline{\psfig{file=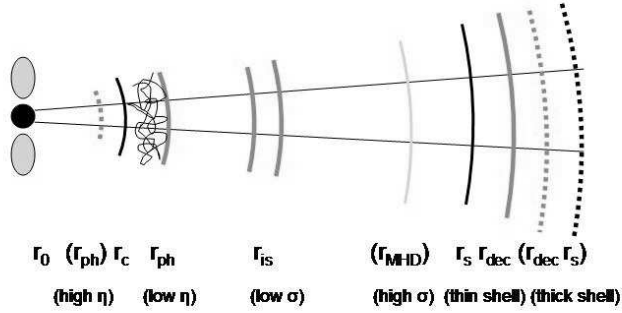,width=8cm}}
\vspace*{8pt}
\caption{Various characteristic radii in a generic relativistic
fireball.}
\label{fig:radii}
\end{figure}

The evolution of generic fireballs expanding into an ambient medium 
has been extensively
studied\cite{cr78,p86,p90,g86,mlr93,psn93,sp95,kps99,mr00,mrrz02},
assuming a fireball composition of photons, electron/positron pairs 
and a small amount of baryons (but negligible magnetic fields). 
In a simplest toy model, we assume the following input parameters: (a) 
an average constant luminosity of the central engine, $L$; (b) the
duration of the central engine energy injection, $T$, so that the
total energy of the fireball is $E=LT$ and the initial width of the
whole fireball shell in the fixed frame is $\Delta_0=cT$; (c) the
variability time scale, $t_v \sim 1~{\rm ms}~\ll T$, which is
due to the intermittent nature of the fireball central engine,
as reflected by the spiky, irregular GRB lightcurves (\S\ref{sec:grb1}).
The fireball shell therefore actually consists of many mini-shells;
(d) the average mass loading rate, $\dot M$, so that the so-called
dimensionless entropy is $\eta=L/\dot Mc^2$; and (e) the particle
(usually hydrogen) number density of the ambient medium, $n$ (in the
simple toy model here, $n$ is taken as a constant, which is typical
for an interstellar medium, ISM). The evolution is characterized by
several characteristic radii (see Fig. \ref{fig:radii} for a
cartoon picture), which we will discuss in turn below. 

\begin{itemlist}
\item Initial state ($r=r_0$): The base of the fireball 
flow is connected to the GRB central engine, a black hole - torus
system or a rapidly rotating magnetar. Assuming a $10 M_\odot$ black
hole, the initial length scale (taken 3 Schwarzschild
radii)\cite{mr00} is $r_0 \sim 6 GM/c^2 \sim 10^7$ cm. This radius is
also the typical width of the mini-shells, i.e. $\delta_0 = c t_v 
= (3\times 10^7~{\rm cm})~(t_v/1~{\rm ms})$. The initial
state of the fireball is hot, with photons and pairs in equilibrium
with a temperature ${\cal T}_0=(L/4\pi r_0^2\sigma)^{1/4} \sim
(10^{10}~{\rm K})~ L_{51}^{1/4} r_{0,7}^{-1/2}$. Hereafter the
convention $Q=10^n Q_n$ will be adopted in c.g.s. units (e.g. $L_{51}$
means luminosity in unit of $10^{51}~{\rm erg~s^{-1}}$). In the
initial state, the baryons are essentially at rest with respect to the
central engine; 
\item End of ejection ($r = \Delta_0$): After the central engine
constantly ejects energy (with luminosity $L$) for a time $T$, the
fireball radius is $r =\Delta_0 = cT = (3\times 10^{10}~{\rm
cm})~(T/1~{\rm s})$;
\item Coasting radius ($r=r_c$): As the fireball shell expands, the
baryons will be accelerated by radiation pressure. The fireball bulk
Lorentz factor increases linearly with radius, until reaching the
maximum Lorentz factor $\Gamma_0$, which is usually the dimensionless
entropy $\eta$, but sometimes lower than $\eta$ due to the limited
radiation pressure\cite{mr00,mrrz02}. The fireball finally coasts with 
a constant Lorentz factor $\Gamma_0$ at the coasting
radius\cite{mlr93,psn93,kps99}, $r_c = \Gamma_0 \Delta_0 \sim
(10^{13}~{\rm cm})~(T/1~{\rm s})(\Gamma_0/300)$ (this is the radius at which
the entire mass of the fireball has achieved the coasting $\Gamma$).
During acceleration, since all the materials essentially move with
speed of light, the shell width in the fixed 
frame remains constant, $\Delta=\Delta_0$. The comoving shell width
$\Delta'=\Gamma \Delta_0$, on the other hand, increases with radius
linearly. The above conclusions hold 
assuming the fireball shell evolves as a whole. However, if the mini-shells 
evolve separately, i.e., for well-separated GRB pulses, these mini-shells 
are likely to evolve independently, and coast at a smaller radius $r_c =
\Gamma_{sh} \delta_0$, where $\Gamma_{sh}$ is the Lorentz factor of
that particular mini-shell, and $\delta_0 = ct_v$ is the width of the
mini-shells; 
\item Photospheric radius ($r=r_{ph}$): As the fireball shell
expands, the photon number density and the typical photon energy drop.
At a certain radius, the photons become optically thin to both
pair production and to Compton scattering off the free electrons
associated with baryons entrained in the fireball. 
At this radius, although much of the initial energy is
converted to the kinetic energy of the shell, some energy will be
radiated away as emission from $r_{ph}$ with an approximately blackbody 
spectrum (the fireball in the Big Bang has similarly a blackbody spectrum,
seen now as the cosmic microwave background emission from the last 
scattering surface). This is the first electromagnetic signal detectable 
from the fireball. This photosphere radius $r_{ph}$
is usually above the coasting radius $r_c$, with a temperature ${\cal
T} ={\cal T}_0 (r_{ph}/r_c)^{-2/3}$, but could be below $r_c$ if the
initial fireball is clean enough (i.e. a large enough $\eta$), in
which case ${\cal T} = {\cal T}_0$. 
The typical value of the photosphere is $r_{ph} \sim
(10^{12}-10^{13})$ cm for $\eta \sim (100-1000)$. A full discussion
about various regimes of this ``baryonic'' photosphere is presented in 
Ref. \refcite{mrrz02}; 
\item Internal shock radius ($r=r_{is}$): For an intermittent central
engine with typical variability timescale of $t_v$, the typical 
distance between adjacent mini-shells is usually also characterized as
$d = ct_v$. Suppose that a rear shell moves faster than a leading shell, 
i.e., $\Gamma_{(2)} \gg \Gamma_{(1)}\sim \Gamma_0$, the (fixed frame) 
time at which the fast shell catches up with the slow shell is $\hat 
t_{is}=d/(v_{(2)} - v_{(1)}) \simeq 2\Gamma_{(1)}^2 d/c$, and the 
distance is $r_{is} \simeq \hat t_{is} c \simeq 2\Gamma_{(1)}^2 d 
\simeq 2\Gamma_0^2 ct_v \simeq (6\times 10^{13}~{\rm
cm})~(\Gamma_0/100)^2 (t_v/0.1~{\rm s})$. At such distances,
mini-shells collide with each other and typically form strong
``internal'' shocks\cite{rm94,kps97,dm98};
\item Shell spreading radius ($r=r_s$): For a shell with initial width 
$\Delta_0$, assuming the Lorentz factor scatter $\Delta \Gamma$ within the 
shell is of order $\Gamma_0$, the shell starts to spread at a radius $r_s 
\simeq \Gamma_0^2 \Delta_0$, based on a similar argument as used to derive 
the internal shock radius\cite{mlr93}. The spreading radius of the mini-shells 
is roughly the internal shock radius. The spreading radius of the whole 
GRB shell is however much longer, typically $r_s = (3\times 10^{15}
~{\rm cm})~ (\Gamma_0/100)^2 (T/10~{\rm s})$. Beyond the spreading radius 
$r_s$, the shell width starts to spread in the fixed frame, i.e., 
$\Delta = r/\Gamma^2$ for $r>r_s$. 
\item Deceleration radius ($r=r_{dec}$): The fireball shell is 
eventually decelerated by the ambient medium (e.g. ISM). During the
intial fireball-medium interaction, a reverse shock propagates into
the fireball to stop it. Usually a deceleration radius ($r_{dec}$) is
defined as the radius where the reverse shock crosses the fireball
shell. For a prompt fireball or a fireball whose duration is short
enough, we have $r_{dec}=r_\Gamma$, i.e., the radius where the ISM
mass collected by the fireball is equal to $(1/\Gamma_0)$  
of the fireball rest mass, i.e., $M_{\rm ISM} \simeq \Delta M/\Gamma_0$,
where $\Delta M=\dot M T$ is the total baryon loading of the fireball. 
For constant density medium, this radius is $r_\Gamma=
(3 E_{iso}/4\pi n m_p c^2 \Gamma_0^2)^{1/3} \simeq (2.6\times
10^{16}~{\rm cm})~ (E_{iso,52}/n)^{1/3} (\Gamma_0/300)^{-2/3}$. 
As long as the shell spreading radius $r_s$ is less than $r_\Gamma$
(or $t_\Gamma=r_\Gamma /c\Gamma_0^2=>T$), the fireball decelerates at 
$r_{dec} = r_\Gamma$. This occurs at an observer-frame time\cite{rm92}
$t_{dec}\equiv t_\Gamma \simeq 5~(E_{iso,52}/n)^{1/3}
(\Gamma_0/300)^{-8/3} (1+z)$ s.
In literature this is usually termed the ``thin shell'' case\cite{sp95,kps99}. 
Alternatively, if the shell is thick enough (e.g. for a long duration
of fireball ejection so that $T>t_\Gamma$), the deceleration radius 
moves further out to $r_\Gamma < r_{dec} < r_s$. This is the ``thick  
shell'' case\cite{sp95,kps99}. As the fireball starts to decelerate, a
strong external shock also forms and propagates into the medium. So the
deceleration radius is essentially the initial external shock radius.
\end{itemlist}

Since the fireball moves essentially at the speed of light, the relevant
fixed-frame (or central-engine frame) times corresponding to the fireball 
reaching various radii are simply derived by dividing the relevant distance 
with $c$. For example, the time for internal shocks to happen is about several 
hours while the deceleration time is about 10 days. If we imagine a GRB 
occurring in our neighborhood, say 10 pc from us, whose relativistic jet beam
is exactly perpendicular to our line of sight, we would be able to follow 
the real time advance of the GRB emission and trace when the jet head reaches 
various radii. [For $\theta=\pi/2$ and $\mu=0$, we have $d
\hat t=dt$ in eq.(\ref{time})]. However, when we see a GRB from 
cosmological distances, by definition, the jet is beamed towards us. 
The relativistic propagation effect (eq.[\ref{time}]) squeezes all the 
timescales to within seconds.
For example, the time delay between the the onset of the
internal shock emission and the launch of the fireball
(which may be due to the collapse event whose trigger time may be
recorded by future gravitational wave detectors) is only $t_{is} \sim 
(r_{is}/c) /2\Gamma_0^2 \sim (18~{\rm ms})~r_{is,14}
(\Gamma_0/300)^{-2} (1+z)$, while the delay for the external shock
emission is only $t_{dec} =t_\Gamma \simeq (r_\Gamma/c) /2\Gamma_0^2
\sim (5~{\rm s}) (E_{iso,52}/n)^{1/3} (\Gamma_0/300)^{-8/3} 
(1+z)$ (for the thin shell case). In both expressions, the factor 
$(1+z)$ takes into account the cosmological time dilation effect.  

\subsubsection{How common is the thick shell case?}

Although the thick shell case has been widely
discussed\cite{sp95,kps99,k00,kz03a,kz03b}, it is worth questioning
whether a thick shell description is indeed relevant in reality. In
all the current discussions, it is conventionally assumed that the
fireball shell width is defined by the duration of the GRB itself,
i.e. $\Delta_0 = cT/(1+z)$ (where the factor $(1+z)$ is again to
correct the cosmological time dilation). By comparing $r_s =
\Gamma_0^2 \Delta_0$ with $r_\Gamma$ (or alternatively comparing $T$
with $t_\Gamma$), one can define a critical Lorentz
factor\cite{k00,kz03a,zkm03} $\Gamma_c = 125 (E_{iso,52}/n)^{1/8}
(T/100~{\rm s})^{-3/8} [(1+z)/2]^{3/8}$. For $\Gamma_0 > \Gamma_c$,
which is not difficult to satisfy, the fireball is in the thick shell
regime. Hydrodynamically, the reverse shock becomes relativistic in
the thick shell regime, while in the thin shell case, the reverse
shock keeps Newtonian until reaching mildly relativitic at the
deceleration radius.

However, the above analysis is based on the assumption that the
central engine has kept an essential constant luminosity through out
$T$. In reality, the GRB energy injection is likely to be intermittent,
sometimes with a broad gap between emission episodes. In such cases,
it makes no more sense to take $\Delta_0 = cT$. Instead, one should
separate the whole duration into several well-defined emission
episodes, i.e., broad pulses, and treat these sub-shells to decelerate 
independently. One therefore has several consecutive thin shells, but
rarely a thick shell. This situation has actually been indicated in GRB
990123, the only burst from which the rising lightcurve of the optical 
flash was caught\cite{a99}. In the standard afterglow model the
optical flash is attributed to the reverse shock emission 
(see \S \ref{sec:reverse}). The optical
flash peak time (which corresponds to the time when the reverse shock 
cross the shell, $t_\times=t_{dec}$) is at 50 s after the burst trigger. 
The GRB duration, however, was 63 s, which exceeds $t_\times$. This already
is incompatible with the thick shell regime, since in theory, $t_\times$ 
should not be less than $T$. Furthermore, the rising lightcurve is very steep,
which is consistent with a thin shell model, but inconsistent with a
thick shell model which predict a much shallower rising slope\cite{k00}. 
Inspecting the GRB lightcurve in detail, we find that 
the gamma-ray lightcurve of GRB 990123 is well represented as consisting
of  two components, a first more intense one which lasts less than 50 s, 
and a second less intense one with even shorter duration. Thus, it is
reasonable to regard 50 s as the deceleration time of the first shell
in the thin shell regime. More generally, we suspect that in most
cases when a GRB runs into an ISM, the thin shell case is very
common. The thick shell description, may be more relevant in
the case that a GRB runs into a pre-stellar wind or a constant dense
medium. In such a case, the critical time for the thick shell case is
much shorter due to the high density of the wind at the deceleration
radius\cite{kz03b,kmz03}. 

\subsection{Relativistic shocks}

The collision between the relativistic fireball and the ISM leads to 
a relativistic forward shock and possibly a relativistic reverse shock 
as well. The internal shocks are usually at least mildly relativistic
due to the large Lorentz factor contrasts between the colliding
shells. 

For a relativistic shock, when the upstream matter is cold, the shock
jump condition gives\cite{bm76,sp95}
\ba
n_2 &= &(4\gamma_{21}+3) n_1 \simeq 4\gamma_{21} n_1, \nonumber \\
e_2 &= &(\gamma_{21}-1) n_2 m_p c^2 \simeq \gamma_{21} n_2 m_p c^2
\simeq 4\gamma_{21}^2 n_1 m_p c^2.
\label{shock}
\ea
The subscripts ``1'' and ``2'' denote the unshocked (up stream) and
shocked (down stream) materials, respectively; $n$ is number density,
$e$ is internal energy, both measured in the comoving frames of the
fluids, and $\gamma_{21}$ is the relative Lorentz factor
between the fluids 2 and 1, while the Lorentz factor of the shock
front itself is $\gamma_{s}=\sqrt{2} \gamma_{21}$.

In discussing the interaction between two fluids, two shocks are
formed simultaneously at the instant of contact, which propagate into the 
two fluids, respectively. There are four regions for the two fluids
separated by the forward shock, the contact discontinuity, and the
reverse shock. 
For the GRB external shock case, these four regions are (1) unshocked
ISM, (2) shocked ISM, (3) shocked shell, and (4) unshocked
shell\cite{sp95}. The shock jump condition (eq.[\ref{shock}]) 
also applies at the reverse shock. A final condition $e_2=e_3$ (since
$p=e/3$ for relativistic fluids, and equal pressure is required at 
the contact discontinuity) and the fact that the fluids move at
the same speed across the contact discontinuity finally close the
problem, and a simple 
analytical description about the system is available\cite{sp95}. 
Similar analyses can be applied to collisions between the
mini-shells. Each collision is accompanied with a pair of internal
shocks propagating into the two colliders.

There can also be more complicated cases, e.g. involving three (or more)
shell interactions. This happens in the phase when a fireball shell is
already decelerated by the ISM, while a trailing shell (which could be 
ejected by the central engine at a later time, or be ejected by
the engine at essentially a same time as the leading shell 
but with a much lower Lorentz factor) catches up with the decelerated
leading shell and collides with it. This is a typical problem in the
study of GRB afterglow evolution when there is additional energy
injection (see \S \ref{sec:inj}), and 
the recent observed ``step-like'' optical afterglow lightcurve of GRB
030329\cite{gcn} provides a strengthened motivation to study such a
theoretical problem. We have performed a detailed analysis of the
three-shell-interaction hydrodynamics\cite{zm02a}. In this case,
there are altogether six regions separated by three shocks and two
contact discontinuities. By applying shock jump conditions to all 
three shocks (and noticing that the leading shell is hot), one can
derive a self-consistent solution only when 
the relative Lorentz factor between the trailing and the leading shell 
exceeds a critical value defined by the energy ratio between the two shells. 
Otherwise, the injection is only mild, with the injected material connecting 
to the initial shell as a single long-lasting thick shell.

\subsection{Synchrotron emission\label{sec:syn}}

The GRB prompt emission is clearly non-thermal, and so is the afterglow 
emission. The most natural mechanism for  non-thermal emission is
synchrotron emission, i.e. emission from relativistic electrons
gyrating in random magnetic fields. The question of whether the GRB 
prompt emission (e.g. the $\gamma$-rays) is definitely due to synchrotron 
is still subject to debate (see \S\ref{sec:mech}). However, the synchrotron 
shock model is widely accepted as the major radiation mechanism in the 
external shock, which is thought to be responsible for the observed 
broad-band afterglows.

There are three major assumptions that are adopted in almost all the
current GRB afterglow models. Firstly, electrons are assumed to be
``Fermi'' accelerated at the relativistic shocks and to have a power-law 
distribution with a power-law index $p$ upon acceleration,
i.e. $N(E_e) d E_e \propto E_e^{-p} d E_e$. This is consistent with
current shock acceleration numerical
simulations\cite{a01,ed02,lp03}. Secondly, a fraction $\xi_e$ (generally
taken to be $\siml 1$) of the total electrons associated with the ISM 
baryons are accelerated, and the total electron energy is a fraction
$\epsilon_e$ of the total internal energy in the shocked region. 
Thirdly, the strength of the magnetic fields in the shocked region is 
unknown, but its energy density ($B^2/8\pi$) is assumed to be a fraction 
$\epsilon_B$ of the internal energy. These ``micro-physics''
parameters, $p$, $\epsilon_e$ ($\xi_e$) and $\epsilon_B$, reflect our
inability of tackling the problem, whose values are usually fitted
from the data\cite{wg99,pk01,pk02,harrison01,yost03}. 

There are several critical energies in the power-law distribution of
the electrons. For $p>2$ (which is consistent with numerical
simulations, and seems to be consistent with most of the observational 
data), a lower limit is set by the requirement that the average energy 
density in the shock-heated region is $\gamma_{21} n_2 m_p
c^2$ (eq.[\ref{shock}]), which reads
$\gamma_m = g(p) (m_p/m_e) (\epsilon_e/\xi_e) \Gamma \sim 310
[g(p)/(1/6)] (\epsilon_e/\xi_e) \Gamma$, where $g(p)=(p-2)/(p-1)$ with 
$p=2.2$ adopted, and we have simply redefined $\gamma_{21}$ as $\Gamma$.
According to the standard synchrotron emission theory\cite{rl79}, the
radiation power of an electron is $P_e=(4/3) \sigma_T c \gamma_e^2
(B^2/8\pi) \propto \gamma_e^2$, so that high energy electrons ``cool'' 
more rapidly. For a continuous injection, as is the case in an
afterglow (i.e. the forward shock keeps plowing into the ISM), there 
is a break in the electron spectrum at $\gamma_e=\gamma_c$, above
which the electron energy spectrum is steepened due to cooling,
i.e. $N(E_e) d E_e \propto E_e^{-p-1} d E_e$. This energy is
time-dependent, which is defined by equating the comoving dynamical
timescale of the blastwave ($t'\sim \Gamma t$, eq.[\ref{doppler}]) to
the cooling timescale of the electron ($t'_c=\gamma_e m_e c^2/
P_e$)\cite{mrw98,spn98}. Finally, the maximum energy of the
electrons ($\gamma_{M}$) is defined by equating the typical comoving
acceleration timescale, $t'_{acc} \sim 2\pi r_L/c$ (where $r_L$ is the
Larmor radius), with the shorter one of the dynamical timescale and the
cooling scale. For electrons, the latter is relevant,
which results in $\gamma_{M} \sim (3e/\sigma_T B)^{1/2} \sim 5\times
10^7 (B/1~ \mbox{G})^{-1/2}$, where $e$ is electron charge here.

\begin{figure}
\centerline{\psfig{file=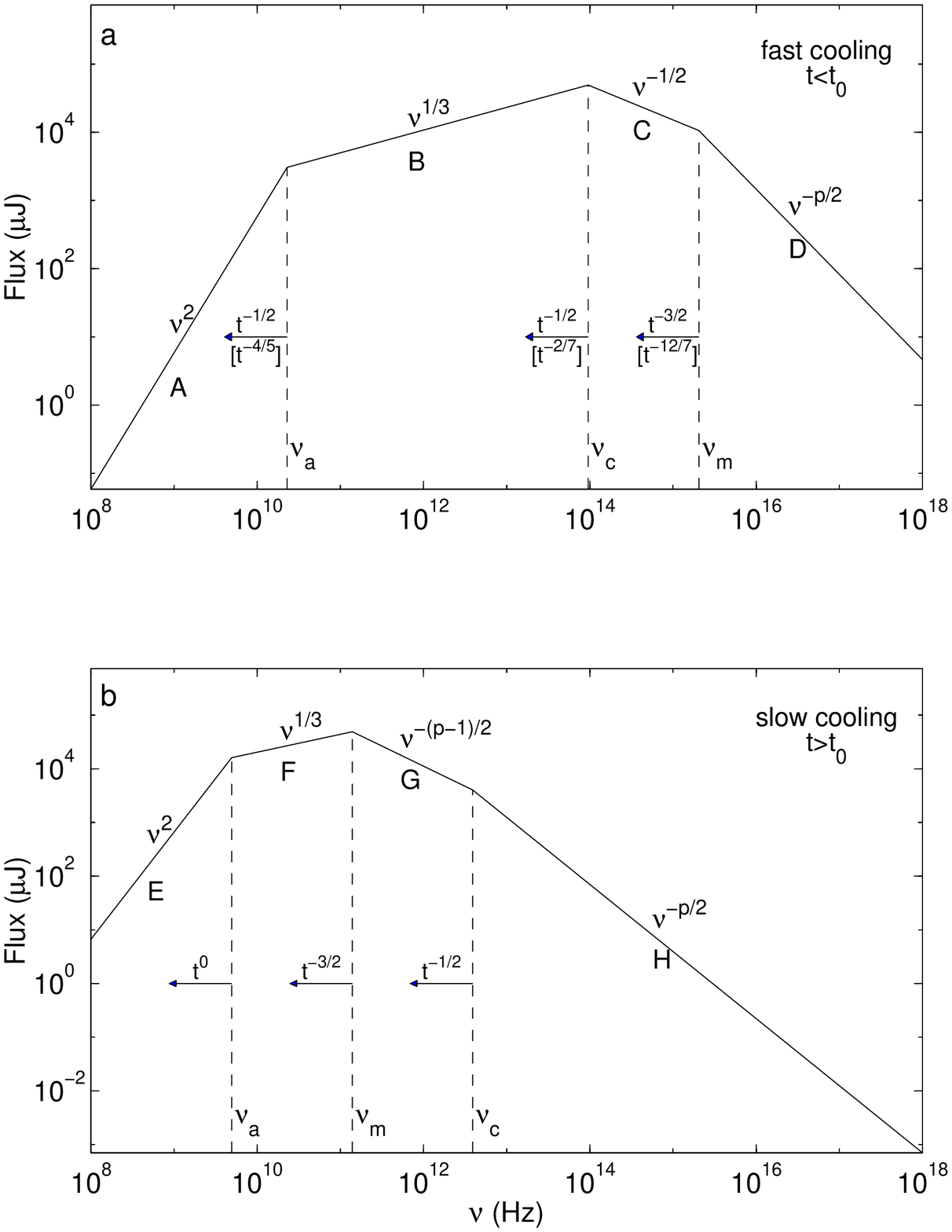,width=5cm}
            \psfig{file=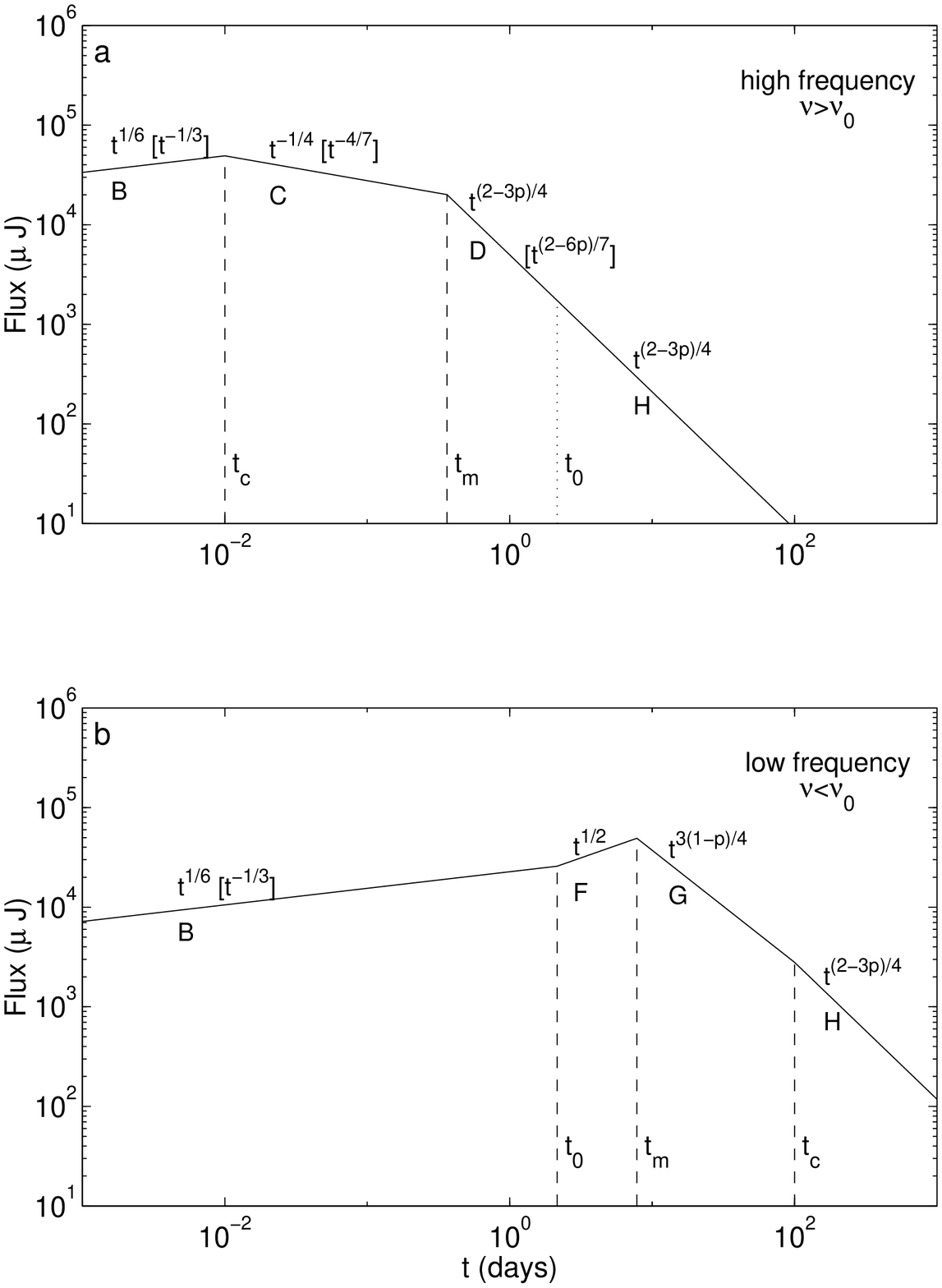,width=4.8cm}}
\vspace*{8pt}
\caption{The afterglow synchrotron spectra and lightcurves for the
simplest fireball blastwave model (from Ref.86).}
\label{fig:sari}
\end{figure}

The typical observed emission frequency from an electron with (comoving) 
energy $\gamma_e m_e c^2$ and with a bulk Lorentz factor $\Gamma$ is 
$\nu=\Gamma \gamma_e^2 (eB/2 \pi m_ec)$. Thus, three critical frequencies 
are defined by the three characteristic electron energies. These are
$\nu_m$ (the injection frequency), $\nu_c$ (the cooling frequency),
and $\nu_M$ (the maximum synchrotron frequency). In the afterglow
problem, there is one more frequency, $\nu_a$, which is defined by
synchrotron self-absorption at lower frequencies. So the final GRB
afterglow synchrotron spectrum is a four-segment broken power
law\cite{spn98,mrw98} separated by the typical frequencies $\nu_a$, $\nu_m$,
and $\nu_c$. Depending on the order between $\nu_m$ and $\nu_c$, there
are two types of spectra\cite{spn98} (Fig. \ref{fig:sari}). For $\nu_m <
\nu_c$, which is 
called the ``slow cooling case'', the spectrum is
\begin{equation}
 F= F_{\nu,m}\left\{ \begin{array}{l@{\quad \quad}l}
              (\nu_a/\nu_m)^{1/3}(\nu/\nu_a)^2  & \nu < \nu_a \cr
              (\nu/\nu_m)^{1/3} & \nu_a \le \nu<\nu_m \cr
              (\nu/\nu_m)^{-(p-1)/2}  &  \nu_m \le \nu < \nu_c \cr
              (\nu_c/\nu_m)^{-(p-1)/2}(\nu/\nu_c)^{-p/2} & \nu_c \le
		\nu \le \nu_M
          \end{array} \right.
\label{slowc}
\end{equation}
For $\nu_m > \nu_c$, which is called the ``fast cooling case'', the
spectrum is
\begin{equation}
 F= F_{\nu,m}\left\{ \begin{array}{l@{\quad \quad}l}
              (\nu_a/\nu_c)^{1/3}(\nu/\nu_a)^2  & \nu < \nu_a \cr
              (\nu/\nu_c)^{1/3} & \nu_a \le \nu < \nu_c \cr
              (\nu/\nu_c)^{-1/2}  &  \nu_c \le \nu < \nu_m \cr
              (\nu_m/\nu_c)^{-1/2}(\nu/\nu_m)^{-p/2} & \nu_m \le \nu
		\le \nu_M
          \end{array} \right.
\label{fastc}
\end{equation}
There are several more complicated regimes, involving 
self-absorption\cite{gps99b,kmz03}. In the above expressions, the
normalization factor is calculated by multiplying the total number of
radiating electrons $4\pi r^3 n_1/3$ by the peak flux from a single
electron\cite{spn98}, which is only the function of $B$ and is
independent of the energy ($\gamma_e$) of the electron\cite{spn98,wg99}. 

One direct consequence of synchrotron emission is that the emission
from an individual particle is polarized\cite{rl79}. Theoretical
predictions about the polarization degree of afterglow emission have
been made\cite{gw99,ml99,gruzinov99}. Due to the random nature of the
post-shock magnetic fields, the polarization is likely to be largely
averaged out, and only a small degree of polarization is left. 
The most likely situation which can result in detectable linear 
polarization in an afterglow is thought to be around the jet-break
time\cite{gl99,sari99}, in which case a collimated jet and an
off-axis line of sight conspire to produce an asymmetry which can lead
to net polarization. In particular, for a large enough offset of the
line-of-sight, a 90$^{\rm o}$ change of polarization angle is
predicted\cite{gl99,sari99} around the jet break time for the simplest 
uniform conical jet model. Current afterglow (optical) polarization
observations detect low-degree ($<5\%$) of polarization
emission\cite{bjornsson03,covino03}, but the data are too sparse to 
adequately test the concrete models. 

\subsection{Simplest afterglow model}

The simplest afterglow model\cite{mr97,wrm97,w97a,w97b,mrw98,spn98}
is based on several assumptions, which minimize the complications. 
These assumptions include (a) isotropic fireball; (b) constant ambient
density (ISM); (c) impulsive injection in the fireball; (d) relativistic
fireball; (e) synchrotron emission of the electrons; and (f) constraints
on the microphysics parameters (e.g. no evolution, $p>2$, etc). Under such
conditions, the fireball Lorentz factor evolves with radius $r$ (or $\hat t$)
and with observer's time $t$ as 
\ba 
& \Gamma \propto r^{-3/2} \propto t^{-3/8}, & r \propto t^{1/4}, 
\nonumber \\ 
& \Gamma \propto r^{-3} \propto t^{-3/7}, & r \propto t^{1/7}~. 
\label{ag} 
\ea 
The first scaling is valid for an adiabatic evolution of the fireball 
in which the energy $E \propto n r^3 \gamma^2$ is constant, 
which is generally valid at late epochs (later than hours) in all
afterglows, and is also valid at earlier epochs for many afterglows. The
second scaling is valid for the evolution of a ``radiative'' fireball,
in which the total energy in the fireball decreases prominently due to
radiation loss\cite{mrw98,pm98b,cps98,bd00a}, while the momentum
$\propto n r^3 \gamma$ conserves. This extreme radiative regime is
only valid for the fast-cooling regime when $\epsilon_e \sim 1$. This
last condition appears unlikely, according to the current afterglow
fits\cite{wg99,pk01,pk02,harrison01,yost03}. More generally, a
blastwave may not be strictly adiabatic or radiative. A reasonable
treatment is to adopt a quasi-adiabatic evolution with small radiative 
correction\cite{sari97}.

With a certain dynamics (e.g. adiabatic evolution), one can quantify
the time evolutions of the critical frequencies, $\nu_m$, $\nu_c$,
$\nu_a$, as well as the normalization $F_{\nu,m}$, parameterized in
terms of the burst properties (i.e. $E_{iso}$, $n$ and luminosity
distance $d_L$) and the micro-physics parameters (i.e. $p$, $\xi_e$,
$\epsilon_e$, and $\epsilon_B$). Although the scalings on all these
parameters are the same for different works, different treatments of
normalization easily cause a factor of a few difference in the
coefficients. Here we take the latest value of the
coefficients\cite{hsd02}, in which a typical value $p=2.2$ has been
adopted. The dependence of unknown $\xi_e$ is also incorporated.
\ba
\nu_m &=& (6\times 10^{15} ~\mbox{Hz})~ (1+z)^{1/2} E_{52}^{1/2}
(\epsilon_e/\xi_e)^2 \epsilon_B^{1/2} (t/1~{\rm day})^{-3/2} \label{sd1}\\
\nu_c &=& (9\times 10^{12} ~\mbox{Hz})~ (1+z)^{-1/2} \epsilon_B^{-3/2}
n^{-1} E_{52}^{-1/2} (t/1~{\rm day})^{-1/2} \label{sd2}\\
\nu_a &=& (2\times 10^9 ~\mbox{Hz})~ (1+z)^{-1} (\epsilon_e/\xi_e)^{-1}
\epsilon_B^{1/5} n^{3/5} E_{52}^{1/5} \label{sd3}\\
F_{\nu,m} &=& (20~\mbox{mJy})~ (1+z) \epsilon_B^{1/2} n^{1/2} E_{52}
d_{L,28}^{-2}. \label{sd4}
\ea
At a given time a comparison of $\nu_{m}$ and $\nu_c$ shows whether the
flow is in the slow cooling or fast cooling regime, and using eqs.(\ref{slowc})
and (\ref{fastc}) one can calculate afterglow lightcurves for a particular
band (i.e. fixing a particular observation frequency). The different segments
on the lightcurves have different temporal decay indices that
correspond to different spectral regimes\cite{spn98}. The commonly
used temporal indices include the following. In X-rays, shortly after
the trigger, the temporal index is essentially constant,
i.e. $(2-3p)/4$, which is steeper than $-1$ for typical $p$ values. 
This is consistent with the observations. In the optical band, the
lightcurve first rises as $\propto t^{1/2}$, peaks around hours, and
then decays with an index $-3(p-1)/4 \sim -1$. Although the rising
lightcurve has not been firmly detected (but has been
inferred\cite{kz03a}) due to the technical limitations, the falling
part of the lightcurve is generally consistent with the 
observational data. The rising lightcurve may be buried by the
contribution of the reverse shock emission at earlier times (as might
have been the case of GRB 990123 and GRB 021211), but there is a large 
region of parameter space for which this part is not expected to be buried 
and should be observable\cite{zkm03}. 

A convenient way to test various model regimes is to perform 
simultaneous measurements of both the temporal and the spectral
indices. Writing $F_{\nu} (t,\nu) \propto t^{\alpha} \nu^{\beta}$,
the relations of $\alpha$ and $\beta$ in various regimes are listed in 
Table 1. Also listed in the table are the jet, wind, and $p<2$ models, 
discussed below.
\begin{table}[ht]
\tbl{Temporal index $\alpha$ and spectral index $\beta$ in various
afterglow models, the convention $F_\nu \propto t^\alpha \nu^\beta$ 
is adopted, from Refs. 85, 86, 110, 289, 326. 
The assumption $\nu_a < {\rm min} (\nu_m,\nu_c)$ is made. (Under certain 
conditions, e.g. for the wind fast cooling case in some limited regime, 
the higher $\nu_a$ case is relevant$^{289}$, so that the values 
collected here are no longer valid). The jet model applies for the sideways
expanding phase, which is valid for both ISM and wind cases and is
usually in the slow cooling regime.}
{\begin{tabular}{@{}lcclcl@{}} \toprule
 & $\beta$ & $\alpha$ ($p>2$, $p\sim 2.3$) & $\alpha(\beta)$ & $\alpha$
($1<p<2$, $p\sim 1.5$) & $\alpha(\beta)$ \\ \colrule
ISM, slow cooling & & & \\ \colrule
$\nu < \nu_a$ & 2 & $\frac{1}{2}$ & & $\frac{17p-26}{16(p-1)} \sim
-0.06$ & \\
$\nu_a<\nu<\nu_m$ & $\frac{1}{3}$ &  $\frac{1}{2}$
& $\alpha=\frac{3\beta}{2}$ & $\frac{p+2}{8(p-1)} \sim 0.9$ & \\
$\nu_m<\nu<\nu_c$ & $-\frac{p-1}{2}$ &  $\frac{3(1-p)}{4} \sim -1.0$
& $\alpha=\frac{3\beta}{2}$ & $-\frac{3(p+2)}{16} \sim -0.7$
& $\alpha=\frac{3(2\beta-3)} {16}$\\
$\nu > \nu_c$ & $-\frac{p}{2}$ &  $\frac{2-3p}{4} \sim -1.2$
& $\alpha=\frac{3\beta+1}{2}$ & $-\frac{3p+10}{16} \sim -0.9$
& $\alpha=\frac{3\beta-5} {8}$\\ \colrule

ISM, fast cooling & & & \\ \colrule
$\nu < \nu_a$ & 2 & 1 & & 1 & \\
$\nu_a<\nu<\nu_c$ & $\frac{1}{3}$ &  $\frac{1}{6}$
& $\alpha=\frac{\beta}{2}$ & $\frac{1}{6}$ & $\alpha=\frac{\beta}{2}$ \\
$\nu_c<\nu<\nu_m$ & $-\frac{1}{2}$ &  $-\frac{1}{4}$
& $\alpha=\frac{\beta}{2}$ & $-\frac{1}{4}$
& $\alpha=\frac{\beta}{2}$ \\
$\nu > \nu_m$ & $-\frac{p}{2}$ &  $\frac{2-3p}{4} \sim -1.2$
& $\alpha=\frac{3\beta+1}{2}$ & $-\frac{3p+10}{16} \sim -0.9$
& $\alpha=\frac{3\beta-5} {8}$\\ \colrule

Wind, slow cooling & & & \\ \colrule
$\nu < \nu_a$ & 2 & 1 & & $\frac{13p-18}{8(p-1)} \sim
0.4$ & \\
$\nu_a<\nu<\nu_m$ & $\frac{1}{3}$ &  0
& $\alpha=\frac{3\beta-1}{2}$ & $\frac{5(2-p)}{12(p-1)} \sim 0.4$ & \\
$\nu_m<\nu<\nu_c$ & $-\frac{p-1}{2}$ &  $\frac{1-3p}{4} \sim -1.5$
& $\alpha=\frac{3\beta-1}{2}$ & $-\frac{p+8}{8} \sim -1.2$
& $\alpha=\frac{2\beta-9} {8}$\\
$\nu > \nu_c$ & $-\frac{p}{2}$ &  $\frac{2-3p}{4} \sim -1.2$
& $\alpha=\frac{3\beta+1}{2}$ & $-\frac{p+6}{8} \sim -0.9$
& $\alpha=\frac{\beta-3} {4}$\\ \colrule

Wind, fast cooling & & & \\ \colrule
$\nu < \nu_a$ & 2 & 2 & & 2 & \\
$\nu_a<\nu<\nu_c$ & $\frac{1}{3}$ & -$\frac{2}{3}$
& $\alpha=-\frac{\beta+1}{2}$ & -$\frac{2}{3}$
& $\alpha=-\frac{\beta+1}{2}$ \\
$\nu_c<\nu<\nu_m$ & $-\frac{1}{2}$ &  $-\frac{1}{4}$
& $\alpha=-\frac{\beta+1}{2}$ & $-\frac{1}{4}$
& $\alpha=-\frac{\beta+1}{2}$ \\
$\nu > \nu_m$ & $-\frac{p}{2}$ &  $\frac{2-3p}{4} \sim -1.2$
& $\alpha=\frac{3\beta+1}{2}$ & $-\frac{p+6}{8} \sim -0.9$
& $\alpha=\frac{\beta-3} {4}$\\ \colrule

Jet, slow cooling & & & \\ \colrule
$\nu < \nu_a$ & 2 & 0 & & $\frac{3(p-2)}{4(p-1)} \sim -0.8$ & \\
$\nu_a<\nu<\nu_m$ & $\frac{1}{3}$ &  $-\frac{1}{3}$
& $\alpha=2\beta-1$ & $\frac{8-5p}{6(p-1)} \sim 0.2$ & \\
$\nu_m<\nu<\nu_c$ & $-\frac{p-1}{2}$ &  $-p \sim -2.3$
& $\alpha=2\beta-1$ & $-\frac{p+6}{4} \sim -1.9$
& $\alpha=\frac{2\beta-7}{4}$\\
$\nu > \nu_c$ & $-\frac{p}{2}$ &  $-p \sim -2.3$
& $\alpha=2\beta$ & $-\frac{p+6}{4} \sim -1.9$
& $\alpha=\frac{\beta-3}{2}$\\ \colrule
\end{tabular}}
\label{table:alpha-beta}
\end{table}

More sophisticated model takes into account the radiation of the 
whole bulk of the blast wave, which is usually modeled as a
Blandford-McKee\cite{bm76} self-similar profile, as well as the
emission contributions from the so-called equal-arrival-time surface,
which is typically is egg-shaped or pear-shaped\cite{w97c,pm98c,sari98,gps99a}. 
These effects result in smooth transitions around the spectral and 
temporal breaks, which provide a better model fit to the afterglow data. 
A complete description is presented in Ref. \refcite{gs02}.

Another important ingredient of the standard afterglow model is the 
emission component from the reverse external shock. 
During the early phase of the fireball shell - ISM interaction, a
reverse shock propagates into the fireball shell itself while the 
forward shock propagates into the ISM\cite{mr97,mrp94}. The reverse shock
heats up the shell and accelerates electrons which emit synchrotron
radiation as well. The shock is short-lived and ends when it crosses
the shell. After this shock crossing time the electrons cool, leaving
a rapidly decaying emission component. Due to the high particle density 
in the shell compared to that in the external medium (e.g. the ISM), the 
peak frequency of the reverse shock is in the optical/IR band at the 
crossing time $t_\times$. Thus the reverse shock emission usually leads 
to an optical flash\cite{mr93b,mr97,sp99a,sp99b,mr99}. The early optical
afterglow data of GRB 990123, GRB 021004 and GRB 021211 are consistent 
with the reverse shock
interpretation\cite{sp99a,mr99,ks00,kz03a,zkm03,f03b,wei03}. 
We discuss in more detail the reverse shock and early afterglow signatures
in \S\ref{sec:reverse}.

\subsection{Additional features of realistic afterglow models}

The six assumptions made to define the simplest afterglow model are
introduced mainly to simplify the problem. In reality, they are not
necessarily satisfied, or not in all cases (although in many cases, 
they work surprisingly well). Violations of one or more of these 
assumptions lead to second-generation or modified fireball models. 
Such effects are in fact naturally expected, and they can describe the 
physical problem more realistically without the need for other ad-hoc 
assumptions. In this sense, they are also part of the standard fireball 
paradigm. The studies of the following effects comprise a majority of 
the recent GRB afterglow theoretical modeling effort.
Because of space limitations, we will not go into the details of
each of them, but rather refer the readers to the original papers that
discuss these effects.

\subsubsection{Jets\label{sec:jet1}}
The ``isotropic'' assumption is one of the likeliest to be unrealistic, 
and investigations of the consequences of assuming that GRBs are collimated 
relativistic flows, i.e., jets\cite{r97,r99}, have been fruitful. There 
are two quantities colloquially (and confusingly) referred to as the 
``beaming factor" in the GRB problem. One is the geometric beaming factor, 
which in the simplest jet model is just the opening angle of the jet, 
$\theta_j$. The second is the relativistic beaming factor, i.e., the 
emission of a particle or an object that moves with a Lorentz factor $\Gamma$ 
is beamed into a cone with opening angle of $1/\Gamma$. 
More accurately, it is useful to refer to the first one as the collimation
factor (or angle), and to the second one as the relativistic beaming factor.
Jets give rise to an interesting interplay between the two effects. 
Initially the jet is ultra-relativistic, with $1/\Gamma < \theta_j$. 
An observer on the beam only receives information from within the
relativistic light cone and has no knowledge about whether outside this cone
the emitter is radiating or not. The description of the dynamical
evolution is therefore equivalent to the isotropic case. As the jet
slows down, eventually the relativistic beam becomes wider than the 
geometric beam or collimation angle, i.e. $1/\Gamma > \theta_j$. Two effects 
come into play. First is the edge effect, i.e., the observer starts to feel 
a deficit of energy per solid angle (which is the key parameter in 
afterglow theories, denoted as $E_{iso}/4\pi$ in the above notations). 
Second, the causally connected region starts to extend to the whole
jet cone around the same time, and can keep expanding sideways. This 
means the jet can start to expand sideways\cite{r99}. The times for the 
two effects to take effect are close to each other\cite{pm99}, or may
coincide\cite{sph99}, depending on the assumption of the unknown
expansion speed. In the asymptotic regime for times much longer than the
jet break time, the dynamics is 
\ba
\Gamma \propto \exp(-r/l) \propto t^{-1/2}, & r\propto t^0
\ea
(where $l=[E_j/(4\pi/3) n m_pc^2]^{1/3}$ is the so-called Sedov
length at which the collected ISM rest mass energy is equal to the jet 
energy $E_j$ itself), so that the temporal dependence of various critical
frequencies as well as the temporal decay indices all differ from the
isotropic case (Table 1). In the asymptotic regime the post-jet-break 
optical lightcurve should have $\propto t^{-p} \sim t^{-2}$, much
steeper than the isotropic case ($\propto t^{-1}$). Therefore a jet
break is the natural interpretation for the apparent steepening
observed in many GRB optical afterglows.

Although the behavior in the asymptotic regimes are well known, the
detailed behavior near the jet break involves various complex effects
including the jet spreading hydrodynamics. Many efforts have been made 
to model the smoothness of the jet
breaks\cite{pm99,msb00,hgdl00,gpkw02}, but consensus is yet to be
achieved even for the simplest uniform jet case. Hydrodynamical
numerical simulations are needed, but so far such studies are still
preliminary\cite{granot01,cannizzo03}. Furthermore, GRB jets may not
be simply uniform, but may be structured\cite{rlr02,zm02b}. This could 
bring more interesting effects in modeling (see \S\ref{sec:jets} for
more discussions).

\subsubsection{Non-uniform external medium: winds and bumps}

The GRB ambient density may not be uniform. A well-discussed scenario 
involves an external medium produced by a stellar wind from the massive 
star progenitor, such as, e.g., a Wolf-Rayet star. The wind can be modeled
to first approximation as ejecting material with a constant mass loss 
rate ($\dot M$) and velocity ($v_w$). Conservation of mass gives a 
$\rho=A r^{-2}$ density profile\cite{cl99}, with 
$A=\dot M/4\pi v_w=5\times 10^{11} A_*~ \mbox{g~cm}^{-1}$. Such a
density profile changes the fireball dynamics into
\ba
\Gamma \propto r^{-1/2} \propto t^{-1/4}, & r \propto t^{1/2}, 
\ea
so that the lightcurves for the 
homogeneous fireball\cite{mrw98,dl98,cl99,cl00a}, for the collimated
jets\cite{kp00a,gdhl01}, as well as for the early afterglow involving
reverse shock emission\cite{wdhl03,kz03b}, are all modified
accordingly (Table 1). It is noticeable that most of the current GRB
afterglow data are consistent with a constant density external
medium\cite{pk01,pk02,f01}, although a handful of bursts could be well
modeled by the wind model\cite{cl00a,lc03,lc02}. 

An external density jump would be expected at the interface of the stellar 
wind and the ISM outside of it\cite{ramirez01}, which causes a distinct 
afterglow signature\cite{dl02,dw03}. The ISM itself may also have density
fluctuations, which would also add imprints on the afterglow
lightcurves\cite{wl00,lazzati02,npg03,hp03}. The consequence of a
sudden drop in the density profile has also been discussed\cite{kp00b}. 

\subsubsection{Post-injection and variable injection\label{sec:inj}}

The fireball deceleration time depends on the ambient density, but the
duration of the energy injection into the fireball is determined by the
central engine behavior. These two timescales are independent of
each other. Thus it is possible, or likely, that the central engine 
is still active while the afterglow starts. This gives rise to a 
post-injection of energy into the fireball, causing 
``refreshed shocks''\cite{rm98}, so called because the kinetic energy 
of the late-arriving material (catching up with the decelerated initial
material) revitalizes the external shock. The injection could be either 
in a continuous form\cite{rm98,dl98b,pmr98,sm00,zm01a} with modulations
superimposed, or discrete, e.g. through disrupted, collisional 
events\cite{kp00,zm02a}. The injection energy, both early and late, could be
either in the form of a Poynting-flux-dominated flow or
kinetic-energy-dominated 
shells\cite{zm02a}. However, the main characteristic of the post injection 
signature is that the flux level is systematically increased after each 
injection event without resuming the pre-injection level\cite{zm02a}, 
a signature that has been seen in GRB 030329\cite{gnp03}. 

Post-injection effect changes the energy per solid angle along the line of 
sight, and influences the fireball dynamics. If the fireball is not
uniform in the angular sense but is patchy, the fireball deceleration will 
cause bright or dim spots to enter the relativistic beam, which equivalently 
changes the energy per solid angle of the fireball along the line of
sight\cite{kp00b}. Both this model, and the varying density profile
model, were proposed to interpret the recent observed optical
lightcurve wiggles in GRB 021004\cite{npg03,hp03}. X-ray data could
disentangle the two 
scenarios, since above the cooling frequency $\nu_c$ emission
is in the density-independent regime. For GRB 021004, variability was
also found in X-rays\cite{f03a}, which lends supports to the patchy
beam model. 

\subsubsection{Relativistic to Newtonian transition}

A decelerating relativistic fireball eventually becomes
non-relativistic (Newtonian) at the Sedov radius $l=(3E/4\pi
nm_pc^2)^{1/3} = (1.2\times 10^{18}~{\rm cm})~ (E_{52}/n)^{1/3}$, when 
the collected ISM rest mass energy is equal to the fireball energy.
In the Newtonian phase, the fireball dynamics evolves
as\cite{wrm97,dl99,hdl99,fwk00,lw00,hc03} 
\ba
v \propto r^{-3/2} \propto t^{-3/5}, & r \propto t^{2/5}.
\ea
The temporal decay indices are\cite{dl99} $-(15p-21)/10$ for
$\nu_m<\nu<\nu_c$, and $-(3p-4)/2$ for $\nu>\nu_c$, which for typical
values of $p$ is steeper than $-1$ as expected for the relativistic
isotropic fireball, but is flatter than $-2$ as expected for the
post-jet-break case. Therefore the evolution of a collimated jet
involves a steepening around the jet break (typically days to weeks)
and a later flattening as the fireball transmits into the Newtonian
phase\cite{lw00}. This later transition time is $t_{N} \sim l/c
= (450~ \mbox{days}) (E_{52}/n)^{1/3}$, which is longer than a year
for typical parameters. Usually this is of observational interest in
the radio band\cite{fwk00}. In order to interpret some shallow
lightcurve steepenings with relativistic-to-Newtonian transition
effect (rather than the jet effect), a very dense medium ($n \sim
10^6$) is needed\cite{dl99}. A self-consistent analytical description
of the relativistic-Newtonian transition is proposed in
Ref. \refcite{hdl99}. 

\subsubsection{High energy spectral components\label{sec:highe}} 

Besides synchrotron emission, there are other mechanisms giving rise to
high energy spectral components which must be operative at some level, 
and which may have interesting observational consequences. A 
straightforward and widely-discussed component is the synchrotron 
self-inverse Compton (IC) component\cite{mr93b,mrp94,mr94,pm96,snp96,w97a,pl98,wl98,dbc00,dcm00,pk00,pm00,se01,zm01b,wdl01a}.
The IC effect plays two roles in studying GRB afterglows. First, electrons cool
both via synchrotron and IC, so that IC potentially influences the
value and evolution of $\nu_c$. Defining $Y=L_{IC}/L_{syn}$, it is
found\cite{se01} that $Y=[-1+(1+4\eta
\epsilon_e/\epsilon_B)^{1/2}]/2  \simeq (\eta
\epsilon_e/\epsilon_B)^{1/2}$ for $Y \gg 1$, where $\eta$ is the
overall (synchrotron plus IC) radiation efficiency. The condition for
IC cooling to be important is essentially $\eta \epsilon_e >
\epsilon_B$. The second role of IC is that it forms a second spectral
component, extending beyond the high end of the synchrotron spectrum. 
To first order, it can also be approximated as a four-segment broken power 
law, separated by three critical frequencies $\nu_a^{\rm IC}=\gamma_m^2
\nu_a$, $\nu_m^{\rm IC} = \gamma_m^2 \nu_m$, and $\nu_c^{\rm IC} =
\gamma_c^2 \nu_c$. In reality the spectrum is more rounded and the
power-law breaks are not as sharp\cite{se01}. The high energy end of
the spectrum is usually the Klein-Nishina limit\cite{zm01b}. The flux 
normalization can be derived through $F_{\nu,m}^{\rm IC} /
F_{\nu,m} \sim (16/3) \sigma_T \xi_e n r$, which is $<10^{-6}$ for
typical parameters and most times of interest\cite{zm01b}. The
condition for the IC component to stick out above the synchrotron
component is defined by $F_{\nu}^{\rm IC}(\nu_c^{\rm IC}) >
F_\nu({\nu_c^{\rm IC}})$ for slow-cooling, and $F_{\nu}^{\rm
IC}(\nu_m^{\rm IC}) > F_\nu({\nu_m^{\rm IC}})$ for fast-cooling, both
relating to the same physical condition\cite{zm01b}. Both this
IC-emission-dominated condition and the above IC-cooling-dominated
condition require $\epsilon_B \ll \epsilon_e$, although they are
in principle different from each other. It is worth noticing
that current afterglow modeling suggests that the shock condition
$\epsilon_B \ll \epsilon_e$ is common among various
bursts\cite{pk01,pk02,harrison01,yost03}, so that IC is potentially
important in afterglow physics.

When observing in a particular fixed band, the IC component peak would 
sweep across the band at a later time than the synchrotron. This leads 
to a distinct bump signature in the lightcurve. The IC peak time 
is\cite{zm01b} $t_{IC} =
(3.4~\mbox{days}) (\epsilon_e/0.5)^{0.89} \epsilon_{B,-2}^{0.08}$
$\xi_e^{1.63} E_{52}^{-0.06} n^{-0.66} (1+z)^{0.32}
\nu_{18}^{-0.68}$. The higher the energy band, the earlier the bump
shows up, and a denser medium helps to ease the bump condition needed
to detect the IC component. A GeV afterglow bump should be common for 
the currently favored shock parameters\cite{zm01b}. In the X-ray band, an 
IC bump is expected to emerge on the power-law decaying lightcurve around 
a couple of days if the medium density is moderately dense (say,
$n>5$)\cite{pk00,se01,zm01b}, and such a bump has already been detected in
GRB 000926\cite{harrison01}.

Besides the IC component, several other high energy spectral components 
are also expected, which are related to hadronic processes. It is believed 
that the relativistic shocks also accelerate protons besides
accelerating electrons. The protons also radiate via the synchrotron
mechanism, and can also interact with photons (e.g. synchrotron photons 
from the electrons) to produce pions and muons\cite{vietri97,bd98,totani98}.
The neutral pions decay into gamma-rays directly, and
charged pions and muons also emit gamma-rays via synchrotron
radiation\cite{bd98}. However, these components may not contribute
significantly to the afterglow emission for the currently-favored shock
parameters, and the parameter space regime for these components to be
important is small\cite{zm01b}. 

\subsubsection{Microphysics}

Although an electron distribution index $p>2$ is consistent with numerical 
simulations of shock acceleration\cite{a01,ed02,lp03}, and is generally 
consistent with the observational data\cite{pk01,pk02}, in some bursts (e.g. 
GRB 010222) the temporal decaying slopes both before and after the ``jet 
break'' are too shallow, so that one seems to require
$1<p<2$\cite{masetti01,stanek01}. In such a regime, the maximum energy 
power is distributed towards the high energy end of the electron spectrum,
bringing some new features for afterglow emission. A possible mechanism
for producing such distributions is given in Ref. \refcite{bm96}, and a study
of this afterglow regime is presented in Ref. \refcite{dc01}, the relevant
conclusions being also collected in Table 1.

Essentially all current afterglow model fits assumed non-evolution of all
the shock parameters, $p$, $\epsilon_e$, $\epsilon_B$. In principle,
these may change, but this is hard to quantify, and the effect may be
degenerate with other effects so that one might never be able to
disentangle them. Detailed afterglow fitting seems to be compatible
with the model that one or more such parameters evolve with
time\cite{yost03,pk04}. 
There might also be a gradient of magnetic fields behind the shock,
but the data suggest that fields do not decay very
rapidly\cite{rr03}. 

\subsubsection{Pair formation, neutrons, grains and other effects}
\label{sec:other}

There are a number of other physical processes that would modify the simplest
afterglow model. 

First, the prompt gamma-ray form a radiation front which moves ahead of
the blast wave and interacts with the ISM before the blastwave starts to
decelerate\cite{mt00,tm00}. The gamma-rays back-scattered by the medium
can interact with outgoing gamma-rays and generate electron-positron pairs. 
The pairs enhance the opacity, and for a moderately dense external medium, 
the process leads to a run-away pair-loading process that modifies the 
blastwave dynamics considerably\cite{tm00,db00,mrr01,b02}.  

Second, it is likely that there are free neutrons entrained in the
fireball\cite{dkk99,b03a,b03b}. If neutrons and protons coast at
essentially a same speed (which is not the case under certain
conditions\cite{bm00,mr00b}), the neutron shell would lead the proton
shell when the latter is decelerated by the ISM. The comoving neutron
decay time scale is $\tau'_n \sim 900$ s, so in the fixed frame, the
distance where neutrons decay is $R_\beta=c \tau'_n \Gamma_n \simeq
(0.8\times 10^{16} ~\mbox{cm})~ (\Gamma_n/300)$. This leads to a
neutron-decay trail extending to a distance $R_{trail} \sim 10 R_\beta 
\sim 10^{17} ~\mbox{cm}$. The trailing proton blastwave would interact 
with this neutron trail and form a possible bump on the
lightcurve\cite{b03a}. 
 
Third, if GRBs originate in star forming regions, the existence of 
dust grains would cause a new emission component, in the form of a  
``dust echo''\cite{eb00}. Conversely, dust may be destroyed by the
reverse-shock-induced UV-optical flash\cite{wd00}, or by the X-ray
afterglow\cite{fkr01}. The small-angle scattering of X-rays off the
dust grains, on the other hand, may cause a soft X-ray bump on the
afterglow lightcurve\cite{mg00}.

Finally, as supported by accumulating evidence, at least some GRBs are 
associated with supernova\cite{g98,k98b,sta03,h03}. Therefore, a
supernova lightcurve, usually signaled by a reddened bump, 
is superposed on the optical afterglow lightcurve (as well as, in some 
bright cases, by a distinctive supernova spectrum\cite{sta03,h03}).

\section{Problems\label{sec:prob}}

A cartoon of a GRB fireball is shown in Fig. \ref{fig:radii}. The
theoretical model 
we have been discussing so far only includes those aspects that
have been extensively tested, i.e., the afterglow part. This arises in
the outer part of the fireball, from which most of the information is 
retrieved. This picture is widely accepted in the GRB community, and 
is used  as a standard theoretical framework to confront with the 
afterglow data.  In this section, we zoom in towards the GRB central 
engine, and discuss several aspects of the model on which consensus 
has not been quite reached. The closer towards the central engine, 
the less we know about the physical processes going on. Nonetheless, 
theorists have utilized the limited information available to construct  
toy models to tackle the problems. For some of the ``problems'' discussed 
below there exists a leading model to interpret the phenomenon (e.g. the 
internal shock model for GRB prompt emission). However, there is (are) 
other competing model(s) that is (are) still widely discussed, and 
there are continuing debates on some of these issues.

A first major problem is the nature of the GRB prompt $\gamma$-ray 
emission itself. The origin of the gamma-rays has been debated since
their discovery. Although the discovery of the afterglows settled the 
GRB distance issue, eliminated many GRB models, and achieved a consensus
on attributing the afterglow radiation mainly to synchrotron radiation
from a blast wave, there are still several variants within the current 
generic fireball scheme which are argued about for interpreting the 
limited GRB prompt emission data\cite{zm02c}. The major uncertainties include 
(a) the unknown fireball content (e.g. how important are magnetic fields in
generating GRBs) and the unknown fireball 
energy dissipation mechanism (shocks or magnetic dissipation such as
reconnection), (b) the unknown location where the GRB prompt
emission is produced (i.e. internal or external), and (c) the
radiation mechanism (e.g. synchrotron 
radiation or other mechanisms such as Comptonization). These three
aspects of the 
problem will be discussed below in \S\ref{sec:mag}-\ref{sec:mech}. We then 
move on to discuss the global GRB jet structures (\S\ref{sec:jets}),
progenitors (\S\ref{sec:proge}) and central engines
(\S\ref{sec:engine}). Finally we discuss GRB environments
(\S\ref{sec:environ}) and uncertainties in the shock physics
(\S\ref{sec:shock}). 
Some of these questions have also been discussed in
Refs. \refcite{ghis03,zmw02}.

\subsection{Fireball content: kinetic energy or magnetically dominated?}
\label{sec:mag}

\subsubsection{Internal shock model}

The leading model for the GRB prompt $\gamma$-ray emission is the internal 
shock model\cite{rm94}, although there are various problems with it which
are unresolved. The main motivation for this model is to explain short time 
(down to ms) time variabilities\cite{mr94}, more efficiently than
e.g. external shocks\cite{sp97a} (c.f. Ref. \refcite{dm99}).
In this model, as discussed in \S\ref{sec:prog2}, it is 
assumed that the fireball behavior is essentially hydrodynamical, i.e.
baryon or kinetic energy dominated, and magnetic fields are neglected when 
considering the dynamics, and are only introduced when discussing 
synchrotron radiation through the parameterization 
of $\epsilon_B$.  As discussed in \S\ref{sec:radii}, if the central engine
ejects energy intermittently in the form of mini-shells, due to
non-uniformity of the shell Lorentz factor, these mini-shells would
collide at a typical distance of $r_{is} \sim 2 \Gamma_0^2 c t_v
\sim (6\times 10^{13}~{\rm cm})~ (\Gamma_0/100)^2 (t_v/0.1~{\rm s})$. 
The collisions produce a pair of shocks propagating into both shells,
heating them and accelerating electrons (and protons). With some magnetic
fields in the shell (either carried from the central engine or
generated in situ, e.g. by turbulent dynamos or instabilities), these 
electrons emit synchrotron radiation in the gamma-ray band after Doppler 
boosting. This is taken to be responsible for the observed GRB emission.  
In order to have the gamma-rays from an optically thin medium (to avoid 
thermalization), only internal shocks whose radii are above the baryonic 
photosphere ($r_{ph}$, see \S\ref{sec:radii}) contribute to the observed 
emission. There is another relevant issue currently widely considered. The 
gamma-rays generated from some closer-in internal shocks may have a large 
opacity to $\gamma\gamma$ interactions leading to pair production. The pairs 
generated at small radii form a Thomson scattering screen for the gamma-rays, 
and the radius where this Thomson scattering optical depth drops to unity
defines another ``pair photosphere'', $r_{pair}$, which is usually
above $r_{ph}$, limiting the minimum variability timescale of the
lightcurves\cite{gsw01,krm02,mrrz02}. The optical depth in this pair
photosphere may be self-regulated to be around a moderate value of a few,
so that variabilities shorter than the scale defined by $r_{pair}$
are only smoothed rather than completely washed out\cite{mrrz02}.

The internal shock model has been extensively studied by various
groups\cite{kps97,dm98,psm99,spm00,gsw01}. Its most successful aspect
is its ability to model the complex temporal profiles of GRB prompt
emission lightcurves. Numerical simulations indicate that the observed 
temporal behavior essentially reflects the temporal behavior of the
central engine\cite{kps97}. A caveat is that the central engine needs 
to be ``bare'', or have a channel leading out to a lower density
environment where optically thin shock radiation can be observed . 
In the currently favored ``collapsar'' scenario (\S\ref{sec:proge}) for 
long bursts, the central engine is located deep inside a collapsing star,
whose envelope may act as an additional agent to regulate the variability 
of the relativistic flow\cite{zwm03}. In any case, as long as the
relativistic jet emanates intermittently from the central engine, internal
shocks are likely to develop (a different situation arises if the jet is
strongly Poynting- dominated, as discussed next). The combination of the 
internal shock emission and the pair photosphere screen can reproduce
the -5/3 slope and the 1 Hz break in the power density spectra
(PDS)\cite{psm99,spm00,krm02} which are deduced from the 
data\cite{bss98,bss00}. 
An important characteristic of the internal shock model is that 
outflows with lower Lorentz factors have higher $E_p$ values\cite{gsw01,zm02c}.
This is contrary to most other models, and against the simple intuitive 
expectation that lower $\Gamma$ (dirty) shells have a smaller Lorentz boost 
so the emission should be softer. The reason for this behavior in internal
shocks is mainly that the low $\Gamma$ shells collide at closer distances 
relative to the central engine, where the magnetic fields are stronger. 
A major reason why this model is widely studied is that there is
a clear operational theoretical framework (e.g. shock jump condition,
equipartition parameters, synchrotron radiation, etc.) upon which
model simulations could be performed (in contrast to the
Poynting-dominated model discussed below).

There are several problems or caveats about the internal shock model. 
First, the emission energy has to be extracted from the
{\em relative} kinetic energy between the colliding shells, but
the radiation efficiency is typically small, e.g. around
$1\%-5\%$\cite{kumar99,psm99,spm00}. In order to achieve a high GRB
efficiency, as required by the fact that the prompt gamma-ray
energy is or the same order of the afterglow kinetic
energy\cite{f01,pk01}, some novel suggestions have been made, 
including non-linear dissipation\cite{b00} and quasi-elastic
collisions\cite{ks01}. In any case, a large relative Lorentz factor
dispersion in the flow is required\cite{b00,ks01,gsw01}. A recent
study indicates that by taking into account neutron-decay in the
fireball, the efficiency of the internal shock mechanism is further
reduced\cite{rossi04}. Second, the
BATSE bright burst spectral sample\cite{preece00} suggests that, at least
within the same burst, the peak photon energy $E_p$ distribution is narrow. 
This is hard to achieve within the internal shock model unless one
invokes a strong bimodal distribution of the Lorentz
factors\cite{gsw01,ak03}, which lacks straightforward physical origins. 
Although none of the above criticisms has displaced the internal shock model
from its position as the leading paradigm, these are important issues 
to consider. For this reason, other models are worth investigating as well,
some of which are discussed below.

\subsubsection{Fireball with strong magnetic fields}
\label{sec:magfireball}

There are several motivations for considering strong magnetic fields 
in the GRB problem, as opposed to the weaker fields needed to produce the
observed radiation. (1) Electromagnetic energy is ``clean" in the baryon 
load sense, and can propagate in vacuum. Since a GRB fireball typically 
requires a very small baryon loading to achieve the high Lorentz factor 
needed to solve the compactness problem, a Poynting flux potentially
makes it easy to 
transport a large amount energy without carrying much baryons\cite{le03}; (2)
Current GRB central engine models commonly invoke a rapidly rotating
black hole circulated by a debris torus, or a millisecond neutron
star. During collapse, magnetic flux conservation naturally gives a
field of order $10^{12}-10^{13}$ G, as is observed in radio pulsars. 
With rapid rotation near the breakup frequency, magnetic
fields are likely to be magnified via an $\alpha-\Omega$
dynamo\cite{dt92} to achieve $\sim 10^{15}$ G or higher. Such high
fields are also argued to be possible for direct 
collapse of a magnetized white dwarf\cite{u92}. A strongly-magnetized
rapidly-rotating central engine is a likely scenario; (3) Magnetic
fields are a possible agent to tap energy from the two energy
reservoirs in the engine, i.e., the gravitational energy of the torus, 
and the rotational energy of the black hole or the neutron
star\cite{bz77,mr97b,u92,lwb00,van01}. Another energy extraction
mechanism, i.e., the neutrino pair production process ($\nu\bar\nu
\rightarrow e^{+}e^{-}$)\cite{e89}, is found to just barely power GRBs
when a beaming correction is taken into
account\cite{ruffert97,ruffert99}, so that the magnetic power is at
least helpful in meeting the GRB energetic needs\cite{mrw99}; (4) Magnetic
fields are helpful in collimating jets\cite{vk03a,vk03b}. Other
arguments in favor of the magnetic mechanism include its ability to
alleviate the inefficiency problem of the internal shock
model\cite{sdd01,ds02}, and the possibility of achieving narrow $E_p$
distributions\cite{zm02c}. There are also three observational facts
that support a strongly magnetized central engine (although the flow
is not necessarily completely Poynting flux dominated). (1) The strong
gamma-ray polarization\cite{cb03} (cf. Refs. \refcite{rf03,bc03}) may
indicate a strongly magnetized 
central engine, either in pure Poynting-flux-dominated
form\cite{lyutikov03}, or in conventional hydrodynamical form but with
a globally organized magnetic field
configuration\cite{waxman03,granot03}; (2) Modeling the reverse shock
emission of GRB 990123 indicates that the reverse shock region should
anchor a stronger field (by a factor of 15 in strength) than in the
forward shock 
region\cite{zkm03,fan02}. A similar conclusion may also apply to GRB
021211\cite{zkm03,kp03}. These may indicate that the fireball contains 
a large portion of ``primordial'' magnetic fields carried from the
central engine, although the flow is {\em not} completely
Poynting-flux dominated (otherwise there should be no strong reverse
shock\cite{kc84}). (3) The non-detection of a bright photospheric 
thermal component may need a significant fraction of energy being
stored in the magnetic form, so that the photosphere temperature could
be reduced (i.e. by a factor of $(1+\sigma)^{-1/4}$) to evade
detection\cite{dm02,zm02c} (but see Ref.\refcite{gcg03}). 

The evolution of a fireball with large magnetic content, and how 
such a fireball may give rise to prompt gamma-ray emission has been
reviewed by various authors\cite{u99,wheeler00,blandford02,lb02,sd03},
but so far there is not a standard framework, as is the case for 
non-magnetic fireballs. The magnetic fireball evolution is now an interplay
among three components, an electromagnetic component, an internal
energy component, and a kinetic (bulk) energy component. On the other hand,
traditional fireballs only invoke the latter two, with the internal
component dominating in the beginning, converted into the kinetic
form via radiation-pressure-driven acceleration, and partially
re-converted back to the internal form via shock dissipation.  The total 
amount of energy of these two components is essentially conserved except 
for the radiation losses. To simplify the magnetic case, one can regard the 
sum of these two components (internal and kinetic) as one single component, 
through a parameter $\sigma$ which is the ratio between the energy densities
in the electromagnetic component and in the internal+kinetic 
component\cite{zm02c}.  At the beginning of the fireball evolution when
the kinetic component is negligible, $\sigma$ is just the ratio of the
cold component (in the form of Poynting flux) and the hot component
(in the form of photon-pair fireball generated via neutrino
annihilation)\footnote{We note that the coexistence of both components
is natural when a cataclysmic collapse event forms a rapidly rotating, 
strongly-magnetized compact object.}. During the coasting  
regime, when radiation-driven acceleration is completed, the ratio
$\sigma$ is simply the Poynting-flux-to-kinetic-energy ratio, as has
been widely discussed in pulsar wind nebula theories\cite{kc84}. In
the GRB problem that we are interested in, $\sigma$ is likely to decrease 
with radius, since the rapid spin of the central engine greatly eases the
conditions for magnetic reconnection and other instabilities to
develop, so that the magnetic field energy is dissipated\footnote{In
pulsar wind nebula theories, on the other hand, it has been a
long-standing problem to solve the so-called $\sigma$-problem, i.e.,
there is no obvious reason to reduce $\sigma$ from a pulsar wind (slow 
pulsars compared with GRB central engine) from a very high value
($10^4$) to a low enough value ($<1$) required in the reverse shocks
as inferred from the observations.}. About a half of the dissipated
magnetic energy is converted to internal energy, and the other half
energy is used to accelerate the fireball through magnetic pressure
gradient\cite{dren02,vk03a,vk03b,sd03}. If the dissipation radius is
smaller than the photosphere radius, the internal energy is also
converted to the kinetic energy eventually. However, if the
dissipation radius is above the photosphere radius, the internal
energy would be radiated eventually via the non-thermal electrons
accelerated during the reconnection
event\cite{sdd01,ds02,sikora03}. This is 
the main source to interpret the GRB prompt emission in this model. 

The inclusion of electromagnetic fields greatly complicates the
GRB problem. For not very high $\sigma$ flows, one simplified
treatment is magnetohydrodynamical (MHD). 
The adoption of such an approximation is justified when the
electromagnetic fields are fully coupled to the fluid, i.e., no
dissipation is allowed. As discussed above, a GRB magnetic fireball 
needs to be intrinsically dissipative in order to interpret bursts, 
thus casting the MHD approach in question. Nonetheless,
the MHD method is still used to solve the first-order
problem\cite{vk03a,vk03b}, assuming that the radiation loss is only
minor, and that $\sigma$ is not very high. For a high-enough $\sigma$
(e.g. $\sigma \simg \sigma_{c2} \sim 200$), the MHD condition breaks
down globally at a distance $r_{_{\rm MHD}}=(1.8\times 10^{18}~{\rm
cm}) ~ L_{52}^{1/2} [\sigma (1+\sigma)]^{-1/2} (t_v / 1~{\rm ms})
\Gamma_{0,2}^{-1} \simeq (1.8\times 10^{15}~{\rm
cm}) ~ L_{52}^{1/2} \sigma_3^{-1} (t_v / 1~{\rm ms})
\Gamma_{0,2}^{-1} $, which is defined by the condition that the real
plasma density is lower than what is required for
corotation\cite{zm02c,u94}, i.e. the Goldreich-Julian
density\cite{gj69}. Here $\sigma_{c2}$ is a critical value (at the
relevant radius, here it is $r_{_{\rm MHD}}$) above which 
$r_{_{\rm MHD}} < r_{dec}$, i.e., the MHD condition globally breaks
down before the fireball starts to decelerate\cite{zm02c}. Although
the condition for this to happen is stringent\cite{sdd01,zm02c}, when
it happens there might be a global magnetic field dissipation region
within the radius range $r_{_{\rm MHD}} < r < r_{dec}$, in which
electrons are randomly accelerated by turbulent electromagnetic fields,
giving rise to high energy radiation (although the typical
energy may not coincide with the observed GRB emission)\cite{lb01}.
In reality, such a high $\sigma$ value may not persist out to such a 
large radius, especially when reconnection-induced dissipation
presumably take place from the very beginning of the
flow\cite{sdd01,ds02,dren02}. Thus the MHD description may be generally
valid to reveal the first order physics\cite{vk03a,vk03b}. However,
the formalism is still rather complicated even in the MHD
formalism\cite{vk03a,vk03b}. When the 
fluid inertia is negligible, as is the case in the high-$\sigma$ limit, 
an even simpler treatment, i.e., the force-free approximation, could
be introduced\cite{blandford02,lb02}. Using this simple formalism, the 
evolution of a magnetic bubble is delineated, and is found
(surprisingly) to be similar\cite{lb02} to the self-similar evolution
of the hydrodynamical blast waves\cite{bm76}. An interesting finding
is that a point explosion gives rise to structured angular
distributions of energy and Lorentz factor\cite{lb02}, i.e.,
$\epsilon(\theta) \propto (\sin\theta)^{-2}$ and $\Gamma \propto
(\sin\theta)^{-1}$, which nicely matches one version of the structured
jet model\cite{rlr02,zm02b} (see more discussions in \S\ref{sec:jets}).

An important characteristic of a high-$\sigma$ flow is that strong
shocks cannot develop\cite{kc84}. This is because the shock frame
pressure is dominated by the magnetic field pressure, and not much
internal energy is available for radiation. The condition for strong
shocks is $\sigma < 0.1$\cite{kc84}. As a consequence, the internal
shock mechanism has to be replaced by something else if the GRB
outflow is strongly magnetized (e.g. $\sigma > 1$). Besides
reconnection-induced magnetic dissipation\cite{sdd01,ds02}, other
mechanisms include Comptonization of the photospheric emission via
Alfven turbulence\cite{t94} and synchro-Compton radiation in a large
amplitude electromagnetic wave as the magnetic fireball is decelerated 
by the ambient ISM\cite{su00}. In all these models, however, the
fireball is assumed to be converted back to hydrodynamical in the
deceleration phase in order to match with the achievements of the
hydrodynamical afterglow theories\cite{u99,lb02}. Such a transition,
although seemingly ad hoc at first sight, is in fact reasonable since such 
low-$\sigma$ flows have been observed in pulsar wind nebula, even though
the initial pulsar wind is definitely a high-$\sigma$ flow.

The development of magnetic-field-powered GRB models is still
preliminary. The difficulties that hamper its progress lie in the
intrinsic complication introduced by electromagnetic fields so that
it lacks an operational scenario to work on. The situation has
been changing recently\cite{sdd01,ds02,dren02,vk03a,vk03b,lb02}. One
potential problem for the magnetic model is that an extremely clean
fireball involves much fewer baryon-associated electrons. Giving the
same amount of dissipation energy, the typical energy of emission
tends to be much harder than the sub-MeV band\cite{zm02c}, unless
secondary pairs are involved in the problem, which on the other hand
tend to smear out sharp variabilities observed in GRB
lightcurves\cite{zm02c,mrrz02}. 

\subsection{GRB location: internal or external?\label{sec:site}}  

A second uncertainty in understanding GRB prompt emission lies in its 
location within the fireball. The non-thermal spectra suggest that 
the location should be above the photosphere radius $r_{ph}$, below
which the emission is opaque. Afterglow radiation limits the prompt
emission radius to be smaller than the deceleration radius
$r_{dec}$. Generally there are three suggestions in 
the literature\cite{zm02c}: (1) the external
models\cite{mr93,pm98,cd99,dm99,dbc99,su00} that suggest GRB prompt
emission occurs upon strong deceleration of the fireball at $r_{dec}$; 
(2) the internal models\cite{rm94,kps97,dm98,ds02,lb01} that suggest 
GRB occurring at a radius $r_{ph} < r < r_{dec}$; and (3) the
(innermost) models\cite{t94,mr00,mrrz02,krm02} that suggest GRB occurring
right above the photosphere ($r \simg r_{ph}$). The current consensus
is that the third emission component at most contributes to the final
emission only partially, while a debate between the first two
components (external or internal) is still going on.

The main focus of the debate is the GRB variability. For those bursts
containing chaotic, spiky pulses, an internal scenario seems more
natural to interpret the data. The variability could be due to the
intermittent behavior of the central engine\cite{kps97}, including
that induced from black hole accretion instability,
the intrinsic instabilities involved in the jet propagation
within the stellar envelope\cite{zwm03}, and the intrinsic
chaotic behavior of magnetic reconnections and instabilities for the
high-$\sigma$ scenario\cite{sdd01}. However, whether an external model 
can be ruled out is still uncertain. For a burst running into a
homogeneous ISM, if the central engine lifetime ($t_{eng}$) is
shorter than the deceleration time $t_{dec}$ (thin shell), the
observed duration of the burst should be simply\cite{rm92} $t_{ang}
\sim r_{dec}/\Gamma^2c$, which is caused by the effect of
angular spreading (\S\ref{sec:times})\cite{fmn96}. 
In such a case, an external shock
model only produces one single smooth pulse. The caveat is that the
external medium could be clumpy, and if the scale of the clumps $d$ 
is smaller than $r_{dec}/\Gamma$, variable lightcurves could be
produced. An argument against this scenario is that the process is 
likely to be inefficient\cite{sp97a}, the efficiency being estimated
as $\eta \sim t_v/T \ll 1$ to first-order.
However, taking into account the angle-dependent flux
effect\cite{dm99,dm03}, the observed variability may still be
reproduced with a higher efficiency. So solely from this fact, the two
scenarios can not be differentiated. The attractive feature of the
internal scenario is that the variability arises more naturally, while 
the external shock model has to introduce an additional assumption, i.e.,  
a (very) clumpy medium. It would then be worth checking the further
consequences of such a clumpy medium. Evidence for non-uniform medium has 
been suggested for GRBs with X-ray line features\cite{amati00,p00}, and for
GRB 021004 which shows a wiggling afterglow
lightcurve\cite{lazzati02} and multi-component absorption
features\cite{s03}. However, in most afterglows, a clumpy medium is
not required, and limits may be posed on the clumpiness. It is worth
modeling the prompt GRB emission as well as the afterglow emission within
the external shock model in a unified manner, and to make use of any
statistical correlations between the GRB variability and afterglow
irregularity, so as to validate or falsify the external shock GRB model. 

Other clues can also help to differentiate the two scenarios. (1) The
correlation between the ``waiting time'' and the amplitude of the next
pulse\cite{rr01} favors an engine-dominated (internal shock) model; 
(2) In the external shock model, one expects pulse widths spreading with
time\cite{frrw99}, which is not apparent from the data. However, this
objection may be circumvented by taking into account the
noise\cite{dm03}; (3) The absolute value of $\Gamma$ enters the
problem in the external shock model, while only the relative value of
$\Gamma$ enters the problem in the internal shock model. As a result,
some correlations are expected in the external shock model but not
necessarily in the internal shock model. For example, a
brightness-spectral hardness correlation is expected in the external
shock model\cite{pm98}, which is consistent with the data. A
brightness-duration anti-correlation is also expected, but this is not
apparent from the data. A statistical study about various correlations
among duration, flux and $E_p$ in the external shock model\cite{bd00b}
suggests that the model predictions are compatible with the data, but
the results do not necessarily rule out the internal shock model. For the
internal shock model, the burst duration is expected to be independent
with flux and spectral behavior. 

Eventually, close monitoring of gamma-ray and X-ray emission at the
transition time of prompt emission and early afterglow would be able
to clearly identify whether or not there is a distinct afterglow
component to stick in at a later time. This would eventually settle
down the internal-external debate. This is in principle doable in the
{\em Swift} era. Preliminary evidence of two distinct components is
available for some bursts\cite{giblin02}.

One point (which is sometimes overlooked) is that the internal and 
external shock scenarios are {\em not} mutually exclusive.
Both are part of a generic cosmological fireball\cite{rm94,mr94,ps98} model. 
It is likely that both components coexist in all fireballs, with different 
components dominating in different bursts. For example, in some parameter 
regimes, the internal shock radii are beyond the external shock 
(deceleration) radius, so in this case, no internal shock is expected. 
There is a small fraction of BATSE bursts that show single-peak FRED-like 
temporal profiles\cite{fm95}. These events are most naturally interpreted 
as of external shock origin\cite{kp03}. 

In the magnetically-dominated fireballs, there are also an internal
emission component\cite{ds02,lb01} and an external emission
component\cite{su00,lb02}. A highly variable lightcurve is 
possible for both cases. 

\subsection{GRB emission mechanism: synchrotron, or others?\label{sec:mech}}

The leading radiation mechanism introduced to interpret GRB prompt
emission is synchrotron emission\cite{mr93,mr93b,mrp94,rm94,k94}. 
This is the most natural mechanism, which is found successful to
interpret afterglows and many other astrophysical phenomena. Many of
the observed GRB spectra are found to be consistent with this
interpretation\cite{tavani96a,tavani96b,cohen97}. However, the
synchrotron model encounters difficulties in explaining several
observational facts. The most notable one is sometimes called the
synchrotron ``line-of-death''\cite{preece98}, i.e., the low energy 
photon number spectral index should not exceed -2/3 (or the $F_\nu$
spectral index should not exceed 1/3). A good portion of
GRBs seem to violate this limit\cite{preece98} in the BATSE database. 
Proposals to solve this apparent inconsistency include critical
discussions of the sensitivity of BATSE to give accurate slopes at  
such low energies\cite{lp00}, introducing small-angle ``jitter''
radiation\cite{medvedev00}, anisotropic electron pitch angle
distribution and synchrotron self-absorption\cite{lp00,lp02}, as well
as photospheric emission\cite{mr00}. Another argument raised against
the synchrotron mechanism is the cooling problem\cite{gcl00}. The
synchrotron cooling time scale is very short, so that the prompt
emission should be in the ``fast-cooling'' regime with very low
cooling frequency, and 
an $F_\nu$ spectral index of -1/2 (or photon number spectral index
-3/2) should obtain, either for the low or the high energy
spectral indices, but this is not apparent in the data. This
criticism may be circumvented by taking into account more complicated
acceleration and cooling processes\cite{lp00,lp02}. 

Although the synchrotron model continues to be the main paradigm, some 
other alternative radiation mechanisms have been proposed to interpret 
GRB prompt emission. The most natural extension is to involve
Compton scattering. If the typical synchrotron emission frequency is well
below the sub-MeV band\footnote{In fact, in order to raise the synchrotron
typical energy to the BATSE band, some parameters need to be pushed to
extremes. For example, $\epsilon_B$ needs to be close to unity, and
$\xi_e$ (the fraction of electrons that are shock accelerated from the 
shocks) needs to be sometimes less than unity\cite{bm96,dm98}, so that the
energy per electron is raised.}, then the GRB prompt emission could be 
due to synchrotron self-inverse-Compton (IC)
emission\cite{pm00}. Caveats about this suggestion include, (a) a
strong IC component indicates a higher radiation energy density than
the magnetic field energy density, so that a positive feedback results 
in increasingly stronger higher-order IC components before the
Klein-Nishina cutoff, and consequently greatly increases the total
energy demand of GRBs\cite{dkk01}; (b) a strong IC component requires
$\epsilon_B/\epsilon_e \ll 1$\cite{pk00,se01,zm01b}, while in internal 
shock scenarios, $\epsilon_B$ can not be too small given a
strongly-magnetized central engine\cite{zm02c}; (c) an IC mechanism
tends to further increase the $E_p$ dispersion due to the high power
dependence of $E_p$ on $\Gamma$\cite{zm02c}, which within the internal
shock synchrotron model is already problematic without introducing
bimodal distributions of $\Gamma$'s\cite{gsw01,ak03}.

Alternatively, models involving Comptonization of thermal or
quasi-thermal particles have been proposed. A saturated Compton cooling 
model seems to be able to fit the observed GRB spectra
well\cite{liang97a,liang97b}. It is unclear how the model parameters 
needed for the fitting could be generated naturally in the fireball model, 
although attempts to transplant this mechanism to the
internal shock scheme has been made\cite{gc99}. Comptonization off a
very dense photon bath as the GRB mechanism was also
proposed\cite{lgcr00}, with the soft photon field being provided by 
the trapped photons within the massive envelope funnel of the
progenitor\cite{glcr00}. A proton-induced pair-photon synchrotron
cascade model, in analogy to a nuclear pile reaction, was proposed
recently\cite{kgm02}, which can reproduce a narrow $E_p$ distribution 
around 0.5 MeV (electron rest mass). For the usual luminosity and bulk 
Lorentz factors, and when cosmological redshift correction is
incorporated, this seems to be consistent with the data\cite{lr02}.

The report of a measurment of strong linear polarization in gamma-ray 
prompt emission\cite{cb03}, though subject to debate and
confirmation\cite{rf03,bc03}, potentially holds important clues for 
understanding the GRB radiation mechanism.
Models producing high gamma-ray polarization include Poynting-flux 
dominated models\cite{lyutikov03}, models involving hydrodynamical 
shells with entrained globally structured magnetic 
fields\cite{waxman03,granot03}, and models involving off-beam 
observers for a narrow jet\cite{waxman03,npw03}. Models involving inverse
Compton scattering with offset beaming angles can also give rise to
large degrees of polarization\cite{sd95,el03,darder03,lazzati04}. 
Additional information is needed to differentiate these possibilities.

Within the standard synchrotron model, there is a straightforward way
to analyze the $E_p$ distribution within various fireball
models\cite{zm02c}. In general, one can write
\be
E_p \sim \Gamma \gamma_e^2 (\hbar e B/m_e c) (1+z)^{-1} 
\sim (2\times 10^{-8}~\mbox{eV})~ (\Gamma B) \gamma_e^2 (1+z)^{-1},
\ee
where $\Gamma$ is the bulk Lorentz factor of the fireball, $\gamma_e$
is the typical comoving Lorentz factor of the electrons, $B$ is the
comoving magnetic field, and $z$ is the redshift. In principle, $B$
could have two origins, one carried from the central engine, and
another generated locally (e.g. in the shocks), although recent
evidence suggests that the engine-origin magnetic field may be the
dominant component\cite{cb03,zkm03}. Both field components, once formed, 
have the same dependence on distance from the central engine, 
i.e., the transverse part goes as $B \propto r^{-1}$, if both 
$\sigma$ and $\epsilon_B$ do not
change substantially with radius\cite{zm02c}. The $r-$dependence of
$\gamma_e$ essentially determines the $E_p$ distributions for the
models. For the 
internal shock model, since $\gamma_e$ is determined by the relative
Lorentz factor between the two colliding shells, it should not have a
strong dependence on $r$, so that $E_p \propto (\Gamma B) (1+z)^{-1}
\propto L^{1/2} r^{-1} (1+z)^{-1}$. The dependence on $r$ reflects
a strong dependence of $\Gamma$ ($E_p \propto
\Gamma^{-2}$)\cite{zm02c,rl02}, which is the intrinsic reason why the
internal shock model cannot reproduce a very narrow $E_p$ distribution
within a same burst (unless bimodal distribution of $\Gamma$ is
introduced\cite{gsw01,ak03}). For a magnetic-dominated acceleration model,
on the other hand, $\gamma_e$ should only depend on the local field
$B$, generally as $\gamma_e \propto B^{-1/2}$ if Lamor
acceleration is involved. This nicely cancels out all the
$r$-dependences, so that $E_p \propto \Gamma (1+z)^{-1}$. Only the
dispersion of $\Gamma$ and $z$ contribute to the final dispersion of
$E_p$. This is one of the strengths of the magnetic-dominant internal
models\cite{zm02c}. For the external models in both the low- and
high-$\sigma$ regimes, the case is more complicated with the interplay 
of the ambient density. The $E_p$ distribution is found to be wide if
independent distributions of luminosity and other parameters are
assumed\cite{zm02c}. A narrower $E_p$ distribution is possible if some
intrinsic parameter correlations are taken into account\cite{bd00b}.

Different dependences of $E_p$ on various parameters for different
models provide a framework on which the models may be tested with future 
data (Table 2)\cite{zm02c}. The parameters on which $E_p$
depends in various models are in principle measurable with current
and future observational facilities. Two major unknown parameters
include the redshift and the bulk Lorentz factor. The former is
measurable in large numbers based on expected localizations
by the next generation GRB mission, {\em Swift}\cite{swift}, 
scheduled to be launched in  2004. The latter could be also
measured or at least constrained in the {\em Swift} era using the
early afterglow data\cite{zkm03} (see more in \S\ref{sec:reverse}), or 
could be measured using the high energy cutoff of GRB spectra\cite{baring} 
when the next generation gamma-ray mission, {\em GLAST}\cite{glast}, is
launched in 2006. So there are good prospects for pinning down the 
location and mechanism of the GRB prompt emission as well as the
content of GRB fireball through statistical analyses of future
data.

\begin{table}[h]
\tbl{The dependences of the GRB spectral break energy ($E_p$) on
various parameters in the synchrotron emission model of various
fireball variants$^{342}$. Model parameters: $L$: initial
total luminosity (not just the radiated luminosity) of the fireball;
$E$: initial total energy of the fireball; $\Gamma$: bulk
Lorentz factor at the radius of GRB radiation; $t_v$: typical
variability time scale; $n$: ISM density; $z$: redshift.}
{\begin{tabular}{@{}ll@{}} \toprule
Model & $E_p$-dependences \\ \colrule
Internal shock model & $E_p \propto L^{1/2}\Gamma^{-2} t_v^{-1}
(1+z)^{-1}$ \\
Internal magnetic dissipation model & $E_p \propto \Gamma (1+z)^{-1}$
\\
External shock model & $E_p \propto \Gamma^4 n^{1/2} (1+z)^{-1}$ \\
External magnetic dissipation model & $E_p \propto \Gamma^{8/3}
L^{1/2} E^{-1/3} n^{1/3} (1+z)^{-1}$ \\
Pair photosphere model & $E_p \propto \Gamma (1+z)^{-1}$ \\
Baryonic photosphere model (wind coasting) & $E_p \propto L^{-5/12}
t_v^{1/6} \Gamma^{8/3} (1+z)^{-1}$ \\
Baryonic photosphere model (shell coasting) & $E_p \propto L^{-1/12}
t_v^{-1/6} \Gamma (1+z)^{-1}$ \\
Baryonic photosphere model (shell acceleration) & $E_p \propto L^{1/4}
t_v^{-1/2} (1+z)^{-1}$ \\ \colrule
\end{tabular}}
\label{table:Ep}
\end{table}

\subsection{GRB jet: uniform or quasi-universal?\label{sec:jets}}

The geometrical configurations are an essential ingredient in 
characterizing and understanding astrophysical phenomena (e.g. 
pulsars, AGNs, etc.). Evidence
suggests that GRBs are very likely collimated. Understanding the
degree of collimation as well as the possible structure of the
collimated flow would be very essential to understand the burst
mechanisms and the true event rates. 

\begin{figure}
\vspace*{10pt}
\centerline{\psfig{file=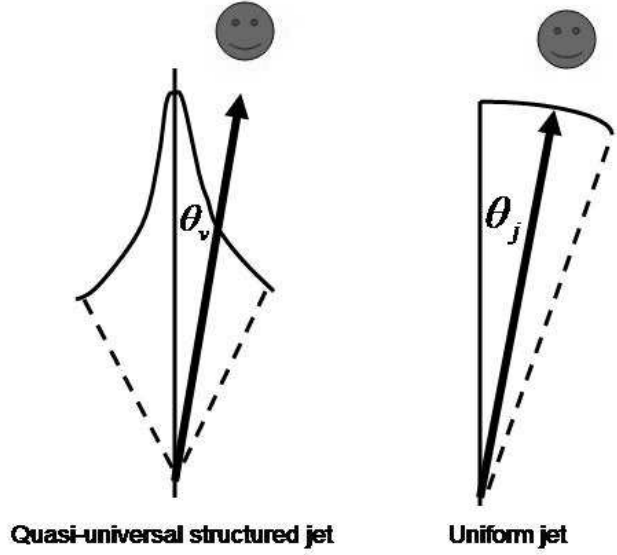,width=8cm}}
\vspace*{20pt}
\caption{Comparison between the quasi-universal structured jet model
and the uniform jet model.}
\label{fig:jets}
\end{figure}

The main evidence for GRB fireball collimation is provided by the
achromatic steepening breaks in some GRB optical afterglow
lightcurves. The simplest model, i.e., a conical jet with a uniform
energy distribution within the cone and sharp energy depletion at the
jet edge\cite{r97,pmr98,r99} (Fig. \ref{fig:jets}), and how this model
interprets the 
lightcurve steepenings, have been introduced in \S\ref{sec:jet1}. 
In this model, the time for the break to occur roughly corresponds to 
a measure of the jet opening angle, if the observer's line-of-sight is
in the jet cone and not too close to the jet edge. Data show that such
inferred jet opening 
angles ($\theta_j$) show a large dispersion among different bursts, and so
do the corresponding isotropic gamma-ray emission energies ($E_{\gamma,iso}$). 
However, both dispersions conspire in such a way that 
$E_{\gamma,iso} \theta_j^2$ is essentially a constant\cite{f01,bfk03}. 
Within the uniform jet model, the total energy released in a GRB event 
can be estimated as $E_\gamma=(E_{\gamma,iso}/4\pi) 2\pi
(1-\cos\theta_j) \times 2 = (1-\cos\theta_j) E_{\gamma,iso}$, where $2 
\pi (1-\cos\theta_j)$ is the solid angle of a conical jet with an
opening angle $\theta_j$, and the factor 2 takes into account the
consideration that the jet is likely bipolar. At small angles, $f_b
\equiv (1-\cos\theta_j) \simeq \sin^2\theta_j/2 \simeq \theta_j^2/2$. 
Therefore the above fact suggests a constant energy reservoir among
long GRBs\cite{f01,bfk03}. One puzzle posed by this is why a standard 
energy is collimated to rather different degrees among different bursts.

An alternative, and in principle more elegant, model is to postulate
a quasi-universal jet structure, i.e., there is a non-uniform
distribution of energy per solid angle within the jet, so that all
burst outflows could more or less retain a similar geometry,
the diversity of observed jet break times and isotropic-equivalent 
energies being caused by a different line-of-sight relative to the 
jet axis\cite{rlr02,zm02b,zdlm03}. If such a model can be constructed, 
GRBs might be considered to have not only a standard energy reservoir, 
but also a standard geometric configuration. A lesson learned from AGN 
studies is that much of the apparent diversity in AGNs is simply caused by 
viewing-angle effects. Current unified AGN models try to find a paradigm 
interpreting various phenomena as being due to a standard configuration which
is viewed at different angles\cite{antonucci93,urry95}. A unified picture 
for the GRB phenomenology would be similarly appealing\footnote{Before the
standard energy suggestion\cite{f01}, it had been 
speculated\cite{lpp01} that GRBs may have both a standard energy
reservoir and a possible standard configuration based on a handful of
afterglow data. The geometric configuration suggested in
that work includes three distinct components, i.e., two
uniform cones (one narrow and one wide) and one quasi-isotropic
component. There was no discussion about whether such a configuration
is consistent with the afterglow lightcurves, and it seems that such a
structure is likely to cause some distinct lightcurve signatures that may
violate most of the data. It is, however, likely to be consistent with the
recent two-component jets as inferred in GRB 030329\cite{b03,sheth03}.}. 
A straightforward speculation is to introduce 
a jet configuration such that $\epsilon(\theta) \propto \theta^{-2}$,
as suggested by the $E_{\gamma,iso}\theta_j^2 \sim$ const empirical
law. Here $\epsilon(\theta)$ is the energy per solid angle along the
direction defined by $\theta$ (the angle between the viewing direction
and the jet axis), which by definition, is equivalent to
$E_{\gamma,iso}/4\pi$. Such a configuration has been discussed earlier
to model afterglow lightcurves\cite{mrw98,dg01}, but the viewing
direction was placed close to the jet axis. By placing the viewing
direction at arbitrary angles\cite{rlr02,zm02b}, it is found that an
achromatic lightcurve steepening is naturally reproduced. However, the
time for the steepening to occur now corresponds to the epoch when the
line-of-sight Lorentz factor $\Gamma(\theta_v)$ is decelerated to a
value below $1/\theta_v$ (rather than below $1/\theta_j$ for the uniform jet 
model, see Fig. \ref{fig:jets} for a comparison between both jet models). 

The $\epsilon(\theta) \propto \theta^{-2}$ model (hereafter $k=2$
power law structure model) is directly motivated by the
$E_{\gamma,iso}\theta_j^2 \sim$ const empirical law. From the point of 
view of reproducing jet steepening with the viewing-angle effect, more 
general types of jet structure, e.g. power law models with more
general indices $\epsilon(\theta) \propto \theta^{-k}$, or even
non-power-law structure, such as a Gaussian profile, may do the same
job\cite{zm02b}. This is because well before the lightcurve ``jet 
break" time, what is relevant for the dynamical evolution as viewed by 
the observer is only the average energy per solid angle within the 
$\sim 1/\Gamma(\theta_v)$ cone,
which essentially remains the same as the relativistic beaming cone
gets larger (given that $[\Gamma(\theta_v)]^{-1} \ll \theta_v$ is
satisfied initially). After the jet axis enters the field of view,
eventually all the initial jet structure is expected to be smeared out
due to energy redistribution and sideways expansion. So in both asymptotic
regimes, a structured jet has the same temporal evolutions as the
uniform jet model\cite{zm02b}. Different jet structures only manifest
themselves differently around the jet break time. The lightcurve of a
$k=2$ power-law structured jet has been modeled by various 
authors\cite{rlr02,kg03,gk03,pk03,wj03,salmonson03}. It is found that
the sideways expansion effect is not prominent, so that a model with
this effect neglected can still roughly reproduce the basic feature of 
the lightcurve\cite{kg03,gk03}. For more general structure functions,
the lightcurves are not adequately modeled due to the complicated
physics involved. Modeling Gaussian-type jet evolution\cite{kg03}
revealed a substantial energy structure redistribution during the
evolution. The resultant lightcurves are consistent with the jet data. 

A structured jet model has several clear predictions to confront
with the observations. (1) Because different luminosities in this
model are caused by different viewing angles, whose probability is well 
defined, the structured jet model a specific prediction about the GRB
luminosity function for bursts in the same redshift
bin\cite{zm02b,rlr02}. For the power-law model, one has $N(L)dL
\propto L^{-1-2/k} dL$, while for the Gaussian model, one has $N(L)dL
\propto L^{-1} dL$\cite{zm02b}. When the parameters are allowed to have
some dispersion (quasi-universal picture), Monte Carlo simulations
show that the above scalings still hold, with turnovers near the
low- or high-luminosity ends caused by the variations of the
parameters\cite{ldz03}. In contrast, the uniform jet model has no
predictive power concerning the luminosity function. Currently,
although the sample of bursts with spectroscopic redshifts is too
small to allow a direct luminosity function study, the luminosity
functions derived using various statistical
methods\cite{schmidt01,schaefer01,lloyd02b,norris02,stern02} are not
inconsistent with the model predictions, especially for the
quasi-universal Gaussian jet model with constant total
energy\cite{zm02b,ldz03}. (2) By taking into account the cosmological
effects, the total number of bursts detected (regardless of
redshifts) as a function of jet angle also has a clear prediction in
the structured jet model\cite{psf03}. For the $k=2$ power-law model, a 
distribution peak is predicted around 0.12 rad, which is in rough
agreement with the current data. The peak is expected to move to
larger angles when the detection sensitivity threshold is
increased\cite{psf03}. Again the uniform jet model has no prediction
power on this. In the {\em Swift} era, both of the above predictions
may be tested.

Besides the preliminary support suggested by the above two tests, the
structured jet model may also be indirectly supported by other theoretical
and experimental results. (1) Progenitor and central engine studies naturally
give rise to jet structures. Simulations within the collapsar model 
naturally predict a jet structure after the jet penetrates through the
stellar envelope\cite{zwm03}. Studies of magnetic-threaded central
engine models\cite{bcl03} as well as the evolution of a Poynting-flux
dominated flow\cite{lb02} result in jet structures in a natural way; 
(2) The interpretation of some GRB prompt emission empirical relations, 
such as the spectral lag - luminosity correlation\cite{norris96}, are 
consistent with the structured jet 
hypothesis\cite{salmonson01,sg02,norris02}, so that a coherent picture 
is achievable for both the prompt emission and the afterglow emission
within this theoretical framework.

Several criticisms and caveats for the structured jet model have been
raised. (1) The uniform jet model is very simple, and one can question
the need to introduce jet structures unless it is
necessary\cite{pk03}. On the other hand, a non-uniform jet structure
is a natural expectation in any jet model which arises from realistic
physical processes. Even if the structured jet model so far does only
the same job as the uniform jet model, it is still worth exploring,
and should be explored, since it is more physical and has more
predictive power; (2) The $k=2$ power-law 
model predicts some anomalous signatures inconsistent with the
data\cite{gk03}. However, this may be caused by the assumption of an
unphysical singular point in the jet structure, and is removable when
more realistic jet structure (e.g. Gaussian) is considered\cite{kg03};
(3) Recent optical polarization data\cite{rol03} indicate a near
$90^{\rm o}$ change of the polarization angle, which is inconsistent
with the model prediction in the $k=2$ power-law jet
model\cite{rossi02}. However, the data accumulated are so far not
conclusive\cite{lazz03}.

An important contraint on the jet structure is posed by the recent
discovery that an $E_p \propto (E_{iso})^{1/2}$ correlation earlier
proposed for GRB appears now to extend all the way down to 
energies characteristic of XRFs\cite{amati02,sakamoto03,lamb03}. 
This relation, together with the $E_{iso} \propto \theta_j^{-2}$
correlation\cite{f01,bfk03}, immediately leads to the inference 
$E_p \propto \theta_j^{-1}$, which poses important constraints on both 
the simple $k=2$ power law model and the on-beam uniform jet 
model\cite{zdlm03}.  This is because $E_p$ varies by two orders of 
magnitude from GRBs to XRFs, so that even if XRFs correspond to events 
viewed at the equator (or isotropic events), GRBs would have to be 
events viewed within the $1^{\rm o}$ viewing angle (or corresponding 
to jets with opening angle smaller than $1^{\rm o}$). For the 
$k=2$ model, this tends to greatly over-generate XRFs\cite{lamb03,zdlm03}, 
which are found to contribute $\sim 1/3$ of the total GRB/XRF population 
in the HETE-2 data\cite{lamb03}. For the uniform jet model, the very 
narrow GRB jets contradict or do not address some important afterglow 
jet break data\cite{zdlm03}. All current GRB/XRF prompt emission and
afterglow data are however consistent with a quasi-universal
Gaussian-like structured model\cite{zdlm03}, with an angular structure
$\epsilon(\theta)=\epsilon_0 \exp(-\theta^2/2\theta_0^2)$. In this
model, GRB/XRF jets still retain a characteristic angle $\theta_0$,
with a mild structure inside and a rapid exponential decay outside. 
The XRFs are only those events with viewing angles $\theta_v
\sim (3-4)\theta_0$, which greatly decreases the number of XRFs.
Statistically, the jet parameters (e.g. the typical angle, total
energy in the jet, etc.) are allowed to have some scatter\cite{ldz03}. 
Monte Carlo simulations indicate that with reasonably small scatter, 
a quasi-universal jet model not only solves the GRB/XRF population
problem, but can also reproduce the $E_{iso}\propto \theta_j^{-2}$
correlation\cite{ldz03,zdlm03}. 
More complicated jet structures are in principle possible, at least
for some events\cite{b03,sheth03,huang03}. These models along with the 
simplest Gaussian model should be more extensively confronted with the 
various data, including luminosity function, redshift
distribution, opening angle distribution, etc.\cite{dz04,lwd03}

To conclude, evidence in support of long GRBs having a quasi-universal 
structured jet configuration is mounting, but issues and contradictions 
remain. This paradigm will be more fully tested with the advent of 
the extensive data sets expected in the {\em Swift} era.

\subsubsection{Other jet models}

Besides the uniform conical and the structured jet models, there are
some other GRB geometric suggestions discussed in the literature. 

One suggestion is that GRB jets might be
cylindrical\cite{chl01,mhdl03}, as observed in some AGNs. Current
modeling assumes that the line-of-sight is on the jet beam, and no
apparent jet break is predicted from the model, so that the model can
not interpret the whole GRB phenomenology, but may account for a
sub-category of the bursts. A cylindrical jet is however consistent
with the MHD description of a Poynting-flux-dominated
flow\cite{vk03a,vk03b}. 

Another suggestion is that GRBs arise not from fluid fireball jets, 
but are rather ``cannon balls'' ejected from the central
engine\cite{ddd02,ddd03}. The picture is taken from analogy with some
observations in micro-quasars and supernova remnants, although it is 
unclear how cannon balls could survive as compact entities against
instabilities if they are accelerated over a large dynamical range
of radii and Lorentz factors.  The model has 
several distinct predictions (e.g. superluminal motion of the source, 
non-resolvable source image, and so on) that are different from those 
from the standard fireball jet models, which may be easily tested 
with the current and future data. The association of GRBs with supernovae 
has been quoted as a strong support for this model\cite{ddd02b,ddd03b}, 
but this piece of evidence is not exclusive, since GRB-SN associations 
are widely expected within the standard model as
well\cite{p98,bloom99,reichart99,galama00,sta03,h03}. 

\subsubsection{Orphan afterglows\label{sec:orphan}}

An important phenomenon associated with jets is the issue of ``orphan 
afterglows'', i.e., the implication of low energy (X-ray, optical, radio)
decaying transients which are not be detectable in $\gamma$-rays. 
The idea initially comes from the uniform jet model\cite{r97}. The 
starting point is that the initial GRB jet is sufficiently misaligned 
respect to the line-of-sight so that the observer misses the bright 
gamma-ray emission. Later as the jet decelerates, it expands sideways 
and enters  the line-of-sight. The observer is expected
to see a steeply rising afterglow lightcurve followed by a normal
decaying lightcurve\cite{pmr98,gpkw02}. The event rates of orphan
afterglows in various bands have been
estimated\cite{tp02,levinson02,npg02}, but so far there is no firm
detection of orphan afterglows of any kind. 

The issue of orphan afterglows is complicated by two other
considerations. First, there might be cosmological ``dirty
fireballs''\cite{dcb99} or ``failed GRBs''\cite{hdl02} whose Lorentz
factors are too small to allow 
transparent gamma-rays to be detected. These fireballs, when
decelerated by ISM, can also give rise to orphan afterglows, whose
signatures are not easy to differentiate from those of the jet
orphan afterglows\cite{hdl02}. Some possible differences between
these two types of transient events have recently been
proposed\cite{rhoads03}. Second, if GRB jets are
structured\cite{rlr02,zm02b}, the orphan afterglows
of the first kind would be greatly reduced since there is no sharp jet
edge cutoff. The second-type (dirty fireball) may exist due to higher
baryon loadings at larger angles from the jet axis. So far there is no
detailed discussion about the orphan afterglow rates for the
structured jet model. 

\subsubsection{Nature of X-ray flashes\label{sec:XRF}}

So far there have been essentially four types of interpretations for
the XRF phenomenon (see \S \ref{sec:grb1}) in the
literature\cite{zm02c}, i.e. dirty fireballs or failed
GRBs\cite{dcb99,heise,hdl02}, high redshift GRBs\cite{heise},
fireballs dominated by photosphere emission\cite{mrrz02,dren02,krm02},
and geometry-related events, such as offbeam GRBs in the uniform jet
model\cite{wl99,nakamura99,yin02} and large viewing angle events
in the structured jet model\cite{zdlm03,jw03}. The last model is
likely in an anisotropic supernova explosion\cite{woosley99}. 
The first two interpretations attribute the soft-faint nature 
of XRFs to the dispersion of two parameters, i.e. $\Gamma$ and $z$,
respectively\footnote{Note that the dirty fireball interpretation is 
inconsistent with the internal shock model, which predicts a higher 
$E_p$ for a lower $\Gamma$ (\S\ref{sec:mech}). It is however consistent 
with the external models and with the internal model involving a 
Poynting-dominated flow\cite{zm02c}.}. In reality, $E_p$ is determined 
by the combination of many parameters (\S\ref{sec:mech}), so that XRFs are 
likely to be caused by the dispersion of at least several of them, as 
indicated by Monte Carlo simulations\cite{zm02c}. The photospheric
interpretation\cite{mrrz02,dren02,krm02} 
may require the $E_p$ distribution of the combined sample of
GRBs/XRFs to have a separate component in the low energy regime. 
Recent afterglow observations for XRFs\cite{soderberg03} pose
important constraints on some of the above suggestions (e.g. high-$z$
bursts and dirty fireball). In the {\em Swift} era, a large data set
of both GRBs and XRFs, with photometric redshift measurements, should
become available, which will eventually pin down the nature of XRFs. 
At present, the geometric model, especially the one involving
structured jets, appears to us the most promising possibility\cite{zdlm03}.

\subsection{Long burst progenitor: collapsar or supranova?\label{sec:proge}}

\begin{figure}
\centerline{\psfig{file=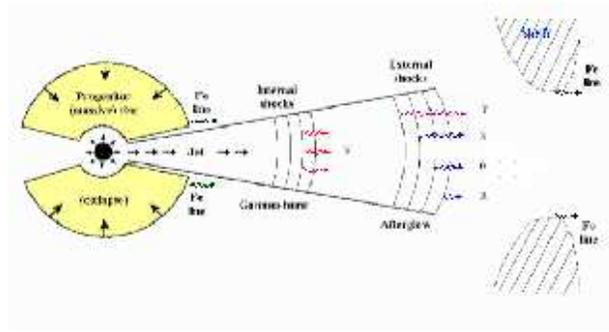,width=8cm}}
\vspace*{8pt}
\caption{Comparison of the geometric configurations of the collapsar
and supranova models (from Ref.5).}
\label{fig:collapsar}
\end{figure}

Accumulating evidence suggests that long GRBs are associated with
deaths of massive stars. The first cosmological progenitor scenario, 
i.e., the NS-NS merger model\cite{p86,e89,npp92}, is now disfavored
for long bursts\cite{mw99,bloom02b}, although it is still a leading
contender to account for the short, hard bursts
(cf. Ref. \refcite{vo01}). 
A variety of GRB progenitor models have been discussed in the 
literature\cite{fwh99,mrw99}, and the requirement for accretion onto
a central object occurring in a neutrino-dominated regime
significantly constrains some of the scenarios\cite{npk01}. 
Although the formation of GRBs from binary systems\cite{brown00} or from
gravitational energy loss in neutron stars\cite{spruit99} 
have also been discussed, the leading scenario for causing a long GRB 
involve the core collapse of a single massive
star\cite{w93,mw99,a00,wheeler00,mwh01,wheeler02,zwm03,matzner03,proga03}.
A supernova (SN) explosion is naturally expected to be associated with the 
GRB (although in its first incarnation\cite{w93} this was thought to be a
``failed" supernova, i.e. a core collapse which did not to achieve ejection 
of its stellar envelope). The launching of a GRB jet is widely believed to 
involve a black hole - torus system resulting from this core collapse, 
hence the GRB is either produced simultaneously with the SN, if the
star collapses to a black hole promptly, or the GRB is delayed with respect to 
the SN (which would give rise to a ``supramassive'', rapidly rotating
neutron star as a first step, the neutron star later collapsing to
a black hole in the second step, after loosing the angular momentum
required to sustain the mass). Either a one-step or a two-step collapse 
are possible, depending on the mass and angular momentum of the 
progenitor core as well as the details of the collapse. If the delay of 
the second collapse is not very long, e.g. from minutes to hours
as suggested by numerical simulations\cite{mw99}, the two-step collapse
scenario is not very different from the one-step collapse, and both of 
these cases can be referred to as ``collapsars''\cite{wzh02}, e.g. of type
I and II. On the other hand, an alternative suggestion is that the delay 
is long, e.g. from days to weeks or even months, and in this case the burst 
environment is very different from the collapsar case, due to the role of the
well-separated supernova shell and a central pulsar wind\cite{vs98,k03}. This
latter case is referred to as a ``supranova''. Figure
\ref{fig:collapsar} is a cartoon picture for the geometric
configurations in both models.

The collapsar model has many merits in reproducing the data. First and
foremost, it naturally involves a GRB-SN association, and predicts that 
GRBs are associated with star forming regions. The existence of the
stellar envelope helps to collimate a jet with angular structure and 
can help to regulate the jet flow intermittently\cite{zwm03}. The duration 
of a burst is set by the fall-back timescale rather than the accretion
timescale\cite{mw99}, which is natural to interpret the durations of
long GRBs. The requirements for the progenitor include a narrow range
of specific angular momentum ($3\times 10^{16} ~{\rm cm^2~s^{-1}} < j
< 2\times 10^{17} ~{\rm cm^2~s^{-1}}$) and poor metallicity. Such a
constraint may lead to a rough standard GRB energy and an event rate
roughly consistent with the true GRB event rate after collimation
corrections\cite{mw99,wzh02}. All of these are consistent with the
main-stream observational and theoretical progress in the GRB field. 
The interaction of such a jet with the stellar envelope should also make
unique signatures in both electromagnetic and non-electromagnetic forms.
For example, prompt X-ray and gamma-ray signals are expected when the
jet breaks through the envelope, leading to precursors to the main
burst\cite{mr01,rcr02,wm03}, or even leading to short, hard
bursts\cite{zwm03}. The breakout of the jet cocoon is also a candidate 
to interpret the X-ray emission line features hours after the burst trigger 
as have been seen in several bursts. Internal shocks within the jet
before penetrating through the envelope would lead to high energy
protons to produce prompt TeV neutrino signals\cite{mw01}. A collapsar
may also harbor a long-lived central engine that continuously pumps
energy into the GRB fireball\cite{rm00,wzh02}. Recently, more
realistic MHD simulations for the collapsar model have
commenced\cite{proga03}, which will reveal the possible magnetic
nature of the fireball. The difficulties of the 
collapsar model include the following. First, the required high
angular momenta of the progenitors are difficult to
achieve\cite{wzh02}. Second, although the top candidates for
collapsars are those stars that have undergone intense mass loss
before collapse (e.g. Wolf-Rayet stars), the expected wind environment 
for GRB afterglows\cite{cl99,cl00a} is not commonly identified in
current afterglow studies\cite{f01,pk01}.

The supranova model was initially introduced to alleviate the
baryon-loading problem, since weeks to months after the SN explosion,
the environment is relatively clean\cite{vs98}. Two other incentives 
were added later. One was its promise in interpreting X-ray line
emission (and absorption) features. The conjectured supernova shell
(or some torus-like or funnel-like remnant) located at $10^{15}-10^{17}$
cm from the central engine (which was ejected by the progenitor days
or months before, moving with a speed of $\sim 0.1 c$) provides
a large mass of heavy elements for the inferred photoionization (by the 
continuum emission of the GRB or the early afterglow) and recombination
processes to produce a strong Fe line\cite{lcg99,bottcher00,vietri01},
which are thought to be responsible for many of the discrete  X-ray 
spectral features detected in several bursts\cite{amati00,p00,r02}. 
(More recently, lower-Z elements such as Ca, N, S etc have also been 
reported in bursts\cite{r02,watson03}).
Such an interpretation of X-ray lines has some distinct energetical 
advantages\cite{lrr02}, but it requires high clumpiness in the shell.
This type of model is not the only one able to explain the lines.
A competitive class of models attributes the spectral features to 
continuous photoionization by a long-lived central engine jet\cite{rm00}, 
or by a jet cocoon in the collapsar scenario, interacting with the outer
layers of the stellar progenitor\cite{mr01}. These models make the line
at smaller distances ($r\sim 10^{12}-10^{13}$ cm, hence they are referred
to as ``nearby" models) and in a much higher density medium, require a 
smaller iron mass. Detailed modeling indicates that they are able to 
reproduce the data\cite{kmr03}. 
Other X-ray line models proposed include irradiation of a pair screen 
caused by the early afterglow back-scattering\cite{kn03} and the 
Cerenkov line emission mechanism\cite{wzy02}. 

A second advantage of
the supranova suggestion is related to the role of the central pulsar
after the SN explosion; this can form a pulsar wind bubble which
can modify the ambient medium from a pre-stellar wind to a
medium with constant (lower) number density, consistent with the observations. 
High equipartition parameters ($\epsilon_e$, $\epsilon_B$) as inferred 
from afterglow modeling\cite{pk01,pk02} are also naturally
interpreted\cite{kg02,k03}. 

The supranova model suffers several criticisms. 
First, it takes fine-tuning to make a GRB months to years after the 
formation of the neutron star\cite{shapiro00,wzh02}. Second, 
lacking an envelope, the collapse may not produce a long
burst\cite{bf01} and the collimation mechanism is not clear. Third,
detailed radiative transfer 
calculations\cite{kmr03} indicate that a supranova model would reproduce 
the reported large equivalent widths only for very shallow incidence angles
of the ionizing continuum, which is naturally expected in nearby models 
(involving ionizing the walls of a stellar funnel), but is less natural 
for an ionizing continuum incident on a distant supernova shell, where 
normal incidence is expected. Finally, the most severe objection to the
supranova model is the fact that the association between GRB 030329
and SN 2003dh allows at most a very short delay (less than two days) 
between the SN and the GRB events, and is compatible with both events being
simultaneous\cite{h03}. This rules out the supranova model at least
for this burst, and supports the collapsar model. Unfortunately, no
high spectral-resolution X-ray observation were obtained in a timely
manner, thus missing an excellent chance to test whether a pre-ejected 
SN shell is a requisite to generate X-ray lines.

Although the GRB 030329/SN 2003dh association greatly changes the
balance between the collapsar/supranova debate, the issue of whether
there is still a sub-category of bursts originating from supranovae 
(e.g. perhaps those which have strong X-ray lines) is still unsettled. 
Future observations can help to distinguish the two scenarios. 
First, both scenarios predict different electromagnetic signals. 
The collapsar scenario predicts X-ray and gamma-ray
precursors\cite{rcr02,wm03}, while the interaction between the
fireball and the SN shell in the supranova model tends to produce
enhanced, delayed GeV emission signals\cite{wdl02,gg03a,igp03};
Second, the supranova model predicts more emission components and
stronger flux 
levels of high energy neutrino emission due to the existence of the SN 
shell which provides more target photons and protons for $p\gamma$ and 
$pp$ interaction\cite{gg03b,gg03c,rmw03,da03}. This leads to stronger
neutrino signals both for individual sources and for the diffuse
background. Detection/non-detection of such an excess emission may
be used to prove/disprove the supranova model. Finally, suggestions to 
directly search for SNs that occurred weeks to years before GRBs have
been proposed\cite{heyl03}, which may pose direct constraints on
the models.

\subsection{Central engine: what is behind?\label{sec:engine}}

The most widely discussed GRB central engine invokes a central black hole
and a surrounding torus. There are three ultimate energy sources, the
gravitational binding energy of the torus, the spin energy of the
black hole and the magnetic energy. Two ways of extracting the
accretion energy and black hole spin energy have been considered,
i.e. the neutrino mechanism\cite{e89,ruffert97,ruffert99,mw99,fm03},
and the Blandford-Znajek\cite{bz77} mechanism. The former mechanism
typically powers the conventional ``hot'' fireball, while the latter
invokes strong magnetic fields threading the black
hole\cite{mr97b,lwb00,van01,li00} which typically would power a 
(at least initially) ``cold'' Poynting-flux dominated flow. 
The identification of the content of the fireball (\S\ref{sec:mag}) 
and the mechanism of GRB prompt emission (\S\ref{sec:mech}) would 
shed light on the mechanism that powers the central engine.

Another type of GRB central engine often discussed in the literature
is a millisecond
magnetar\cite{u92,u94,t94,yb98,by98,kr98,nakamura98,spruit99,wheeler00,rtk00}.
These are rapidly rotating neutron stars with surface magnetic fields
of order $\sim 10^{15}$ G and higher. These objects would be able to
satisfy the essential conditions for a GRB central engine, e.g.,
energy, duration, variability, baryon loading, birth rate, etc,
especially when considering the possibility that initially there could
be some temporary toroidal fields with much stronger strength
($10^{17}$ G)\cite{kr98,spruit99,rtk00}. The ultimate energy
sources include the spin energy of the pulsar and the magnetic energy. 
For a pulsar engine, there could be in principle two energy
components, an initial prompt component (either via neutrino mechanism 
or via destroying the temporary toroidal field) powering the prompt
GRB, and another long-term component due to spindown of the millisecond 
pulsar. The latter component continuously injects energy into the
fireball through Poynting flux\cite{dl98b,zm01a,zm02a}, or in the form 
of electron-positron pairs analogous to the pulsar wind
bubbles\cite{dai03}, leaving well-defined bump signatures in the
afterglow lightcurves\cite{dl98b,zm01a,zm02a,dai03}. 
It also naturally provides a long-lived central engine, which may be
the agent to continuously photo-ionize the Fe in the stellar envelope,
giving rise to the observed X-ray emission features\cite{rm00}. In the 
two-step supranova models, a pulsar is invoked in the first step, which
powers a pulsar wind to drive a magnetic-enriched bubble\cite{kg02}.

A more exotic central engine mechanism involves a phase transition 
from normal neutron matter to strange quark matter\cite{witten84}.
The process is likely in a detonative mode, leaving behind a star
completely composed of strange quark matter, called a strange
star\cite{af086}. The possibility of neutron star - strange star
phase transitions powering GRBs has been discussed within the context
of one-step\cite{cd96,bd00,odd02} or two-step
processes\cite{wdlwh00,bbdfl03}. A special category of such models
lead to intermittent energy injection due to unstable photon decay,
which is arguably a viable GRB central engine\cite{os02}. A common
caveat about the strange star mechanism is that there is no evidence 
yet about the existence of strange quark matter, while the
existence of black holes and pulsars have been widely tested.

\subsection{Environment: what is in front?\label{sec:environ}}

On the galactic scale, GRB host galaxies are broadly similar to the
normal, star-forming faint field galaxies at comparable redshifts and 
magnitudes\cite{d03}. The distribution of GRB-host
offsets from the galactic center\cite{bloom02b} as well as heavy element 
abundance studies\cite{d03} are fully consistent with a progenitor 
population associated with sites of massive star formation. 

On the parsec scale, issues about the GRB immediate environment include
whether the medium is (quasi-) uniform, wind-like or clumpy, and how dense
the medium is on average. Insights about these questions can be obtained 
through afterglow lightcurve modeling\cite{pk01,pk02,harrison01,yost03} and
time-dependent absorption feature modeling\cite{pl02}, but the
situation is still controversial. As mentioned earlier, the constant
density medium model is consistent with most afterglow
data\cite{pk01,pk02,harrison01,yost03}, but the $\rho\propto r^{-2}$ 
wind model also works in some bursts\cite{cl00a,lc03,lc02}. It is puzzling 
how different GRBs could have quite different immediate environments 
if they come from the same type of progenitor. There are so far no studies
on correlations of the inferred GRB environment with other properties of
the gamma-ray prompt and afterglow emission. The inferred medium
density also varies significantly among different modelers, even for the
same burst\cite{f01,pk01,dl99,harrison01,piro01}, although there seems
to be a trend towards favoring a universal moderate-dense medium with 
$n \sim 10~{\rm cm^{-3}}$ as being consistent with (most of) the 
data\cite{frail03c}. Suggestions that GRBs are embedded in 
dense molecular clouds have been made\cite{rp02}, which is
consistent with the high hydrogen column density ($N_H$) inferred from
some X-ray afterglow studies\cite{gw01,pl02}. A high-density afterglow
model, on the other hand, although inferred from some afterglow 
fits\cite{dl99,piro01b}, may not be compatible with broad-band
data\cite{pk01,harrison01} or with the inferred standard energy budget 
for all GRBs\cite{f01}. The dust grains expected in molecular clouds 
may reflect and irradiate the afterglow emission and form a bright dust 
echo\cite{eb00}, or a distinct IR signature detectable for nearby 
bursts\cite{vb01}. Dust extinction is also invoked as the cause of at 
least some optically dark bursts\cite{djorgovski01}. So far direct
evidence of dusts is not 
yet collected, and the possibility of dust destruction\cite{wd00,fkr01}
further complicates the issue. In some bursts (e.g. GRB 021004),
multiple spectral absorption systems and density bumps are identified,
which refer to a rather non-uniform, bumpy environment from close to
the burst to much farther away from the burst\cite{s03,mirabal03}. 
It has been suggested that time-dependent absorption features both in 
optical\cite{pernal98} and in X-rays\cite{pl02} are powerful tools to 
study GRB environments, but the present data are too sparse to allow 
firm conclusions to be drawn.

\subsection{Shock parameters: universal or unpredictable? \label{sec:shock}}

It is almost certain that afterglows are produced by collisionless
relativistic shocks energized by the GRB. The physics involved in 
these shocks is however poorly known. A widely discussed scenario is 
that particles are accelerated via repeatedly crossing a shock front, 
and achieve a power-law energy distribution through the well-known Fermi 
mechanism\cite{fermi49,be87}. The latest numerical simulations 
indicate that the resultant power-law index is universal\cite{a01}, 
i.e., $p \simeq (2.2-2.3)$. This is in sharp contrast with what is 
inferred from afterglow fit data. With the simplest jet model, 
broadband modeling indicates that $p$ is quite unpredictable,
ranging from $\sim 1.4$ to $\sim 2.8$ among different
bursts\cite{pk01,pk02}. An important caveat for the modeling is that
the present model only uses the simplest assumptions, e.g. uniform 
jets, non-evolving equipartition parameters, etc. The electron
power-law index $p$ is usually derived from the temporal decay index
after the lightcurve break or steepening attributed to jet properties
(i.e. $F_\nu \propto t^{-p}$). Whether
the inconsistency between broad-band modeling and shock acceleration
simulations is caused by the simplified GRB model or by the simplified
shock acceleration model is not known, and developments in both
directions are needed. For example, proposals for generating a flat
electron index have been outlined\cite{bm96}, while more
complicated afterglow models (e.g. post injection, or refreshed
shocks) could change the inferred $p$ value considerably\cite{bjornsson02}.
%An elegant picture would be that shock physics
%is universal, while the diversity of the apparent $p$ parameter is
%caused by the more complicated physical properties (e.g. jet structure,
%evolution of the equipartition parameters, and so on).
 
Unlike the magnetic fields invoked in the magnetically dominated
scenarios for GRB emission, which may be carried out from the central 
engine, the origin of the magnetic fields in the internal or external 
shocks needs to be addressed. The inferred fields are typically of 
strength $\epsilon_B \sim 10^{-3}-1$. Loosely, these have been attributed 
to a turbulent dynamo mechanism behind the shocks. More specific 
suggestions include two-stream instability of relativistic plasma\cite{ml99} 
or the presence of an intrinsically magnetized ambient medium\cite{kg02}. 

\section{Prospects\label{sec:pros}}

We have summarized the achievements (\S\ref{sec:prog1} and
\S\ref{sec:prog2}) and uncertainties (\S\ref{sec:prob}) in our
current understanding of the nature of GRBs. Guided by past
experience, we believe that our knowledge of GRBs will be further
advanced in the coming years, driven by future observational breakthroughs. 
In this section, we attempt to foresee some of the possible milestones
in the upcoming new epoch of GRB study led by NASA's two future missions, 
{\em Swift} and {\em GLAST}, as well as by experimental developments in 
new channels of GRB study. We will discuss how our knowledge of GRBs is likely
to be extended in the temporal domain (i.e. early afterglows and the prompt 
emission - afterglow bridge, \S\ref{sec:reverse}), in the spectral domain
(i.e. GeV-TeV emission, \S\ref{sec:GeVTeV}), and into other,
non-electromagnetic regimes, including high energy neutrinos
(\S\ref{sec:nu}) and gravitational waves (\S\ref{sec:gw}). We will
also discuss how future developments could unveil the mystery of short
GRBs (\S\ref{sec:short}) and how GRB studies may become a unique tool
to investigate the early universe (\S\ref{sec:cos}). Some of the topics
covered in this section are also discussed in Refs. \refcite{w03,mkrz03}. 

\subsection{Early afterglow and reverse shock emission\label{sec:reverse}}

The discovery of GRB afterglows\cite{c97,van97,f97} has greatly extended
our knowledge of GRBs, both in the temporal domain (hours to even years 
after the burst) and in the spectral domain (below the gamma-ray band, 
from X-rays down to radio). However, due to the long time scale of the 
mission alerts up to the ground, the slowness in slewing 
instruments and/or bad weather in ground-based instruments, in most
cases afterglow observations have started hours after the burst trigger. 
At this phase, the afterglow blastwave has been decelerated and entered
a self-similar regime, and the behavior is essentially determined by
the total energy per solid angle in the fireball and the properties of 
the ambient medium. Some precious information characterizing the fireball, 
such as the initial Lorentz factor of the fireball and the magnetic content 
of the fireball, has at this stage been lost. In order to retrieve such 
information, very early afterglows need to be studied, which contain 
information on the emission from the reverse shock region during and 
shortly after the reverse shock crosses the fireball shell at the very 
beginning of the shell-medium interaction. As of October 2003, only four 
bursts (GRB 990123, GRB 021004, GRB 021211 and GRB 030418) were caught 
within less than 10 minutes after the triggers\cite{greiner}, thanks greatly 
to the growing number of robotic optical telescopes spreading all over the
world. The situation will change soon following the launch of {\em
Swift} in June 2004, which will automatically record essentially
all the triggered GRB early afterglows starting from 100 seconds after
the triggers with an on-board X-ray telescope (XRT) and a UV-optical
telescope (UVOT). The mission will also issue prompt alerts to the
ground to allow rapid follow-up observations from the ground-based
telescopes. 

There are two important differences between the reverse shock emission 
and the forward shock emission. First, the reverse emission is prompt and
short-lived. The electrons are continuously accelerated only until the 
reverse shock crosses the initial shell. This happens at the
deceleration radius $r_{dec}$, which is $r_\Gamma$ for the thin shell
case, or at $r_\times$ for the thick shell case (see \S\ref{sec:radii}). 
The observer's time at which this occurs is usually defined by the
crossing time, $t_\times = \mbox{max} (t_\Gamma, T)$, where 
$t_\Gamma = [(3 E/4\pi \Gamma_0^2 n m_p c^2)^{1/3} / 2\Gamma_0^2 c]
(1+z)$, and $\Gamma_0$ is the initial Lorentz factor. The fireball
Lorentz factor at the shock crossing time is $\Gamma_\times =
\mbox{min} (\Gamma_0, \Gamma_c)$, where $\Gamma_c \simeq 125
E_{52}^{1/8} n^{-1/8} T_2^{-3/8} [(1+z)/2]^{3/8}$. The reverse shock
emission is therefore generally divided into two segments. For $t<
t_\times$, the synchrotron spectrum is (as in the forward shocks)
characterized by a four-segment broken power law, as discussed in
\S\ref{sec:syn}. 
However, for $t>t_\times$, since no new electrons are accelerated, the
emission above the cooling frequency $(\nu > \nu_c)$ totally
disappears. Second, before and at the crossing time, both the
pressure and the internal energy density across the contact
discontinuity are the same, but the particle density in the shocked shell 
is much larger than that in the shocked ISM. Given similar electron
equipartition 
and injection factors ($\epsilon_e$ and $\xi_e$), the typical energy
per electron is much lower in the reverse shock region than in the
forward shock region. So the synchrotron peak frequency in the reverse 
shock is much lower than that of the forward shock, mainly in the
optical/IR regime\cite{mr93b,mr97,sp99a,sp99b,mr99}, and the reverse shock
emission component is typically characterized by an optical flash and
a radio flare. 

\begin{figure}
\centerline{\psfig{file=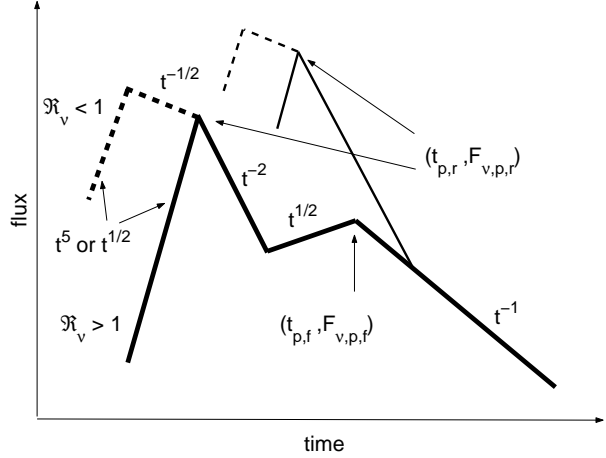,width=8cm}}
\vspace*{8pt}
\caption{Typical GRB optical early afterglow lightcurve that includes
the contributions from both the forward and the reverse shock emission
components (from Ref.120).}
\label{fig:reverse}
\end{figure}

Compared with the forward shock emission, there are more physical cases 
to be considered in the reverse shock lightcurves, depending on issues such 
as whether one is in the thick shell regime (in which case the reverse 
shock becomes relativistic before $t_\times$) or in the thin shell regime 
(in which case the reverse shock only becomes mildly relativistic at 
$t_\times$); what is the  relative position of the
observational band with respect to the break frequencies both before
and after the crossing time; fast or slow cooling; whether the medium 
is an ISM or wind-like, etc. For the ISM
case, a full discussion about various lightcurve cases (12 altogether)
is presented in Ref. \refcite{k00}. A closer investigation taking
reasonable parameters reveals that there are only 4 most relevant
cases\cite{zkm03}, depending on whether the shell is thin or thick and 
another parameter ${\cal R}_\nu \equiv
\nu_R / \nu_{m,r} (t_\times)$, which defines whether the observational
band is above (for ${\cal R}_\nu > 1$) or below (for ${\cal R}_\nu <
1$) the injection synchrotron frequency of the reverse shock emission
at the shock crossing time. A common feature of all four cases is
that the final temporal decay slope is steep, i.e. $F_\nu \propto
t^{-\alpha}$ with $\alpha \sim 2$. This segment of the lightcurve has been 
identified in GRB 990123\cite{a99} and GRB 021211\cite{li03,f03b},
lending credence to the reverse shock scenario. 
The reverse shock lightcurve component eventually joins the forward shock 
component at a later time. A generic expected early optical afterglow 
lightcurve involves two peaks\cite{zkm03} (Fig. \ref{fig:reverse}),
i.e., a reverse shock 
peak with $(t_{r,p}, F_{\nu,r,p})$ at the beginning of the $\alpha \sim 2$ 
lightcurve segment (which usually corresponds to $t_\times$), and a forward 
shock peak with $(t_{f,p}, F_{\nu,f,p})$, which is the transition point 
from the $\propto t^{1/2}$ lightcurve segment to $\propto t^{-1}$ lightcurve 
segment corresponding to the typical forward shock synchrotron frequency 
crossing the band. 

The information provided by the reverse shock emission has been widely 
used to estimate the initial Lorentz factor of the
fireball\cite{sp99a,ks00,wdl00,sr02,fan02,kz03a,wei03}. Usually these
methods rely on the poorly known shock parameters ($\epsilon_e$,
$\epsilon_B$ and $p$) as constrained by the forward shock modeling,
and they implicitly assume that same parameters also apply in the reverse
shock. The absolute values of these parameters are usually used 
to derive $\Gamma_0$. We have recently proposed a straightforward
recipe to derive $\Gamma_0$ by using a combined reverse and forward
shock analysis\cite{zkm03}. The only information needed is the time
and flux level at both the reverse shock and the forward shock
emission peaks. Since there are simple correlations of $\nu_m$,
$\nu_c$ and $F_{\nu,m}$ between both shocks at the crossing
time\cite{kz03a,zkm03}, the {\em ratios} of the two peak fluxes and
the two peak times only depend on $\Gamma_0$ (or a parameter depending 
on both $\Gamma_0$ and $\Gamma_c$ for the thick shell case), the
unknown parameter ${\cal R}_\nu$, as well as the {\em ratios} of the
shock microphysics parameters. Suppose that both lightcurve peaks are
detected in an idealized observational campaign; the two unknown
parameters $\Gamma_0$ and ${\cal R}_\nu$ can be then solved for from 
the peak flux ratio and the peak time ratio, if the microphysics
parameters are the same in both shocks (as implicitly assumed
in almost all previous studies). Otherwise, the derived $\Gamma_0$
(and ${\cal R}_\nu$) can be still derived as a function of the {\em
ratios} of the microphysics parameters. No knowledge about the
absolute values of $\epsilon_e$ and $\epsilon_B$ is needed. 
If the GRB central engine is strongly magnetized, it is natural to
expect that the magnetic field in the reverse shock region could be
stronger than that in the forward shock region. Generally, one can
introduce a free parameter ${\cal R}_B \equiv B_r/ B_f$ into the
problem, which may be solved for along with $\Gamma_0$ if enough
information is available. Case studies suggest that the central engines 
of GRB 990123 and possibly of GRB 021211 are strongly
magnetized\cite{zkm03} (see also Ref. \refcite{kp03}). This provides
an independent clue in addition 
to the gamma-ray polarization\cite{cb03}
(cf. Refs. \refcite{rf03,bc03}) to suggest a magnetized
central engine and fireball, although not necessarily Poynting-flux
dominated. 

For a wind-like environment, the early afterglow lightcurves are
considerably different\cite{kz03b,wdhl03}. Due to the high density of
the shell at the shock crossing time, the reverse shock emission is in 
the fast cooling regime. After shock crossing, the emission above $\nu_c$
disappears. For typical parameters, $\nu_{c,r}(t_\times)$ is well
below the optical band, so that the lightcurve after the crossing time 
is characterized by a steep decay ($\propto t^{-2-\beta}$, $\beta$ is
the spectral index, which is either $1/2$ or $p/2$ for the case we are 
interested in) due to the off-axis angular time delay
effect\cite{kz03b}. This rapidly-decaying reverse shock emission joins 
the forward shock lightcurve later, at a time earlier than the one
found in the ISM case\cite{zkm03}. There is also a simple
correlation between the fluxes and times for the shock crossing and
the crossing of the forward-shock typical frequency across the
band\cite{kz03b}, which is related to $\Gamma_0$ and ${\cal R}_B$ as
well, so that a similar recipe to derive $\Gamma_0$ and ${\cal R}_B$
as in the ISM case\cite{zkm03} can be utilized. When the
self-absorption frequency is above injection frequency in the fast
cooling regime (which is likely in some parameter regimes), balance
between cooling and self-absorption heating 
implies a bump in the electron distribution spectrum, and hence, in
the emission spectrum\cite{kmz03}. This gives rise to interesting
spectral and temporal signatures for the early afterglows that can be
used to diagnose fireball and wind parameters. Such a signature is
also relevant for the case of a constant dense medium\cite{kmz03}.

The significance of $\Gamma_0$ and the forward to reverse field ratio
${\cal R}_B$ in understanding the GRB fireball content and prompt 
emission mechanism has been discussed
in \S\ref{sec:mag} and \S\ref{sec:mech}. In the {\em Swift} era,
an abundance of early afterglow data is expected, and it is desirable 
to systematically analyze these data to retrieve essential fireball 
parameters such as $\Gamma_0$ and ${\cal R}_B$. A follow-up statistical
analysis of these data along with other measurable parameters would
eventually lead to constraints on, or even the identification of, the 
GRB prompt emission site and mechanism\cite{zm02c}.

The early afterglow data may also reveal distinct emission features
from processes such as pair loading\cite{b02} and neutron
decay\cite{b03a}. More detailed studies about these emission
signatures (spectra, lightcurves) are needed in order to differentiate
them from other signatures such as post-injection\cite{zm02a} and the
passage of the forward shock synchrotron injection frequency across
the observational band\cite{kz03a}.

\subsection{Short GRBs and other possible sub-categories}
\label{sec:short}

In a discussion of prospects for future GRB studies, one major
unsolved puzzle concerns the nature of the short, hard bursts, which
comprise of about 1/4 of the total GRB population detected. 
Essentially all the information about (and from) afterglows discussed 
above is for the long bursts. The short bursts still remain as mysterious 
as the long bursts were before 1997. So far only upper limits on the
afterglow emission of a few short bursts are available, although the 
{\em BATSE} archival data contain marginal evidence of weak X-ray
afterglows\cite{lrg01}. Directly scaling the long burst afterglow
models to the short bursts, the calculations are straightforward with
most parameters unchanged except for a smaller total energy (due to the 
short duration if a roughly similar luminosity is assumed), or a smaller
energy per solid angle, if the jets are broader (as might be expected 
e.g. for NS-NS mergers where a large envelope is absent), and possibly 
a lower external medium density (as might be expected for the NS-NS
merger scenario if these wander out of the galaxy). The results indicate 
that the afterglows for short bursts are faint and consistent with the 
current upper limits, and that the afterglows are most easily detected 
in the X-ray band\cite{pkn01}. 

On the theoretical side, simple estimates reveal that NS-NS mergers are
likely to result in central engines with short
durations\cite{npp92,ruffert97,pwf99}. Extensive numerical modeling
has been carried out to reveal how compact merger events can produce
GRBs\cite{rd02,leer02,rrd03}. Significant collimation is required for the
neutrino-driven model in order to account for the detected isotropic-equivalent
luminosity of short bursts (assuming cosmological distances), while a
magnetically-driven model requires fields as high as $10^{17}$ G to
account for the short duration (otherwise, mergers may also produce
long bursts)\cite{rrd03}. There is a suggestion that short bursts may
also originate from collapsars\cite{zwm03}. Other ideas to account
for the long and short bimodal distribution include whether or not strong 
GW losses are incurred in the spin down of a millisecond magnetar 
model\cite{u92}, and whether, in a collapsar scenario, the black 
hole - torus involves hyper-accretion or suspended accretion because of 
the interplay between the disk and the black hole spin\cite{vo01}.
All these models, however, have to be developed further to circumvent the
apparent lack of the afterglows for the short bursts.

{\em Swift}, thanks to its rapid slewing ability, is expected to be
able to catch faint afterglows of some short bursts, if they exist, and
consequently to lead to the identification of their locations,
redshifts as well as other properties such as jet beaming. 
Merger events have been expected to occur in regions with a large 
offset from the host galaxy center (due to the asymmetric kicks during
the formation of the NSs\cite{bsp99}; see, however,
Ref. \refcite{kalogera02}), 
the position information alone relative to the host is already very 
essential for the short bursts' identity and/or merger rate. Furthermore,
ambient density estimates through broadband afterglow modeling will also
help to reveal whether short bursts are located in low-density regions
as expected at least for some fraction of short bursts in the merger models.

Will new sub-categories of GRBs be identified in the {\em Swift} era? 
One can only speculate on this. It may be unlikely to identify new
categories based on one piece of information alone (e.g. burst
duration). However, cross correlations among multiple parameters may
well lead to the identification of new categories, such as XRFs\cite{heise}, 
long-lag bursts associated with the super-galactic plane\cite{norris02}, and
the sub-energetic rapid-decaying bursts\cite{bfk03}. Alternatively, and
equally exciting, it may be that new observational evidence and theoretical 
modeling might reveal that there is a unified paradigm behind all these 
apparently diverse sub-classes. 

\subsection{High energy photon emission\label{sec:GeVTeV}}

A new and so far barely explored window in the electromagnetic spectrum 
is the high energy extension in the GeV-TeV photon energy range,
which holds significant promise for a better understanding of GRB.
After the better studied MeV prompt emission and the broad band low 
energy (X-ray down to radio) afterglow studies, the final (photon) frontier
is the very high and ultra-high energy domain, extending at least up to
TeV, and possibly beyond. In the intermediate range of 10-100 MeV 
observations, while not numerous, have been made with relatively
high significance, e.g. with SMM\cite{hs98}, OSSE\cite{share95}, 
COMPTEL\cite{kippen98}, etc.  In the GeV-TeV energy range, only 
modest to low significance detections of a handful of GRBs have been 
reported, e.g. with the EGRET spark chamber\cite{h94} 
or TASC calorimeter\cite{g03}, or from ground-based air Cerenkov
telescopes\cite{atkins00,GRAND02}. These results have wetted the appetite
for obtaining higher quality data, and have led to a vigorous campaign
of instrumental developments in high energy regimes. Several planned
new missions including {\em GLAST}\cite{glast}, {\em AGILE}\cite{tavani_agile},
and several planned or already operating ground-based air and water 
Cerenkov telescopes (e.g. {\em Milagro}, {\em VERITAS}, {\em HESS}, 
{\em MAGIC} and {\em CANGAROO-II}, etc\cite{weekes00}), are 
expected to open a new era of unveiling the mystery of GRB high energy 
emission. 

On the theoretical side, high energy photons in this energy range
are expected from the leptonic component of the fireball (e.g.
electron IC emission from various emission sites) as well as from
the hadronic component of the fireball (e.g. proton synchrotron, 
$\pi^{+}$ synchrotron emission and $\pi^{0}$ decay from $p\gamma$ 
or $pn,pp$ interactions in various emission sites (see \S\ref{sec:highe} 
for a discussion in the external shock case). Within the standard fireball 
shock scenario sketched in Fig.2, we could in principle have the following 
components emitting in the GeV range and possibly above.

\begin{itemlist}
\item Electron self-IC component from the external forward
shock\cite{mr94,dbc00,zm01b}; 
\item Electron self-IC components from the external reverse shock or
the cross-IC components between the reverse and the forward
shocks\cite{mrp94,wdl01};
\item Electron self-IC component from the internal
shocks\cite{pm96,pl98,gg03d}; 
\item Proton synchrotron emission in the external
shock\cite{vietri97,bd98,totani98,zm01b}; 
\item Photo-meson cascade emission from the external
shocks\cite{bd98,fragile02,da03};
\item Proton synchrotron emission and photo-meson cascade emission
from the internal shocks\cite{rachenm98}; 
\item Cascade emission resulting from $pn$ inelastic collisions during 
the early phase of fireball evolution when the neutron component
decouples from the proton component\cite{dkk99,bm00}; 
\item Baryonic photosphere component (and possibly its Comptonization 
component) extending into the GeV regime (for a low-$\sigma$ high-entropy
fireball)\cite{mr00,t94,mrrz02,zm02c,dm02,gcg03}.  
\end{itemlist}

The above is not an exhaustive list, and not all of the above processes
may be operative at any time or in all models. For example,
if the GRB wind is strongly dominated by a Poynting flux so that 
the GRB prompt emission is not due to internal shocks, the 
internal-shock-related GeV components in the above list would 
be suppressed or absent. Within the collapsar scenario, the precursor
emission tends to be in low energies. But in the supranova scenario,
the presence of the presumed supernova shell and hot pulsar wind
nebula provide copious photon or baryonic targets for additional IC and
hadronic interaction components in the GeV range\cite{wdl02,gg03a,igp03}.
Such emission components may also exist for the type-II collapsars
such that the delay between the SN and GRB is short.

\begin{figure}
\centerline{\psfig{file=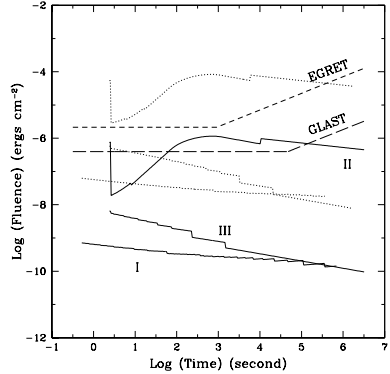,width=5cm}}
\vspace*{8pt}
\caption{Model prediction of GRB high energy (GeV) afterglows as
compared with the GLAST sensitivity. For the IC-dominated parameter
regime, an extended (hour-long) GeV afterglow should be detectable by
GLAST for bursts at typical cosmological distances (from
Ref.186).}
\label{fig:glast}
\end{figure}

An ideal theoretical approach would start with a study of all the above 
emission components within a unified theoretical GRB framework, to
compare the relative importance of each component as well as the parameter 
regimes for each component to dominate. An example of the potential 
benefits of such an approach is provided by the recent observation of
a distinct GeV component identified recently from an archival {\em BATSE} 
burst, GRB 941017\cite{g03}, which evolves separately from the usual
low-energy (sub-MeV) component. So far, only the high energy emission 
components in the external forward shock have been studied in a consistent
manner\cite{bd98,zm01b}. In the external shock scenario, the proton
synchrotron and hadronic cascade emission components are only
important when $\epsilon_B$ is close to equipartition, while
$\epsilon_e$ is very small, i.e. or order of $m_e/m_p$ or below. This
is especially so when the proton power-law index is taken as 2.2-2.3
rather than the special value 2 adopted in most of previous
studies\cite{vietri97,bd98,totani98}. For the more commonly invoked
parameters (e.g. $\epsilon_e \sim 0.1$, $\epsilon_B \sim 0.01$) as
inferred from afterglow studies\cite{pk01,pk02}, the hadronic
components are completely buried beneath the electron IC component,
which forms a distinct bump in the GeV regime. The MeV-GeV lightcurve
predicted within this parameter regime reveals a second broad bump
lasting hours or even a day\cite{zm01b,dbc00}, and the flux level is
detectable by {\em GLAST} for typical bursts at $z\sim
1$\cite{zm01b} (Fig. \ref{fig:glast}). According to this scenario, the
previously detected 
long-lived GeV emission for GRB 940217\cite{h94} is simply a nearby
burst whose GeV afterglow is bright enough to be caught by {\em
EGRET}\cite{mr94,zm01b}, although alternative interpretations have
been also proposed\cite{katz94c,bd99,dl02b}. {\em GLAST} would reveal
whether such kind of long-term GeV emission is common, which would
pose important constraints on GRB shock physical
parameters\cite{zm01b}. 

Another prominent issue is the high energy photon cut-off in the GRB
spectra. During the escape of high energy photons, they are subject to 
absorption through $\gamma\gamma$ interactions with low energy photons 
within the source and in space. These interactions produce electron-positron
pairs and greatly degrade the photon flux level in the original energy 
range received at the detector. At the source, such $\gamma\gamma$ pair 
process would result in a well-defined high energy cutoff\cite{baring} 
that would serve as an important diagnostic of the fireball initial 
Lorentz factor\cite{ls01}. Another consequence of this process is that
the produced pairs are expected to synchrotron radiate again within the
local magnetic fields. Depending on the compactness of the emission
region, the secondary emission process may modulate the emergent
spectrum under certain conditions and possibly smear out the initial
distinct spectral features\cite{pl98,mrrz02}. A self-consistent
pair-modulated numerical model for the GRB prompt emission is needed
for this. Even if a high energy photon is not absorbed, it may be scattered
by the resulting pairs and even by the electrons associated with the
fireball baryons\cite{ls01}. Furthermore, if the source is not compact
enough so that the TeV photons can escape, they may be still absorbed
by the cosmic IR background\cite{des02}. The mean free path of the TeV 
photon depends on both the GRB and the IR background models. Typically a TeV 
source cannot be detected beyond $z\sim 0.1$\cite{des02} so that
the Milagrito event for GRB 970417\cite{atkins00}, if real, has to
come from a nearby source whose compactness is not high (e.g. from
external shocks, or with a high Lorentz factor). For typical
cosmological sources, TeV emission will be almost completely absorbed by 
the IR background. The resultant pairs, if within a not-too-strong
intergalactic medium (IGM) magnetic field, would IC up-scatter the
cosmic microwave background photons to produce a delayed, GeV emission 
component\cite{plaga95,cc96,dl02b}. A well-measured spectrum may be
used to infer the poorly known strength of the IGM magnetic field (see 
Ref. \refcite{dzgmw02} for a detailed modeling within the context of
blazars).

\subsection{Ultrahigh energy cosmic rays\label{sec:UHECRs}} 

Two important non-electromagnetic channels in which GRB may be prominent
sources are cosmic rays and neutrinos, neither of which have so far
been measured. GRB models involving shocks as sites to accelerate 
electrons which produce prompt gamma-rays and long-term afterglows 
naturally suggests that baryons (most likely protons) should be 
accelerated by the same shocks as well. These accelerated ions, 
if not bound in the system and not destroyed during their propagation, 
would arrive at Earth and be detected as cosmic rays. Those trapped 
within the system would interact with photons and other baryons to produce 
high energy neutrinos that might also be detected from Earth.

Discussions about GRBs as cosmic ray accelerators are mainly aimed
at explaining the portion of the cosmic ray spectrum above $10^{18}$
eV, the ``ankle'' region where the spectrum starts to harden, and in
particular the ultrahigh energy cosmic rays (UHECRs) above $10^{20}$ eV,
also referred to as Greisen-Zatsepin-Kuzmin ("GZK") cosmic rays.
The isotropic distribution of their arrival directions and small
magnetic deflection they 
suffer at these high energies suggest their extra-galactic origin, 
and the requirement that they must survive the attenuation by the 
cosmic microwave background through photomeson interaction constrains 
their distances to radius of about 50-100 Mpc, the so-called ``GZK''
volume\cite{g66,zk66}. Two broad classes of models have been suggested 
to interpret the UHECRs, the ``top-down'' scenarios that attribute
UHECRs to decay products of fossil Grand Unification defects, and the
``bottom-up'' scenarios that suggest UHECRs are hadrons accelerated in
astrophysical objects to these high energies. Among the few serious 
viable candidates in the bottom-up scenario, GRBs (and/or perhaps AGNs) 
are considered to be likely sources which may fulfill the constraints 
on the known data\cite{w95,v95,mu95}. 
Two sub-scenarios of the bottom-up GRB cosmic ray origin have been 
suggested, one involving acceleration in the internal 
shocks\cite{w95,w95b,w02} and one involving acceleration in the 
external shock\cite{v95,vdg03}. The suggestion is based on two 
coincidences, i.e., the shock conditions required to accelerate protons
to $\sim 10^{20}$ eV are similar to the conditions required for generating
the observed prompt gamma-rays, and the observed UHECR energy injection 
rate into the universe ($\sim 3\times 10^{44} ~{\rm erg~Mpc^{-3}~yr^{-1}}$) 
is similar to the local GRB energy injection rate\cite{w95,v95}. Both 
coincidences have been questioned on various
grounds\cite{ga99,a01,ob02,stecker00,ss00}, but these have been met
by effective counter-arguments, using new data and additional considerations 
for both the internal shock\cite{w02} and the external shock\cite{vdg03}
scenarios. At this time, GRBs are, and remain, a promising candidate 
for the UHECR origin. However, both the internal shock and the external 
shock scenarios have major caveats, which have nothing to do with their
ability to accelerate cosmic rays but rather with the generic shock 
itself. The internal shock scenario relies on the assumption that the GRB 
prompt emission is due to synchrotron from electrons accelerated in 
internal shocks. Although this is the leading scenario, there is no 
direct proof so far, unlike in the case of the external shock, where this
origin of the radiation is quite convincing. For example, a 
Poynting-flux-dominated GRB model would be incompatible with an internal 
shock origin of both synchrotron $\gamma$-rays and UHECRs. 
The external shock model, on the other hand, may have to rely on a 
magnetized external medium\cite{vdg03} (as expected in pulsar wind 
bubbles\cite{kg02} in the supranova scenario\cite{vs98}) in order to 
reach the desired cosmic ray energy. On the other hand, the radiation 
and acceleration responsible for $\gamma$-rays is itself unexplored
in Poynting-dominated scenarios, and the need for such a scenario is
at the moment not proven, while the supranova scenario is incompatible
with the observations of GRB030329/SN2003dh. The uncertainties surrounding
possible alternative scenarios are significant, and the standard shock 
scenario remains the most amenable to quantitative modeling and testing, 
which should be able to provide useful constraints on its ability to 
explain the data.

A direct proof of the GRB-origin of UHECRs is not easy. The next generation
of cosmic ray detectors, such as the {\em Auger Observatory}, will have a
substantially enhanced effective target area, which will greatly
improve the cosmic ray count statistics. It also combines elements of
the two techniques which currently lead to different results, namely 
air fluorescence telescopes and water Cerenkov surface tanks. This
will help to disentangle the two distinct scenarios (top-down or bottom-up) 
and to reveal whether a GZK feature indeed exists. Within the bottom-up 
scenario, the direction information may prove or significantly constrain 
the AGN model (the close competitor of the GRB model) and eventually shed
light on whether GRBs are indeed the sources of UHECRs.

\subsection{High energy neutrinos\label{sec:nu}} 
Regardless of whether GRBs can accelerate protons to ultrahigh
(GZK) energies, they must be able to accelerate protons to some high
energies. The implication is that high energy neutrinos and high
energy photons (as discussed in \S\ref{sec:GeVTeV}) must accompany
the current emission seen at sub-MeV energies. Widely discussed processes 
for high energy neutrino emission include
\begin{itemlist}
\item $p\gamma$ process: $p\gamma \rightarrow \Delta^{+} \rightarrow
n\pi^{+} \rightarrow ne^{+}\nu_e \bar\nu_\mu \nu_\mu$;
\item $pp$ process: $pp \rightarrow \pi^{\pm}/K^{\pm} \ldots
\rightarrow \mu \nu_\mu \ldots \rightarrow e\nu_e \bar\nu_\mu \nu_\mu
\ldots$; 
\item $pn$ process: $pn \rightarrow \pi^{\pm}/K^{\pm} \ldots
\rightarrow \mu \nu_\mu \ldots \rightarrow e\nu_e \bar\nu_\mu
\nu_\mu\ldots$ 
\end{itemlist}
The dominant $p\gamma$ process occurs at the $\Delta$-resonance, which 
has the threshold condition $\epsilon_p \epsilon_\gamma \simg 0.3
~{\mbox{GeV}^2}$ in the center of mass frame. In the case of GRBs,
this is usually translated in the observer's frame to $\epsilon_p
\epsilon_\gamma \simg 0.3 ~{\mbox{GeV}^2} \Gamma^2$ when both protons
and photons are generated in relativistic shocks (e.g. in the internal 
or external shock scenarios), or to $\epsilon_p \epsilon_\gamma \simg 0.3
~{\mbox{GeV}^2} \Gamma$ when the photons are generated in SN shells
while protons are accelerated from relativistic shocks (e.g. in the
supranova model), where $\Gamma$ is
the bulk Lorentz factor. The threshold condition for both $pp$ and
$pn$ interactions is that the relative drift energy between these
baryons exceed the pion rest mass, i.e., $\epsilon' \geq 140$ MeV. 
Since both $p$ and $n$ have a rest mass close to 1 GeV, the threshold
of $pp$ and $pn$ interaction only demands semi-relativistic relative
motions.

In a GRB event, there are multiple sites where neutrinos with different
energies are generated. Below is an non-exhaustive list which
encompasses most of the processes discussed in the literature, in
a sequence of ascending neutrino energy:
\begin{itemlist}
\item MeV neutrinos: If GRBs are originated from stellar collapses,
they should be associated with strong thermal MeV neutrinos like
supernovae. In most models, the collapse results in a black hole -
torus system, and the thermal neutrino annihilation is one of the 
leading processes for launching the fireball. However, these thermal
neutrinos are extremely difficult to detect from cosmological
distances, due to the very low cross section for $\nu$N interactions
at these energies (the cross section increases $\propto E_\nu^2$ 
at lower energies, and $\propto E_\nu$ at higher energies). 
\item multi-GeV neutrinos: GRB fireballs may be
neutron-rich\cite{dkk99,b03b}. During the fireball acceleration phase, 
neutrons can decouple from protons when the elastic scattering
condition breaks down. The relative drift between both species results 
in inelastic $pn$ interactions giving rise to 5-10 GeV
neutrinos\cite{bm00}. A similar process also occurs within
sub-photospheric internal shocks\cite{mr00b}, which extends
significantly the parameter space for the inelastic neutrino collision 
condition. It was suggested earlier\cite{px94} that $pp$ interactions 
within the internal shocks can also give rise to a 30 GeV neutrino 
burst, although magnetic fields can inhibit the inter-penetration
of charged species streams with different velocities;
\item multi-TeV neutrinos: Within the collapsar scenario, the
relativistic jet launched from the base of the flow (presumably the black
hole and the torus) has to penetrate through the stellar envelope
before breaking out and generating the GRB. The internal shocks below 
the envelope accelerate protons that interact with thermal photons within
the envelope (i.e. $p\gamma$ interaction). Regardless of whether the
jet finally penetrates through the envelope or gets choked, it will
generate strong multi-TeV neutrino signals\cite{mw01}. The signature
is enhanced or even dominated by $pn,pp$ interactions and could be
used as a diagnostic about the type of progenitor stars\cite{rmw03b};
\item PeV neutrinos: $p\gamma$ interactions within the conventional 
internal shocks which produce prompt gamma-rays typically
generate $10^{14}-10^{16}$ eV neutrinos\cite{wb97,gsw01b}; 
\item EeV neutrinos: $p\gamma$ interactions within the external
reverse shock give rise to even higher energy neutrinos. For a
constant density medium the typical energy is $10^{17}-10^{19}$
eV\cite{wb00}, while for a wind medium, the typical energy is $\sim
3\times (10^{15}-10^{17})$ eV and extending above it\cite{dl00,rmw03c}. 
In the forward shock region, assuming the blast wave can accelerate 
protons to ultrahigh energies\cite{v95,vdg03}, a neutrino afterglow is 
expected with the peak energy $\sim 10^{18}$ eV\cite{dermer02b,ldl02}.
\end{itemlist}

In the supranova progenitor scenario, the existence of the
pre-supernova shell provides extra targets for additional neutrino
signals. These signals are broad-band with a distinct 1-10 TeV 
feature with high count rates, which is a useful clue to test the
supranova hypothesis\cite{rmw03,gg03b,gg03c,da03}. The neutrino model
predictions for GRB 030329 indicate that a 2-day delay supranova model
predicts a one order of magnitude higher neutrino event rate in the
TeV-PeV regime than the conventional burst, a signal which is detectable 
by the full ICECUBE as an individual source at the observed redshift $z=0.17$.
If it were to be detected by under-ice neutrino detectors at the South Pole 
or underwater detectors such as Antares, this would further
significantly constrain 
the validity of the supranova model\cite{rmw03c}. 
\begin{figure}
\centerline{\psfig{file=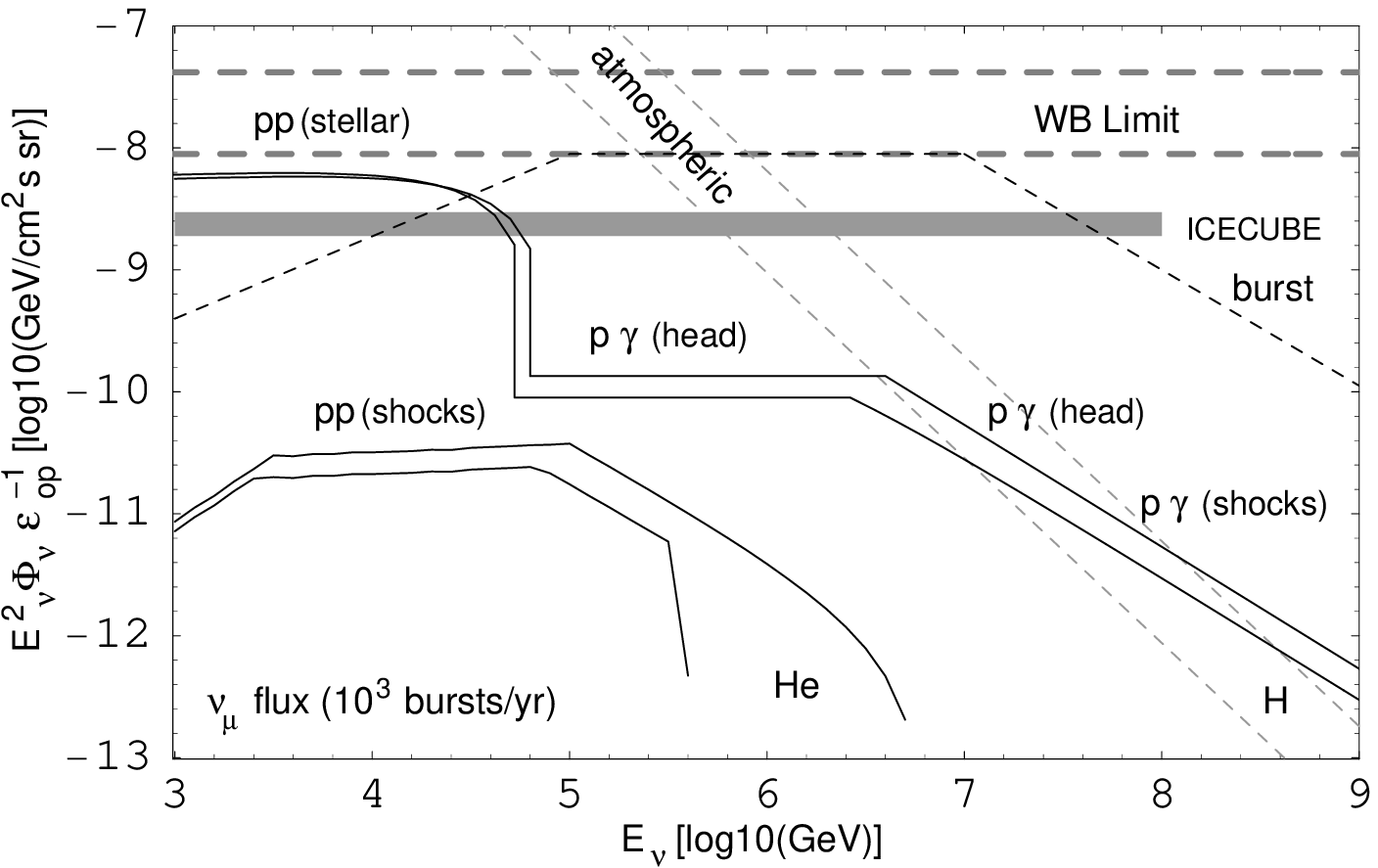,width=5cm}
            \psfig{file=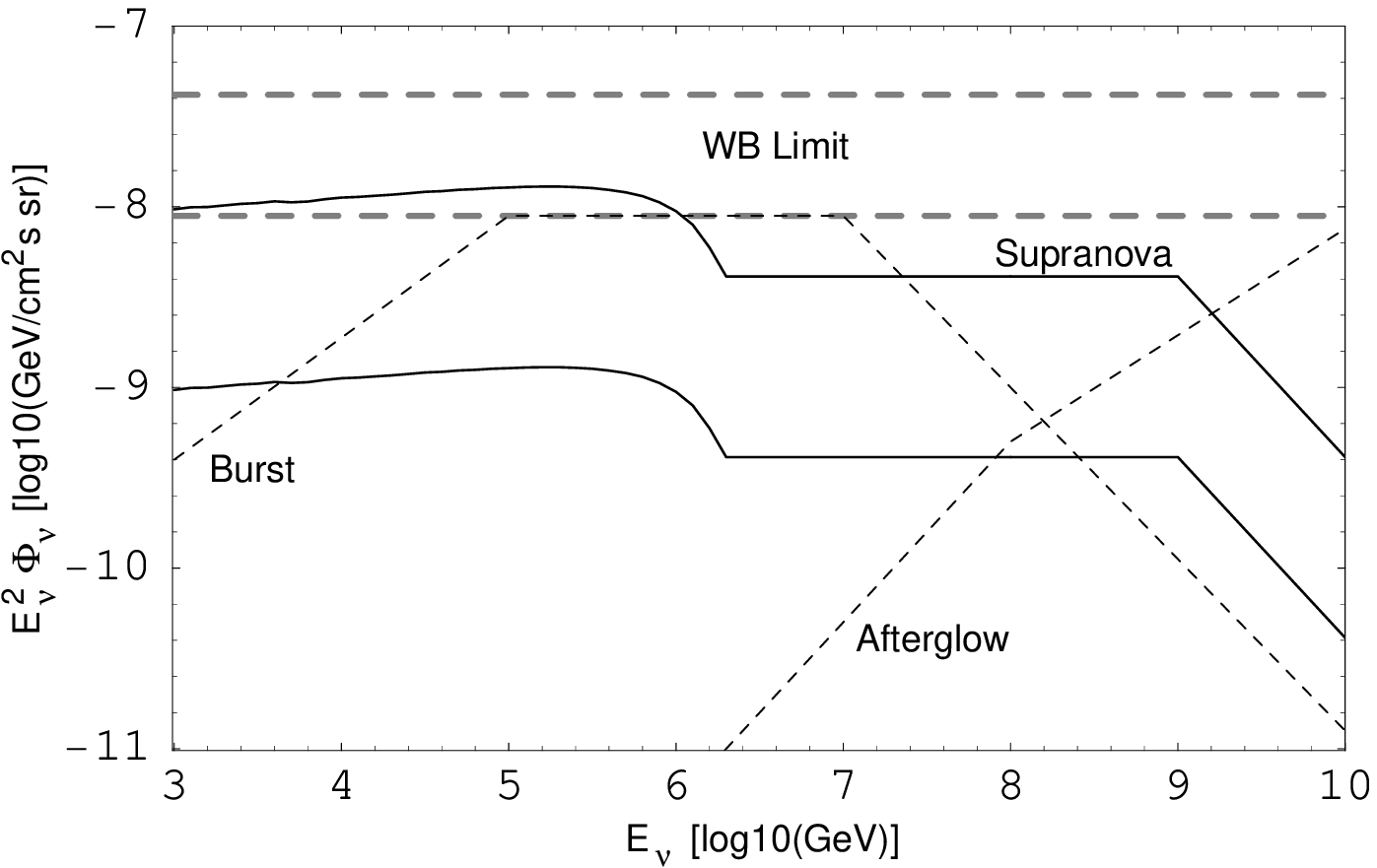,width=5cm}}
\vspace*{8pt}
\caption{The diffuse muon neutrino flux for both the collapsar and the
supranova models (from Refs.463,527).}
\label{fig:soeb}
\end{figure}

All the above are theoretical predictions. So far there is no report about 
the solid detection of neutrino signals from GRBs. Current and future
ice or water Cerenkov neutrino detectors, {\em AMANDA-II}, {\em
ANTARES}, and {\em ICECUBE} will make it possible to detect GRBs from
this new channel. If detected, neutrinos from GRBs could be also
used to study neutrino physics itself as well as to have ramifications for
cosmology\cite{weiler94}.

\subsection{Gravitational waves\label{sec:gw}}

Binary compact-object mergers, such as NS-NS, NS-BH, BH-BH mergers,
as well as BH-WD, BH-Helium star etc., have long been considered
as possible sources of gravitational waves 
(GW)\cite{thorn87,phinney91,cutler93,kp93,ruffert97,janka99,km03}, as 
have core-collapse events\cite{rmr98,fryer02,davies02,km03,van01b,van02,van03}.
This has been regardless of whether they could produce GRBs. However,
since these events are also the leading candidates for being GRB
progenitors, it is natural to expect that a GRB is associated with a GW
burst (although the contrary may not hold). For a selection of
GRB-related GW literatures, see e.g. Ref. \refcite{gwgrb}. A
coincidence between a GW 
signal and a gamma-ray signal would greatly enhance the statistical
significance of the former\cite{fmr99}, making searching for
GRB-associated GW bursts of great interest. A binary coalescence
process can be divided into three phases, i.e., in-spiral, merger, and 
ring-down\cite{fh98,km03}. For collapsars, a rapidly rotating core could 
lead to development of a bar and to fragmentation instabilities that
would resemble similar GW signals as in the binary merger scenarios,
although a larger uncertainty is involved since the fragment masses
and coherence times are unknown. The GW frequencies of the
various phases cover the $10-10^3$ Hz band which is relevant for the
{\em Laser Interferometer Gravitational-wave Observatory (LIGO)} and
other related detectors such as {\em VIRGO}, {\em GEO600} and {\em
TAMA300}\cite{finn}. Because of the faint nature of the typical GW strain, only
nearby sources (e.g. within $\sim 200$ Mpc for NS-NS and NS-BH
mergers, and within $\sim 30$ Mpc for collapsars)\cite{km03} have
strong enough signals to be detectable by {\em LIGO}. When event rates 
are taken into account\cite{fwh99,bbr02}, order of magnitude estimates
indicate that after one-year of operation of the  advanced {\em LIGO}, 
one in-spiral chirp event from a NS-NS or NS-BH merger, and probably
one collapsar event (subject to uncertainties) would be
detected\cite{km03,van03}. Other binary merger scenarios such as 
BH-WD and BH-Helium star mergers are unlikely to be detectable\cite{km03}, 
and they are also unfavored as sources of GRBs according to other
arguments\cite{npk01}. 
\begin{figure}
\centerline{\psfig{file=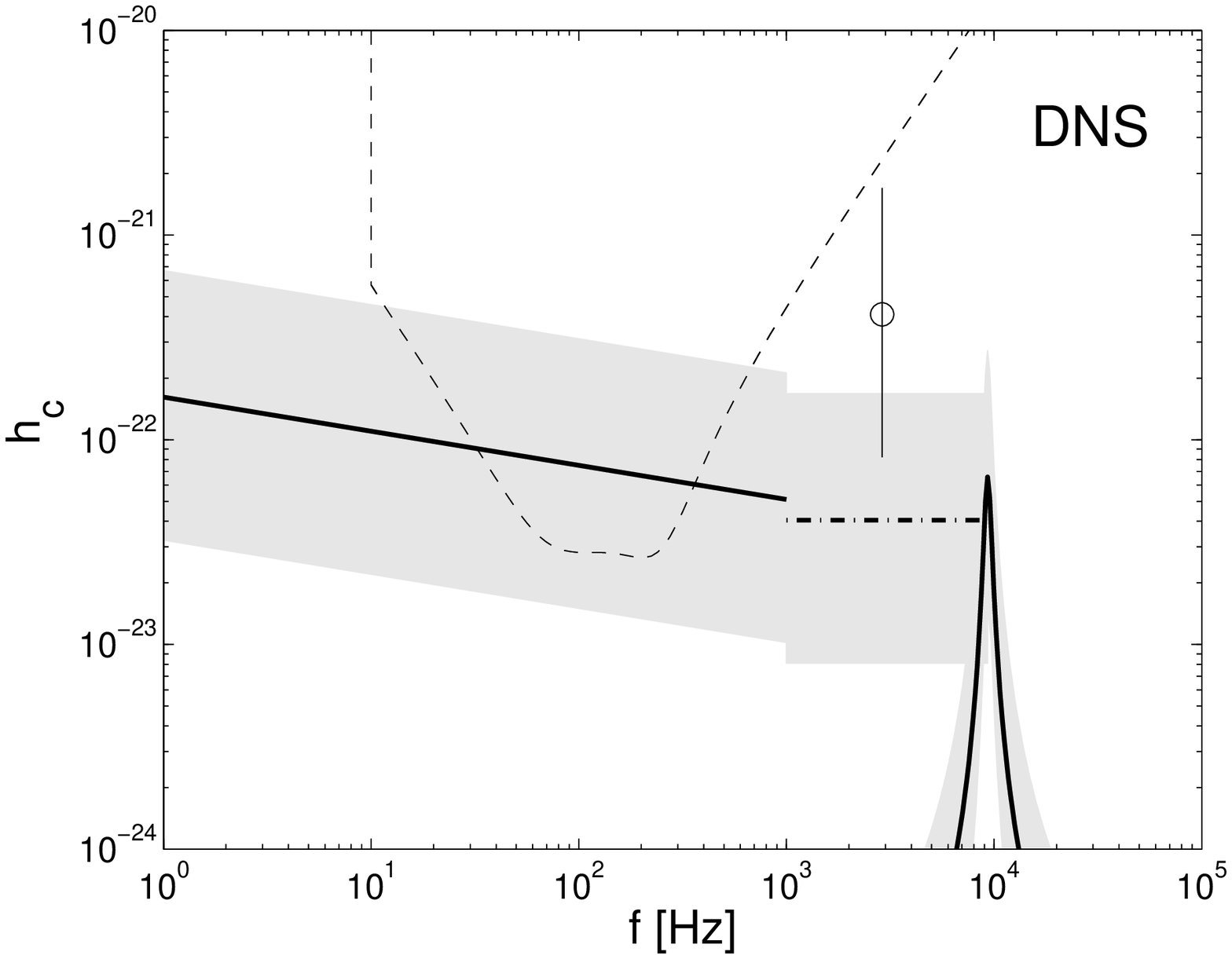,width=5cm}
            \psfig{file=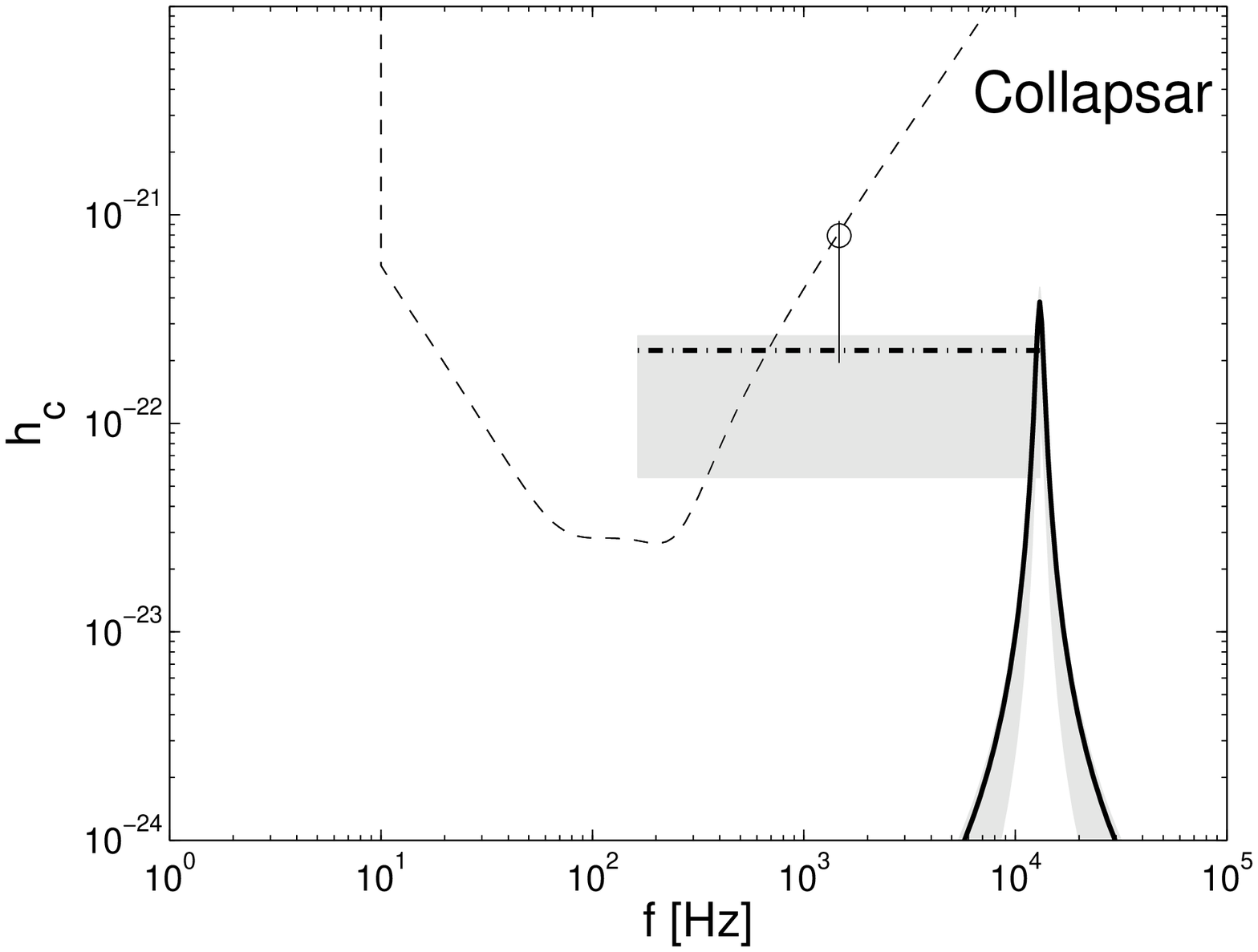,width=5cm}}
\vspace*{8pt}
\caption{The gravitational wave strains for the merging double neutron
star model and the collapsar model as compared with the sensitivity of
the advanced LIGO (from Ref.174).}
\label{fig:shiho}
\end{figure}

The detection of a GRB-GW burst association would have profound
implications on GRB studies, and could shed light on several unsettled 
issues as discussed in \S\ref{sec:prob}. First, it would help to identify
the GRB progenitor. GW signals in the advanced {\em LIGO} band are 
different for the merger and the core collapse scenarios, so that a
coincident GRB-GW detection would unambiguously differentiate both
scenarios. Furthermore, two GW bursts are expected for the supranova
model, with the second one coincident with the GRB. Either detection
or non-detection of the precedent GW signal would provide additional
information on the validity of the supranova model (in addition to 
the negative evidence provided by GRB030329/SN2003dh). Second, it
could help to settle the debate about the location of GRB prompt
emission (i.e. external or internal, \S\ref{sec:site}). Because the GW
wave is associated with the formation of the central engine (and
therefore the launch of the GRB jet), a short time delay ($\siml 0.1$ s)
for the GRB emission with respect to the GW emission would favor
an internal scenario (e.g. the internal shock model), while a longer
delay ($\sim 10-100$ s) would favor an external model (e.g. the
external shock model)\cite{fwh99,fks03} (this however ignores the 
delay incurred by the jet in punching though the stellar envelope, which
could itself be of order 10-100s); Third, it would help to pin down
the GRB jet configuration. It has been suggested that the GW
polarization information could be used to infer geometric information 
such as the angle between the rotation axis (presumably the center of 
the jet) and the line-of-sight direction\cite{km03b}. Such information, 
when combined with other information from prompt gamma-ray and afterglow, 
could help to confirm whether and how the GRB jets are 
structured\cite{rlr02,zm02b,zdlm03}.

\subsection{GRBs \& cosmology\label{sec:cos}}

One of the most exciting prospects for future GRB studies is 
that they could become a unique tool to investigate the high
redshift universe. Although this possibility is currently mainly a 
theoretical expectation, the prospects look bright. The optimism 
is underpinned by two aspects of the evidence obtained so far.

Clearly, since GRBs are stellar events, some GRBs are expected to 
exist at (much) higher redshifts than $z=6.4$, the current high 
redshift record held by a SDSS quasar\cite{fan03}. This is supported
by several lines of reasoning. (1) Preliminary polarization data on the
cosmic microwave background collected by {\em Wilkinson Microwave 
Background Probe (WMAP)} indicate a high electron scattering optical depth, 
hinting that the first stellar objects in the universe should have formed as
early as $z\sim 20$\cite{spergel03}; (2) Independent theoretical
simulations of the formation of the first stars similarly conclude that
these should have formed at redshifts $z \sim (15-40)$\cite{abel02,bromm02}. 
Because there is convincing evidence that at  least long GRBs are 
associated with the deaths of the massive stars, it is conceivable that 
high-$z$ GRBs ($z \siml 15-20$ or even higher) exist. In fact, although 
subject to considerable model uncertainties, theoretical modeling of GRB 
redshift distribution based on the standard $\Lambda$CDM universe suggests 
that $\simg 50\%$  of all GRBs on the sky originate at $z \simg 5$\cite{bl02};
(3) Using empirical luminosity laws such as  gamma-ray variability or
spectral lag correlations with fluence, rough redshifts can be derived for 
a large sample of the BATSE bursts, and it is found that a good fraction of 
these bursts have redshifts in excess of $z\sim 6$, and some even in excess
of 10\cite{frr00,schaefer01,lloyd02b}.

The emission properties of GRBs and afterglows offer several unique 
advantages for studying the high-$z$ universe (see also
Ref. \refcite{loeb03}). 
(1) The luminosities of the prompt gamma-ray\cite{lr00} and the 
afterglows\cite{cl00,gou03} do not fade rapidly with increasing redshifts 
(unlike for quasars). In fact, the highest-redshift burst identified so far 
(GRB 000131 at $z=4.5$) does not show a substantially fainter luminosity 
in both the prompt emission and the afterglow. This is due to a favorable 
combination of the cosmological time-dilation and the flux decay resulting in
a positive K-correction effect. For afterglows, the high-$z$ bursts are 
especially favorable for detection in the IR band, thanks to the
bright reverse shock emission at early epochs\cite{gou03}; (2) GRB
afterglows have intrinsically high luminosities ($10^{51}-10^{54}~{\rm
erg~s^{-1}}$). Even at moderate redshifts, they already greatly
out-shine the host galaxies. This would especially be the case at higher
redshifts, as the host galaxies are expected to become even fainter;
(3) The intrinsic power-law spectrum of GRB afterglows greatly
simplifies the extraction of the IGM absorption features.

The implications of high-$z$ GRB studies for cosmology include the
following. (1) {\em WMAP} only provides an integral constraint on the
reionization history of the universe. Since both the number density
and the intrinsic luminosity of quasars are expected to fade rapidly
beyond $z \sim 6$, only GRBs and afterglows may be able to  act as bright 
beacons to illuminate the end of cosmic dark age\cite{bl01,loebb01,m-e03}, 
and to probe the reionization history of the early universe\cite{iyn03};
(2) Although the apparent GRB luminosity is by no means a standard
candle, the beaming-corrected GRB luminosity appears to be
standard\cite{f01,pk01,bkf03,bfk03}. Also empirical luminosity
laws\cite{frr00,norris00}, if confirmed, could be used as Cepheid-like 
correlations. These raise the possibility of using GRBs to derive a
Hubble-diagram and to constrain the cosmological
parameters\cite{bfk03,sch03b}, although the current data are too
sparse to draw a firm conclusion; (3) Preliminary evidence on the
evolution of GRB properties has been suggested\cite{lloyd02b,wg03}. If 
this is confirmed by future photometric redshift measurements of a
large amount of GRBs (with selection effects properly taken care of),
the data could be used to infer the possible redshift-evolution of the 
GRB progenitors; (4) Future extensive afterglow monitoring for many
bursts would help to constrain the local environments of GRBs as well
as their redshift evolution. In particular, it would tell whether GRB
afterglows are decelerated by the IGM (with an increasingly high density
at higher redshifts) or by a stratified constant medium bubble cleared 
by the progenitor star\cite{gou03}; (5) Insights into cosmological
structure formation and star forming history would be gained through
studying distributions of the GRB host galaxies\cite{mm98}.

A great challenge in the {\em Swift} era is to get direct redshift
information for high-$z$ bursts. The most straightforward approach is
to search for the Ly$\alpha$ break (at 1.216$\mu$m$[(1+z)/10]$) and the
Gunn-Peterson trough\cite{gp65} in the IR band. However, since the 
{\em Swift} UVOT wave-band extends only to $\sim 0.6 \mu$m in the
long wavelength regime, the redshifts of GRBs with $z>5$ can not be 
directly identified with {\em Swift}. This poses however a great opportunity
for ground-based IR cameras to follow up {\em Swift}-triggered GRBs within 
a short period of time. Several IR follow-up teams have developed
such instruments and plan to carry out {\em Swift} follow-up
observations in a timely manner. Besides IR follow-up, it is also possible
that X-ray lines could become important distance indicators at high
redshifts\cite{mr03}. Alternatively,
redshifts for GRBs whose radio afterglows are unusually bright may be
identified with 21 cm absorption features\cite{cgo02,fl02}.

Finally, using cosmological GRB high energy data, one may constrain
a number of ideas in fundamental physics, e.g. related to quantum gravity 
effects, Lorentz invariance violations, etc.\cite{last}. 

\section{Summary}

Our understanding of GRBs has been greatly advanced since their
discovery about 30 years ago. Extensive observational efforts have 
revealed a rich phenomenology of both prompt emission and afterglows
(\S\ref{sec:prog1}), and a successful theoretical framework has been
set up, which is able to interpret most of the observational data so far
(\S\ref{sec:prog2}). However, there remain many questions at the
present stage of GRB studies, which greatly stimulate further
observational and theoretical efforts (\S\ref{sec:prob}). Some of these 
questions will undoubtedly be addressed in the coming years with the 
advent of a number of upcoming space- and ground-based experiments,
both in the electromagnetic and non-electromagnetic channels
(\S\ref{sec:pros}). As suggested by past experience, new challenges
and surprises are bound to emerge, which will stimulate further 
extensive observational campaigns and theoretical efforts. The GRB field 
is likely to remain one of the most active fields in contemporary
astrophysics in the next decades. 

This research was partly supported through NASA ATP NAG5-13286, 
NSF AST0098416 and the Monell Foundation. We thank S. Kobayashi, T. Lu, 
D. A. Frail, J. P. Norris, E. D. Feigelson, Z. G. Dai, Y. F. Huang, and 
S. Razzaque for helpful comments.

% %%%%%%%%%%%%%%%%%%%%%%%%%%%%5

\end{document}